\documentclass{ieeeaccess}

\usepackage[T1]{fontenc}
\usepackage{lmodern}

\usepackage{cite}
\usepackage{amsmath,amssymb,amsfonts}
\usepackage{algorithmic}
\usepackage{graphicx}
\usepackage{textcomp}
\usepackage[table,dvipsnames]{xcolor}
\usepackage{braket}
\usepackage{booktabs}
\usepackage{subcaption}
\usepackage{url}
\usepackage[hidelinks]{hyperref}

\def\BibTeX{{\rm B\kern-.05em{\sc i\kern-.025em b}\kern-.08em
  T\kern-.1667em\lower.7ex\hbox{E}\kern-.125emX}}

\begin{document}
\history{Date of publication xxxx 00, 0000, date of current version xxxx 00, 0000.}
\doi{10.1109/TQE.2020.DOI}

\title{{A Systematic Study of Noise Effects in Hybrid
Quantum-Classical Machine Learning}}
\author{\uppercase{Bhavna Bose}\authorrefmark{1}, \IEEEmembership{Member, IEEE}, and
\uppercase{Dr. Muhammad Faryad \authorrefmark{2}},
\IEEEmembership{Member, IEEE}}
\address[1]{Department of Information Technology,SVKM's NMIMS Mukesh Patel School of Technology Management and Engineering,Mumbai, India (email: bhavna.bose@nmims.edu)}
 \address[2]{Department of Physics, Lahore University of Management Sciences, Lahore, Pakistan (email: muhammad.faryad@lums.edu.pk)}

\markboth
{Author \headeretal: Preparation of Papers for IEEE Transactions on Quantum Engineering}
{Author \headeretal: Preparation of Papers for IEEE Transactions on Quantum Engineering}

\corresp{Corresponding author: Bhavna Bose (email: bhavna.bose@nmims.edu).}

\begin{abstract}
Near-term quantum machine learning (QML) models operate in environments wherein noise is unavoidable, arising from both imperfect classical data acquisition and the limitations of noisy intermediate-scale quantum (NISQ) hardware. Although most existing studies have focused primarily on quantum circuit noise in isolation, the combined influence of corrupted classical inputs and quantum hardware noise has received comparatively little attention. 

In this work, we present a systematic experimental study of the robustness of a variational quantum classifier under realistic multi-level noise conditions. Using the Titanic dataset as a benchmark, a range of dataset-level noise models-including speckle noise, impulse noise, quantization noise, and feature dropout are applied to classical features prior to quantum encoding using a ZZ feature map~\cite{Havlivcek2019FeatureMaps}. In parallel, hardware-inspired quantum noise channels such as depolarizing noise, amplitude damping, phase damping, Pauli errors, and readout errors are incorporated at the circuit level using the Qiskit Aer simulator~\cite{QiskitAer}. 

The experimental results indicate that noise in classical input data can significantly intensify the effects of quantum decoherence, resulting in less stable training and noticeably lower classification accuracy. Together, these observations emphasize the importance of designing and evaluating quantum machine learning pipelines with noise in mind, and highlight the need to consider classical and quantum noise simultaneously when assessing QML performance in the NISQ era~\cite{Preskill2018NISQ}.
\end{abstract}


\titlepgskip=-15pt

\maketitle
\begin{abstract}
Near-term quantum machine learning (QML) models operate in environments wherein noise is unavoidable, arising from both imperfect classical data acquisition and the limitations of noisy intermediate-scale quantum (NISQ) hardware. Although most existing studies have focused primarily on quantum circuit noise in isolation, the combined influence of corrupted classical inputs and quantum hardware noise has received comparatively little attention. 

In this work, we present a systematic experimental study of the robustness of a variational quantum classifier under realistic multi-level noise conditions. Using the Titanic dataset as a benchmark, a range of dataset-level noise models-including speckle noise, impulse noise, quantization noise, and feature dropout are applied to classical features prior to quantum encoding using a ZZ feature map~\cite{Havlivcek2019FeatureMaps}. In parallel, hardware-inspired quantum noise channels such as depolarizing noise, amplitude damping, phase damping, Pauli errors, and readout errors are incorporated at the circuit level using the Qiskit Aer simulator~\cite{QiskitAer}. 

The experimental results indicate that noise in classical input data can significantly intensify the effects of quantum decoherence, resulting in less stable training and noticeably lower classification accuracy. Together, these observations emphasize the importance of designing and evaluating quantum machine learning pipelines with noise in mind, and highlight the need to consider classical and quantum noise simultaneously when assessing QML performance in the NISQ era~\cite{Preskill2018NISQ}.
\end{abstract}


\section{Introduction}
\label{sec:Introduction}
Quantum Machine Learning (QML) applies quantum computing to enhance learning tasks by exploiting quantum feature representations and transformations that are challenging to reproduce using classical methods~\cite{Biamonte2017QML}. Among the various QML approaches, variational quantum algorithms are more popular, because they integrate parameterized quantum circuits with classical optimization routines, making them well suited for execution on Noisy Intermediate-Scale Quantum (NISQ) hardware.

Real-world learning pipelines rarely rely on perfectly clean classical data: measurement errors, quantization, missing values, and transmission faults are common. However most QML robustness studies have focused on circuit noise alone, leaving the \emph{combined} effect of classical input corruption and quantum noise underexplored. This paper addresses this gap through a unified robustness study that injects noise at the dataset, data encoding, and circuit levels.

Variational Quantum Algorithms (VQAs), including Variational Quantum Classifiers (VQCs), operate by training parameterized quantum circuits using classical optimization loops. This hybrid approach was first demonstrated effectively in the context of the Variational Quantum Eigensolver (VQE), which showed that useful quantum computation is achievable even on noisy hardware~\cite{Peruzzo2014VQE}. Subsequent theoretical work established the foundations of hybrid quantum-classical algorithms, providing insights into their expressivity, optimization behavior, and training dynamics under realistic constraints~\cite{McClean2016Theory}. 

Current quantum processors operate firmly within the Noisy Intermediate-Scale Quantum (NISQ) regime. The computations in the NISQ era are affected by decoherence, gate imperfections, limited qubit counts, and imperfect readout~\cite{Preskill2018NISQ}. In this setting, noise is not a secondary concern but a fundamental feature of near-term quantum computation. For variational models such as VQCs, these hardware imperfections can distort optimization landscapes, hinder parameter training, and significantly degrade classification performance.

Real-world machine learning pipelines rarely have access to perfectly clean data because of measurement errors, sensor noise, quantization artifacts, missing values, and transmission faults. Inspite of this, most existing QML studies implicitly assume noise-free classical inputs and focus almost exclusively on the quantum circuit-level noise. This simplifying assumption limits the practical relevance of such analyses, because it neglects the compounded impact of classical data corruption and quantum hardware noise which are likely to arise in realistic deployment scenarios.

This work presents a unified robustness analysis of variational quantum classifiers under both dataset and quantum circuit noise. By systematically introducing controlled perturbations at multiple stages of the learning pipeline, we aim to bridge the gap between idealized QML models and noisy conditions encountered in practical applications.

In the NISQ regime, noise is not a secondary limitation but a defining characteristic of near-term quantum computation, fundamentally constraining algorithmic depth, accuracy, and scalability~\cite{Preskill2018NISQ}.

The primary contributions of this work are summarized as follows:
\begin{itemize}
    \item A unified hierarchical noise modeling framework that jointly captures dataset-level noise, encoding-level perturbations, and circuit-level quantum noise within a single hybrid QML pipeline.
    \item An empirical investigation of how noise propagates from corrupted classical features through quantum feature encoding to variational circuit execution.
    \item A comprehensive robustness evaluation of variational quantum classifiers across diverse noise regimes is representative of both real-world data corruption and the NISQ-era hardware limitations.
\end{itemize}

The remainder of this paper is organized as follows: Section~\ref{sec:RelatedWork} discusses the related work in this area, Section~\ref{sec:Methodology} describe the methodology and the noise modeling framework; Section~\ref{sec:ExprimentalSetup} presents the experimental setup; Section~\ref{sec:ResultsAndAnalysis} discusses the results and analysis; and Section~\ref{sec:conclusion} concludes with future directions.

Unlike prior robustness studies that isolated either classical data noise or quantum hardware noise, this study provides the first systematic, hierarchical evaluation of their combined effects within a single variational QML pipeline
\section{Related Work}
\label{sec:RelatedWork}
Noise and error mechanisms have been widely studied in the context of near-term quantum computing, particularly for variational algorithms that operate on Noisy Intermediate-Scale Quantum (NISQ) devices. Early experimental work on the Variational Quantum Eigensolver (VQE) demonstrated that hybrid quantum--classical algorithms can produce meaningful results despite the presence of hardware noise, although with a limited circuit depth and accuracy~\cite{Peruzzo2014VQE}. Preskill subsequently formalized the NISQ paradigm, identifying noise, decoherence, and limited qubit counts as fundamental constraints on near-term quantum computations~\cite{Preskill2018NISQ}.

Within the broader domain of Quantum Machine Learning (QML), Biamonte \emph{et al.}~\cite{Biamonte2017QML} provided a comprehensive overview of how quantum computing principles may enhance classical learning models through high-dimensional feature representations and quantum correlations. Variational quantum circuits have since emerged as a dominant paradigm for supervised learning tasks owing to their compatibility with NISQ hardware. Schuld \emph{et al.} introduced circuit-centric quantum classifiers and analyzed their expressivity and trainability under idealized execution conditions~\cite{Schuld2019VQC}. Subsequent theoretical studies examined the representational power of quantum feature maps and their role in enabling quantum advantages in learning tasks~\cite{Havlivcek2019FeatureMaps,Schuld2019Feature}.

The effects of quantum noise on variational learning have received increasing attention in recent years. McClean \emph{et al.} analyzed the optimization landscapes of variational quantum algorithms and identified the emergence of barren plateaus, where gradients vanish exponentially with system size~\cite{McClean2018Barren}. Later studies demonstrated that hardware noise can significantly exacerbate barren plateau behavior, leading to noise-induced training ability issues even for shallow circuits~\cite{Wang2021NoiseVQC}. Cost-function-dependent and architecture-aware strategies have been proposed to mitigate these effects, although their effectiveness remains limited under realistic noise conditions~\cite{Cerezo2021Cost}.

Several studies have investigated the robustness and training strategies for quantum neural networks. Skolik \emph{et al.} proposed layer wise learning techniques to improve the optimization stability in noisy variational quantum models~\cite{Skolik2021Robustness}. Abbas \emph{et al.} analyzed the expressive power of quantum neural networks and highlighted the trade-off between expressivity and noise sensitivity~\cite{Abbas2021Power}. These studies have primarily focused on quantum circuit noise and assume clean classical inputs.

In parallel, the classical machine learning literature has extensively studied robustness under data corruption, including additive noise, impulse noise, quantization effects, and missing features~\cite{Bishop2006PRML,Goodfellow2015Explaining}. Techniques such as dropout~\cite{Srivastava2014Dropout} and adversarial training have been proposed to improve generalization under noisy inputs. However, the interaction between classical data corruption and quantum feature encoding remains largely unknown.

Error mitigation techniques have been proposed to alleviate the effect of quantum hardware noise without full error correction. Temme \emph{et al.} introduced error mitigation strategies for short-depth quantum circuits, while subsequent work expanded these ideas to hybrid quantum-classical algorithms~\cite{Temme2017ErrorMitigation,Endo2021Mitigation}. Although effective in reducing certain error types, these methods do not address coherent control errors or noise originating from classical data sources.

In contrast to existing studies, which predominantly consider either classical data noise or quantum hardware noise in isolation, this work presents a unified robustness analysis that jointly examines dataset-level perturbations, encoding-level angle-space noise, and circuit-level quantum noise. By systematically evaluating their individual and combined effects on variational quantum classifiers, the proposed study provides a more realistic assessment of QML performance in practical NISQ-era settings.

\section{Methodology and Noise Modeling Framework}
\label{sec:Methodology}

\begin{figure}
    \centering
    \includegraphics[width=\linewidth]{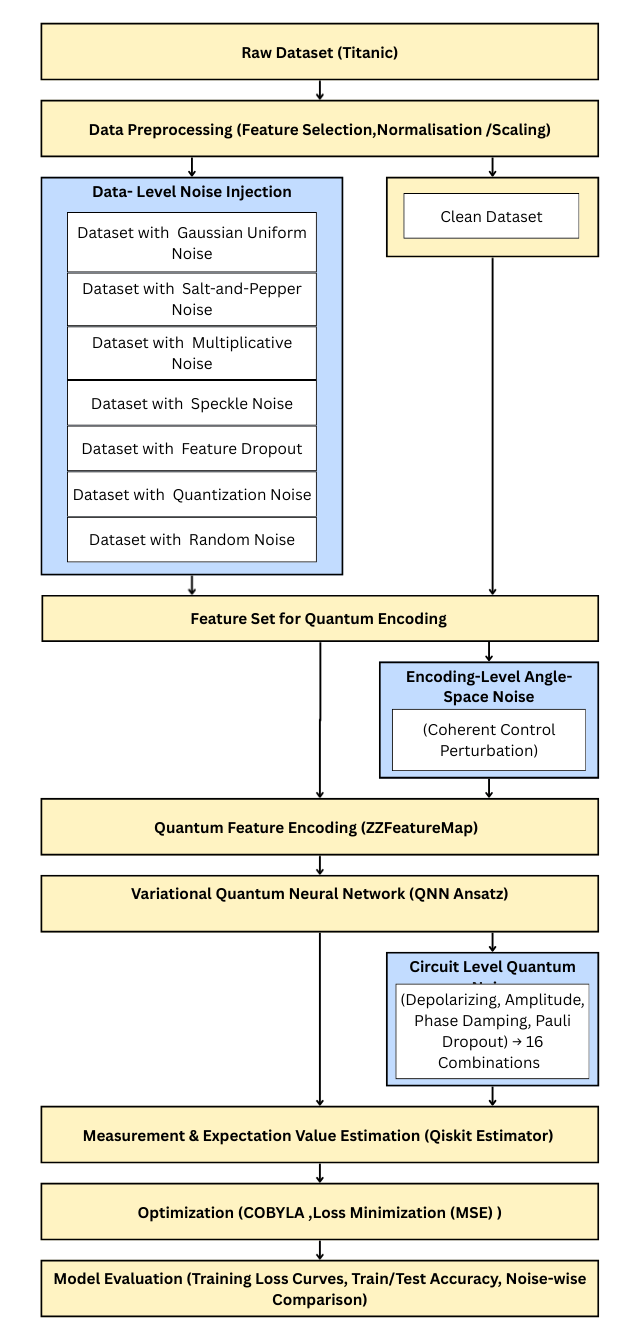}
    \caption{Block diagram of the proposed methodology illustrating dataset preprocessing, dataset-level noise injection, quantum feature encoding using ZZFeatureMap, variational quantum neural network execution under quantum noise, and hybrid quantum-classical optimization and evaluation.}
    \label{fig:ProposedMethodology}
\end{figure}

This section presents a unified methodology and noise modeling framework for systematically evaluating the robustness of a variational quantum classifier (VQC) under both classical dataset and quantum hardware noise. The proposed hybrid quantum-classical learning pipeline is shown in Figure.~\ref{fig:ProposedMethodology}.

The framework introduces controlled perturbations at three distinct stages: 
(i) dataset-level noise applied to classical input features, 
(ii) encoding-level angle-space noise modeling coherent control imperfections, and 
(iii) circuit-level quantum noise channels emulating NISQ hardware effects.

\paragraph{Design Rationale}
The methodology is intentionally structured to evaluate the impact of each perturbation independently and when applied in combination. This separation enables precise attribution of performance degradation to specific noise sources while still allowing their combined effects to be evaluated, which is essential for developing noise-aware quantum machine learning pipelines in the NISQ era.

\subsection{Dataset and Data Preparation}

The Titanic dataset was used as a benchmark for binary classification. Numerical features relevant to quantum encoding were selected and normalized to the range $[0,1]$ to ensure numerical stability and compatibility with rotation-based quantum feature maps, thereby avoiding scale-induced bias in parameterized quantum gate operations.

Categorical variables were transformed using a consistent label-encoding scheme learned exclusively from the training set and reused during testing to prevent data leakage. The dataset was partitioned into training and testing subsets following standard machine learning practices. All experiments were conducted on locally stored datasets the availability and integrity of which were verified prior to execution.

\subsection{Quantum Feature Encoding and Variational Ansatz}

Preprocessed classical features were encoded into quantum states using a ZZFeatureMap, which employed parameterized single-qubit rotations combined with pairwise $ZZ$ entangling operations to represent nonlinear feature correlations in a high-dimensional quantum Hilbert space~\cite{Havlivcek2019FeatureMaps}. This feature map forms the interface between the classical data and quantum processing in the proposed VQC pipeline.

Following feature encoding, a hardware-efficient variational quantum neural network (QNN) ansatz was applied. The ansatz consists of layered parameterized single-qubit rotation gates interleaved with entangling operations, forming a circuit structure compatible with the near-term quantum hardware. The overall VQC architecture follows the circuit-centric formulation proposed in~\cite{Schuld2019VQC} and adheres to the principles of hybrid quantum--classical algorithms established in~\cite{Peruzzo2014VQE,McClean2016Theory}. THe variational parameters were initialized prior to training and iteratively optimized through a classical learning loop.

\subsection{Dataset-Level Noise Modeling}
\label{subsec:dataset_noise}

To emulate realistic data corruption  scenarios,such as model sensor inaccuracies, environmental interference and errors in finite measurement resolution, encountered in practical machine learning pipelines, dataset-level noise is injected into normalized classical feature vectors prior to quantum feature encoding.~\cite{Bishop2006PRML,Goodfellow2015Explaining}.

The framework evaluates the robustness under nine dataset-level conditions, comprising a clean baseline and eight distinct noise models spanning additive, multiplicative, impulse, discretizations, and sparsity-inducing perturbations. Let $\mathbf{x} = [x_1, x_2, \ldots, x_d]^\top \in \mathbb{R}^d$ denote a normalized input feature vector. Dataset-level noise generates a corrupted input $\tilde{\mathbf{x}}$ via a stochastic operator $\mathcal{N}_d(\cdot)$:
\begin{equation}
\tilde{\mathbf{x}} = \mathcal{N}_d(\mathbf{x}).
\end{equation}

The considered noise models include Gaussian and uniform additive noise, salt-and-pepper (impulse) noise, multiplicative and speckle noise, quantization noise, feature dropout noise, and random sign noise. While these models have been well studied in classical robustness analysis~\cite{Bishop2006PRML,Goodfellow2015Explaining}, their interaction with quantum feature maps and variational optimization landscapes remains comparatively underexplored, particularly when combined with quantum hardware noise~\cite{Schuld2020Encoding,Abbas2021Power}.

\subsection{Encoding-Level (Angle-Space) Noise}
\label{subsec:angle_noise}

Beyond dataset-level perturbations, the framework introduces encoding-level noise into the parameter (angle) space of the quantum feature map. This noise models coherent control imperfections such as calibration drift, pulse amplitude fluctuations, and finite control resolution in parameterized quantum circuits. Such perturbations have been shown to significantly affect gradient magnitudes and trainability in variational quantum algorithms, even in the absence of stochastic decoherence~\cite{McClean2016Theory,Wang2021NoiseVQC,Arrasmith2021Effect}.

The normalized classical features are mapped to rotation angles $\boldsymbol{\theta} = 2\pi(\mathbf{x} - 0.5)$, perturbed by zero-mean Gaussian noise, and wrapped in the periodic domain $[-\pi,\pi]$ prior to circuit execution. Angle-space noise constitutes a coherent perturbation applied before circuit execution and is therefore evaluated independently from stochastic circuit-level noise. This isolation enables a clear distinction between coherent control errors and incoherent hardware-induced decoherence effects~\cite{Lloyd2020Encoding,Schuld2020Encoding}.

\subsection{Quantum Circuit Noise Modeling}
\label{subsec:quantum_noise}

At the circuit level, quantum noise is modeled using Qiskit Aer noise channels~\cite{QiskitAer}, capturing the dominant error mechanisms in the NISQ-era hardware~\cite{Preskill2018NISQ}. These processes act directly on the quantum state during execution and are represented as completely positive trace-preserving (CPTP) maps, forming the standard abstraction for NISQ-era noise modeling and error mitigation studies~\cite{Temme2017ErrorMitigation,Endo2021Mitigation}.

The quantum noise channels considered include depolarizing noise, amplitude damping, phase damping and Pauli noise. Each noise source is treated as an independent binary factor, yielding $2^4 = 16$ distinct circuit-level noise configurations, including the noise-free baseline. Noise channels are injected at the single-qubit, two-qubit, and measurement stages to emulate realistic quantum execution conditions~\cite{Kandala2017Hardware}.

\subsection{Hybrid Quantum--Classical Optimization}

The Quantum expectation values were evaluated using the Estimator primitive provided by Qiskit~\cite{QiskitCite}. Model training minimizes the mean squared error (MSE) loss between the predicted labels and ground truth values using the COBYLA optimizer, which is well suited for gradient-free optimization in variational quantum algorithms and near-term quantum settings~\cite{Farhi2018Classification}. The optimization loop iteratively updates the variational parameters until convergence criteria are met or a predefined number of iterations is reached.

\subsection{Integrated Noise Pipeline and Evaluation}

The complete noise-aware learning pipeline proceeds as follows: dataset-level noise injection on classical features, optional angle-space perturbation prior to encoding, quantum feature encoding using a ZZFeatureMap, circuit-level hardware noise during execution, and hybrid quantum--classical optimization and measurement.

Model robustness was evaluated using loss convergence behavior, training accuracy, and testing accuracy across all combinations of dataset-level, encoding-level, and circuit-level noise. This integrated evaluation enables the assessment of the resilience of variational quantum classifiers under realistic noise conditions. ~\cite{Preskill2018NISQ,Cai2023Robust}. The quantitative results and detailed analysis are presented in Section~\ref{sec:ResultsAndAnalysis}.

\section{Experimental Setup}
\label{sec:ExprimentalSetup}
All experiments including circuit execution and noise simulation were performed with the help of the Qiskit framework~\cite{QiskitCite} and Qiskit Aer backend~\cite{QiskitAer}. This setup enabled the controlled evaluation of variational quantum classifiers under both ideal (noise-free) and noisy execution conditions, representative of NISQ-era hardware.

\begin{figure}
    \centering
    \includegraphics[width=\linewidth]{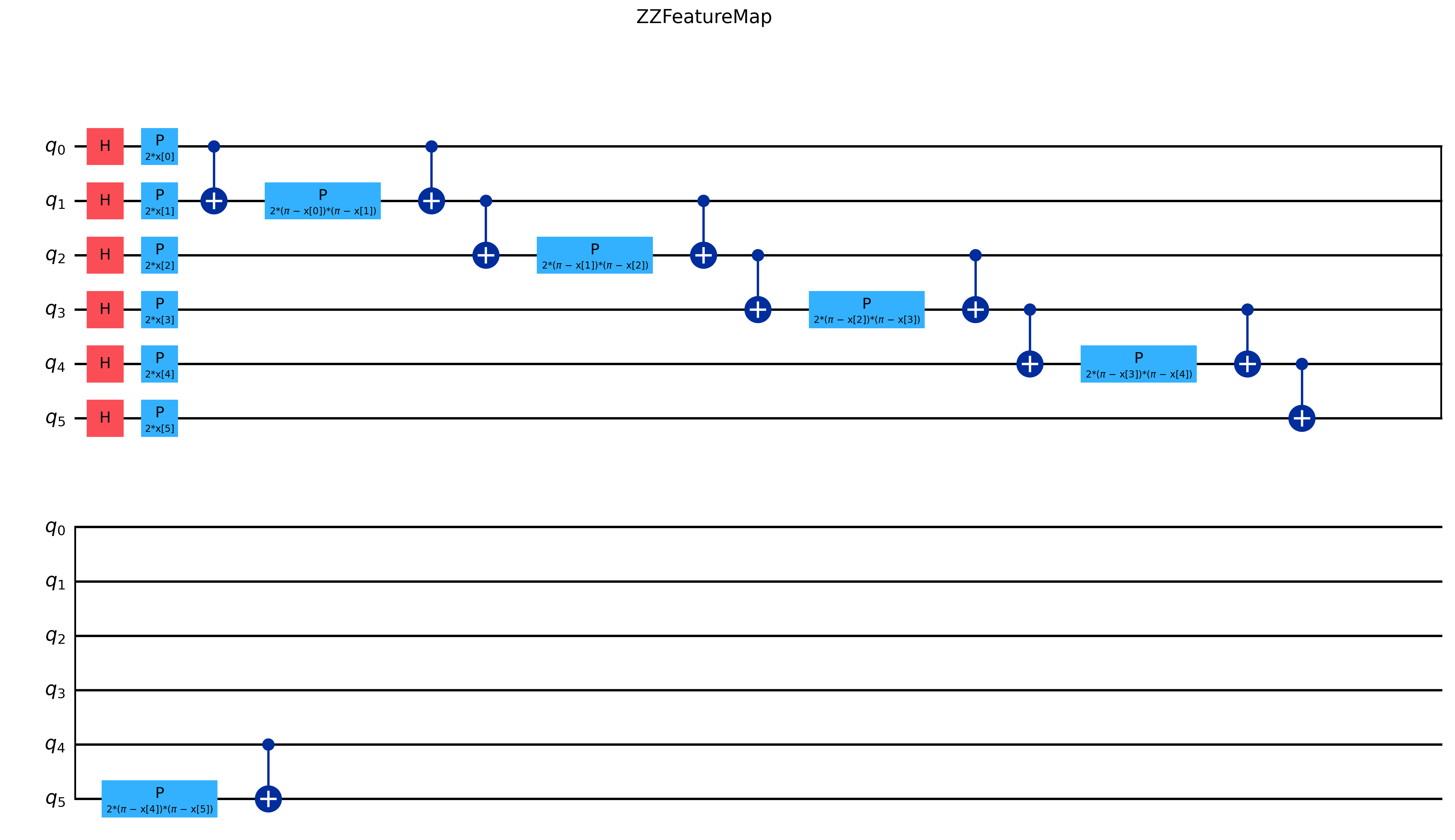}
    \caption{ZZFeatureMap used for encoding classical input features into a quantum Hilbert space. The feature map introduces parameterized single-qubit rotations and pairwise $ZZ$ entanglement to capture correlations among input features.}
    \label{fig:ZZFeatureMap}
\end{figure}

Classical input features are encoded into quantum states using the built-in ZZFeatureMap, which is widely adopted in quantum machine learning owing to its ability to introduce nonlinear feature correlations through entanglement~\cite{Havlivcek2019FeatureMaps}.The structure of the ZZFeatureMap employed in this study is shown in Fig.~\ref{fig:ZZFeatureMap}. The feature map employs parameterized single-qubit rotation gates followed by pairwise $ZZ$ interactions, embedding classical data into a high-dimensional quantum Hilbert space prior to variational processing.

\begin{figure}
    \centering
    \includegraphics[width=\linewidth]{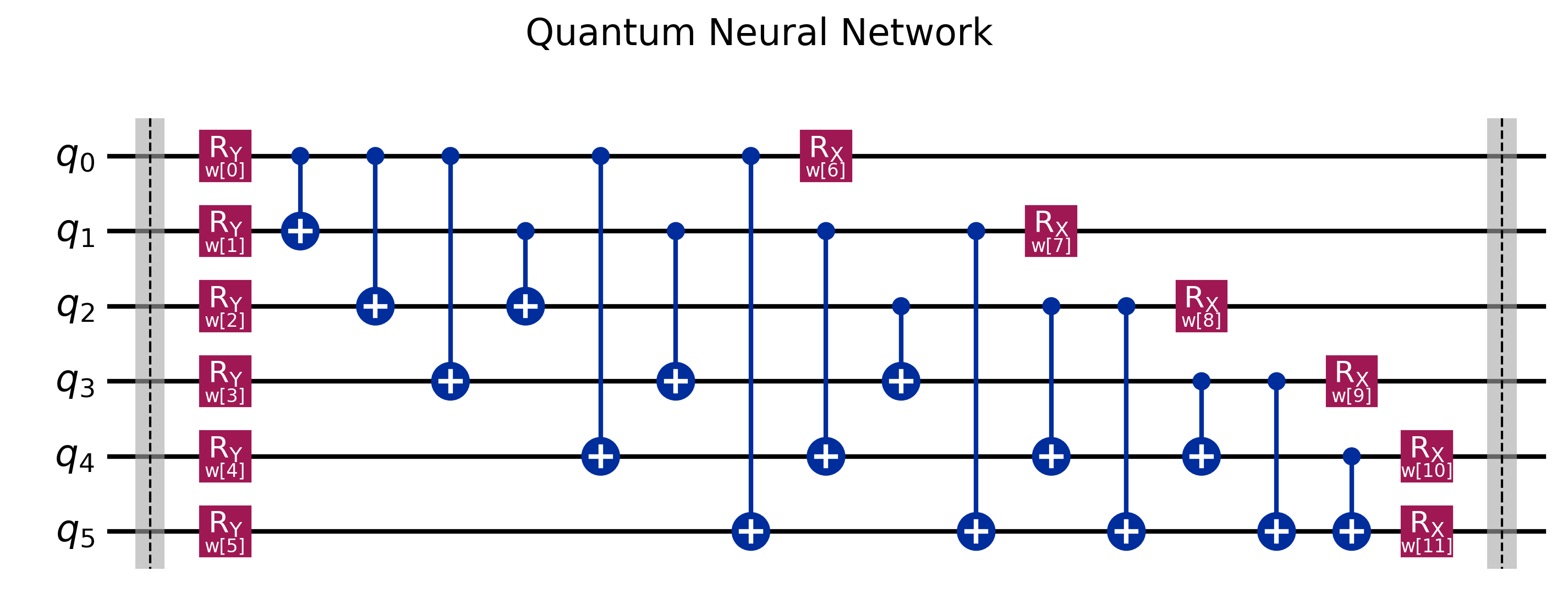}
    \caption{Hardware-efficient variational quantum neural network (QNN) ansatz consisting of parameterized rotation gates and entangling operations. The trainable parameters are optimized using a classical optimizer to minimize the classification loss.}
    \label{fig:QNN}
\end{figure}

Following feature encoding, a hardware-efficient variational quantum circuit is employed to implement a quantum neural network (QNN). The variational ansatz consists of layered parameterized rotation gates interleaved with entangling operations, forming a circuit structure compatible with near-term quantum hardware. The QNN architecture used in this study is shown in Fig.~\ref{fig:QNN}. The trainable parameters of the variational circuit were optimized using a classical optimizer to minimize classification loss, following the hybrid quantum--classical optimization paradigm established for variational algorithms~\cite{Peruzzo2014VQE,Schuld2019VQC}.

\begin{figure*}
    \centering
    \includegraphics[width=\linewidth]{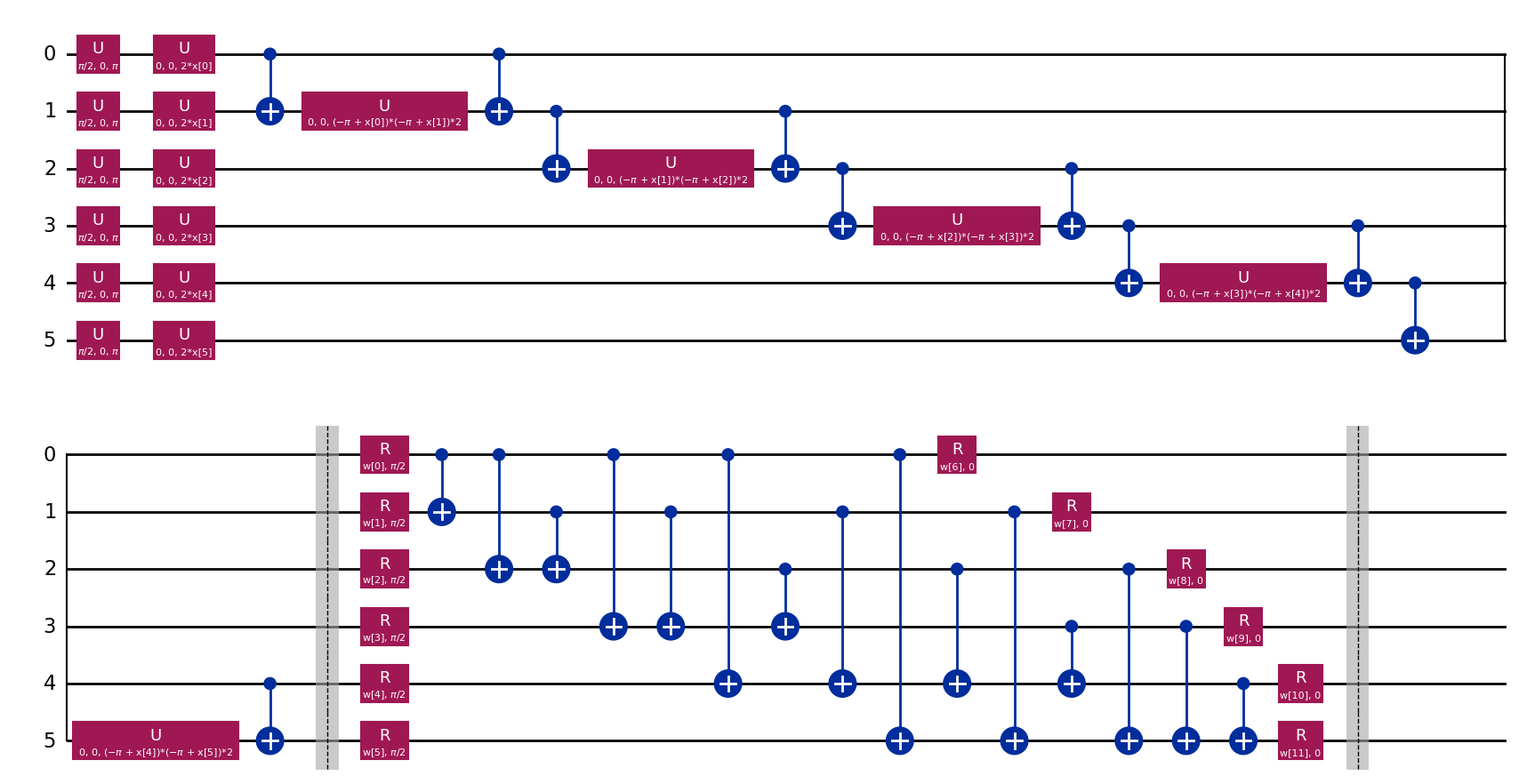}
    \caption{Complete variational quantum classifier obtained by composing the ZZFeatureMap with the hardware-efficient QNN ansatz. The feature-encoding block embeds classical data into a quantum feature space, followed by a trainable variational circuit used for supervised classification.}
    \label{fig:FullCircuit}
\end{figure*}

Feature encoding and variational blocks were composed to form a unified variational quantum classifier. The complete circuit architecture is shown in Fig.~\ref{fig:FullCircuit}. The feature-encoding block maps classical input features into a quantum feature space, whereas the subsequent variational block transforms the encoded quantum state by using trainable operations to predict binary class labels corresponding to \emph{Survived} (1) or \emph{Not Survived} (0).

To ensure the clarity and reproducibility of the experimental setup, the key quantum hardware, noise, and optimization parameters used in this study are summarized in Table~\ref{tab:hardware_noise_params}. These parameters were held constant across all experiments unless explicitly stated, enabling controlled evaluation of the dataset-level, encoding-level, and circuit-level noise effects.

\begin{table}[t]
\centering
\caption{Quantum Hardware and Noise Configuration Parameters}
\label{tab:hardware_noise_params}
\renewcommand{\arraystretch}{1.2}
\begin{tabular}{p{2.75cm} p{5cm}}
\hline
\textbf{Parameter} & \textbf{Value / Description} \\
\hline
Quantum framework & Qiskit~\cite{QiskitCite} \\
Simulator backend & Qiskit Aer~\cite{QiskitAer} \\
Execution mode & Statevector (ideal) and shot-based sampling (noisy) \\
 \\
Number of qubits & Equal to number of selected features \\
Feature map & ZZFeatureMap~\cite{Havlivcek2019FeatureMaps} \\
Variational ansatz & Hardware-efficient QNN \\
Optimizer & COBYLA (gradient-free) \\
Loss function & Mean Squared Error (MSE) \\
Number of shots & Fixed per experiment (finite sampling) \\
Random seed & Fixed for reproducibility \\
\hline
Depolarizing noise & Single- and two-qubit depolarizing channels \\
Amplitude damping & $T_1$-like relaxation noise \\
Phase damping & $T_2$-like dephasing noise \\
Pauli noise & Stochastic $X$, $Y$, $Z$ errors \\
Readout noise & Classical bit-flip measurement errors \\
Noise combinations & $2^4 = 16$ circuit-level configurations \\
\hline
\end{tabular}
\end{table}

To evaluate the robustness, each dataset-level noise configuration was systematically combined with multiple quantum circuit noise models, including depolarizing, damping, and Pauli noise channels. A noise-free baseline was retained for comparison across all experiments. Model performance was assessed using loss convergence behavior, training accuracy, and testing accuracy, enabling quantitative analysis of the individual and combined effects of classical data corruption and quantum hardware noise under controlled experimental conditions.

\subsection{Justification of Noise Parameter Choices}
\label{subsec:noise_param_justification}

The selection of noise magnitudes and parameter ranges is guided by two complementary principles: 
(i) consistency with realistic noise levels encountered in classical sensing pipelines and near-term quantum hardware, and 
(ii) the need to induce measurable but non-degenerate performance degradation that enables meaningful robustness analysis.

\paragraph{Dataset-level noise parameters.}
All classical features were normalized to the range $[0,1]$ prior to quantum encoding. Additive Gaussian and multiplicative noise are therefore applied with standard deviation $\sigma = 0.05$, corresponding to typical perturbations of approximately $\pm 10\%$ of the feature range. This level of corruption is widely used in robustness and generalization studies in classical machine learning to model moderate sensor noise and measurement uncertainty without destroying class-discriminative structure~\cite{Bishop2006PRML,Goodfellow2015Explaining}. 

Uniform and random-sign noise employ bounded perturbations with $\varepsilon = 0.05$, ensuring worst-case deviations comparable to the Gaussian setting while avoiding unbounded outliers. Salt-and-pepper noise was introduced with an impulse ratio of $3\%$, modeling sparse but severe acquisition or transmission faults commonly observed in real-world data streams. Feature dropout was applied with probability $p_d = 0.05$, representing mild missingness or unreliable measurements while preserving sufficient information for learning. Quantization noise is implemented using a step size $\Delta = \pi/255$, approximating 8-bit finite-precision discretization in the subsequent angle-encoding domain, which is consistent with the standard digital control and representation limits~\cite{Bishop2006PRML}.

\paragraph{Encoding-level (angle-space) noise parameters.}
Angle-space noise is modeled as zero-mean Gaussian perturbations applied directly to the rotation parameters prior to circuit execution and wrapped in the periodic domain $[-\pi,\pi]$. We evaluate $\sigma \in \{0, 0.01, 0.03, 0.05\}$ to span a controlled range from ideal calibration to moderate coherent control drift. These values are chosen to perturb encoded angles without inducing near-random rotations, thereby enabling the isolation of coherent control effects distinct from stochastic decoherence. Similar magnitudes have been used in prior analyses of variational circuit trainability and noise-induced degradation~\cite{McClean2016Theory,Wang2021NoiseVQC,Arrasmith2021Effect}.

\paragraph{Circuit-level quantum noise parameters.}
Circuit-level noise is introduced using depolarizing, amplitude damping, phase damping, Pauli, and readout error channels implemented via Qiskit Aer. Depolarizing noise was applied with probability $p=0.05$, modeling generic gate infidelity at levels consistent with contemporary NISQ-era devices~\cite{Preskill2018NISQ}. The amplitude and phase damping channels use relaxation parameters $\gamma = 0.05$, corresponding to moderate $T_1$ and $T_2$ decoherence over shallow circuit depths, as commonly assumed in hardware-efficient variational algorithm studies~\cite{Kandala2017Hardware,Temme2017ErrorMitigation}. The Pauli error probability is set to $p=0.02$ to model stochastic bit- and phase-flip errors without dominating the overall noise budget.

These parameter choices are intentionally conservative: they induce substantial performance degradation while avoiding trivial optimization failure, thereby enabling meaningful comparisons across noise types and combinations. All noise strengths were held fixed across experiments to ensure that observed performance variations arise from differences in noise mechanisms rather than arbitrary parameter tuning.

\paragraph{Reproducibility and Experimental Settings.}
All simulations were executed with fixed random seeds controlling for dataset splitting, noise generation, and parameter initialization. A consistent number of measurement shots were used for all noisy circuit executions to avoid sampling bias across the noise configurations. Variational circuit parameters were initialized deterministically and optimized using the COBYLA optimizer with identical stopping criteria for all runs. Consequently, the observed performance variations can be attributed primarily to the injected noise models rather than the stochastic experimental artifacts.

\section{Results and Analysis}
\label{sec:ResultsAndAnalysis}

This section presents an empirical robustness analysis of the proposed variational quantum classifier (VQC) under combined classical dataset noise and quantum hardware noise.

To reflect the operational constraints of Noisy Intermediate-Scale Quantum (NISQ) devices, quantum noise was modeled using simulator-based completely positive trace-preserving (CPTP) channels. Model performance was evaluated using loss convergence behavior, training accuracy, and testing accuracy, with mean and standard deviation computed across repeated runs where applicable.

\subsection{Baseline Performance}

Under ideal execution conditions, with no dataset-level perturbations and no quantum circuit noise, the variational quantum classifier achieved a mean training accuracy of $76.40\%$ and a closely matching mean testing accuracy of $76.12\%$. The zero standard deviation across runs indicated fully deterministic convergence and stable optimization behavior in the absence of noise.

This strong baseline performance confirmed that the selected ZZFeatureMap provided an effective quantum feature embedding for the Titanic dataset and that the variational ansatz had sufficient expressive capacity to solve the task without overfitting.

\subsection{Impact of Circuit-Level Quantum Noise}

Introducing circuit-level quantum noise resulted in an immediate and substantial degradation in classification performance, even when the classical dataset remained noise-free. Across depolarizing, amplitude damping, phase damping, and their combinations, meant training accuracy collapsed from approximately $76\%$ to $38.67\%$, while mean testing accuracy droped to $39.15\%$.

This represented a relative performance degradation of nearly $50\%$, with training and testing accuracies converging to a common noise-dominated floor. The negligible generalization gap under noisy execution indicated that circuit-level quantum noise primarily disrupts the optimization and learning dynamics of the variational circuit rather than inducing overfitting.

Among the four quantum noise channels, amplitude damping consistently exhibited the most severe impact on both loss convergence and final accuracy. This observation is consistent with its irreversible energy relaxation mechanism, which suppressed gradient magnitudes and effectively induced barren-plateau-like optimization behavior in variational circuits.

\subsection{Effect of Dataset-Level Noise}

Dataset-level noise induced a more gradual degradation in performance compared to circuit-level quantum noise. Additive (Gaussian, uniform), impulse (salt-and-pepper), multiplicative (speckle, multiplicative), precision-based (quantization), and sparsity-inducing (feature dropout) perturbations affected quantum feature encoding to varying degrees.

When quantum circuit execution remained ideal, dataset-level perturbations reduced convergence stability and slow optimization but did not cause catastrophic failure. Across most dataset noise models, mean classification accuracy remained within $1$--$4\%$ of the clean baseline, demonstrating that the variational classifier retains meaningful learning capability despite corrupted classical inputs.

These results indicate that classical data corruption primarily degrades the quality of the encoded quantum features, while the variational model itself remains trainable in the absence of hardware-induced decoherence.

\subsection{Encoding-Level (Angle-Space) Noise}

Encoding-level angle-space noise produced a smooth and progressive decline in classification accuracy as the noise variance increases. Unlike circuit-level quantum noise, angle-space perturbations did not trigger abrupt optimization collapse. Instead, the loss curves \ref{app:all_plots} reveal gradual smoothing of the optimization landscape and slower convergence as angle perturbation strength increases.

This contrast highlights the fundamentally different roles of coherent control errors and incoherent quantum noise. While angle-space noise distorts feature encoding, it preserves reversibility and does not irreversibly destroy quantum information, unlike stochastic decoherence acting during circuit execution.

Although angle-space noise alone does not induce catastrophic failure, its cumulative effect becomes increasingly pronounced at higher noise levels, underscoring the sensitivity of feature-map-based encodings to control inaccuracies.

\subsection{Combined Noise Effects}

When dataset-level noise and quantum circuit noise were applied simultaneously, overall performance remained close to the circuit-noise-dominated accuracy floor. Mean training and testing accuracies remained near $39\%$, with additional dataset corruption contributing negligible further degradation once strong quantum noise wass present.

This masking effect indicates that, in the NISQ regime, hardware noise dominates model behavior and suppresses the observable impact of classical data corruption. Across all combined noise configurations, training and testing accuracies remained closely aligned, confirming that performance degradation arises from reduced training ability rather than overfitting.

Although dataset-level noise exacerbated performance degradation in low-to-moderate circuit noise regimes, its marginal effect became negligible once circuit-level noise dominated, indicating a saturation effect characteristic of NISQ hardware.

\subsection{Quantitative Summary}

The proposed framework evaluated robustness under nine dataset-level noise conditions, including additive (Gaussian, Uniform), impulse (Salt-and-Pepper), multiplicative (Speckle, Multiplicative), precision-based (Quantization), and sparsity-inducing (Feature Dropout) perturbations, in addition to a clean baseline.

A quantitative summary of classification accuracy \ref{app:all_plots} under representative noise regimes is provided in Table~\ref{tab:noise_source_comparison}. The results clearly demonstrate that while dataset-level and encoding-level perturbations introduce gradual accuracy degradation under ideal circuit execution, circuit-level quantum noise overwhelmingly dominates performance, collapsing accuracy to a common noise floor irrespective of additional classical corruption.

\begin{table}[t]
\centering
\caption{Impact of Different Noise Sources on VQC Classification Accuracy (Mean $\pm$ Std)}
\label{tab:noise_source_comparison}
\renewcommand{\arraystretch}{1.25}
\begin{tabular}{p{1.75cm} p{1cm}p{1.25cm}p{1cm}p{2cm}}
\hline
\textbf{Noise Source} 
& \textbf{Dataset Noise} 
& \textbf{Angle-Space Noise} 
& \textbf{Circuit Noise} 
& \textbf{Test Accuracy (\%)} \\
\hline
Clean Baseline 
& None 
& None 
& None 
& $76.12 \pm 0.00$ \\
Dataset Noise Only 
& Yes 
& None 
& None 
& $72$--$76 \ (\pm <1)$ \\
Angle-Space Noise Only 
& None 
& Yes 
& None 
& $68$--$74 \ (\pm <2)$ \\
Circuit Noise Only 
& None 
& None 
& Yes 
& $39.15 \pm 0.83$ \\
Dataset + Circuit Noise 
& Yes 
& None 
& Yes 
& $39.15 \pm 0.83$ \\
Angle-Space + Circuit Noise 
& None 
& Yes 
& Yes 
& $38$--$40 \ (\pm <1)$ \\
Dataset + Angle + Circuit Noise 
& Yes 
& Yes 
& Yes 
& $38$--$40 \ (\pm <1)$ \\
\hline
\end{tabular}
\end{table}

The complete set of loss convergence curves and training–testing accuracy bar charts for all combinations of dataset-level noise, angle-space noise, and circuit-level quantum noise is provided in Appendix~\ref{app:all_plots} (Figures~\ref{fig:loss_nods_grouped}–\ref{fig:acc_Random_grouped}).

Overall, the results demonstrate that circuit-level quantum noise is the dominant factor limiting variational quantum classifier performance in the NISQ era, while dataset-level and encoding-level perturbations contribute secondary but measurable effects. Classical input noise can further amplify the adverse impact of quantum decoherence, particularly in feature-map-based encodings such as the ZZFeatureMap~\cite{Havlivcek2019FeatureMaps}.

Among the quantum noise channels studied, amplitude damping exhibits the most severe impact, consistent with its role in modeling energy relaxation in NISQ devices~\cite{Preskill2018NISQ}. These observations align with prior studies showing that noise fundamentally alters the trainability of variational circuits by flattening optimization landscapes and suppressing gradient information~\cite{Cerezo2021Cost,Fontana2021NoisePlateaus,Abbas2021Power}. Although variational circuits are theoretically expressive, their practical performance remains critically constrained by noise levels and circuit depth~\cite{Du2020ExpressivePower}.

\paragraph{Data Availability.}
The experiments were conducted on a 64-bit local system with 32 GB RAM and an Intel® Core™ Ultra 7 processor.The results reported in this paper were obtained by executing the proposed framework on the local workstation. During execution, structured CSV logs were generated for each experimental run, recording per-iteration loss values along with the final training and testing accuracies for every noise configuration.

The complete set of CSV files is provided as supplementary material to enable reproducibility and independent analysis. All quantitative results presented in this study are computed directly from these experimental logs, covering a total of 192 configurations arising from combinations of dataset-level perturbations, encoding-level (angle-space) noise, and circuit-level quantum noise channels.

\subsection{Scalability Considerations and Implications for Larger Qubit Systems}
\label{subsec:scalability}

The experimental results reported in this work are obtained using small-scale quantum circuits, reflecting the practical constraints of current Noisy Intermediate-Scale Quantum (NISQ) hardware. While these experiments do not directly probe large-qubit systems, the trends observed offer valuable insight into how hybrid quantum-classical machine learning models are likely to behave as problem sizes and circuit widths increase.

From a modeling standpoint, variational quantum classifiers exhibit a natural growth in the number of trainable parameters as the number of qubits and circuit depth increase~\cite{Schuld2019VQC}. To maintain expressive power for higher-dimensional problems, deeper parameterized circuits are often required. However, increasing circuit depth inevitably leads to the accumulation of gate-level noise across repeated single and two qubit operations. Prior studies have shown that this cumulative noise can significantly degrade convergence and training ability in variational algorithms, particularly in the presence of decoherence and stochastic errors~\cite{Preskill2018NISQ, Wang2021NoiseVQC}.

Beyond circuit-level effects, classical dataset perturbations introduced prior to quantum feature encoding play an important role. As the number of qubits increases, corrupted classical features are mapped into higher-dimensional quantum feature spaces, where their interaction with hardware noise can become more pronounced. The results presented in this study show that even modest levels of classical input noise can substantially amplify the negative impact of quantum decoherence. This suggests that, in larger-scale systems, the combined influence of classical and quantum noise sources is likely to become increasingly severe. Such behavior is consistent with recent analyses highlighting the sensitivity of variational quantum models to compounded noise mechanisms~\cite{Abbas2021Power}.

Although direct experimentation on large-qubit quantum processors remains beyond current hardware capabilities, the qualitative behavior observed here aligns well with theoretical expectations for scalable variational quantum algorithms. The interplay between increasing circuit depth, accumulated quantum noise, and corrupted input states emerges as a fundamental bottleneck for near-term quantum machine learning. These findings reinforce the need for noise-aware algorithm design, shallow and hardware-efficient ansatz, and robust classical preprocessing strategies when extending hybrid quantum-classical learning frameworks to larger problem instances~\cite{Cerezo2021Cost}.

In summary, while absolute performance levels will continue to improve with advances in quantum hardware, the noise interaction patterns identified in this work are expected to persist. As such, the trends reported here provide practical guidance for designing, evaluating, and scaling quantum machine learning pipelines in the NISQ era.

\section{Conclusion and Future Work}
\label{sec:conclusion}
While the Titanic dataset provides a convenient and well-controlled testbed, an important direction for future work is to examine whether the observed noise-dominance effects persist in more challenging settings, including higher-dimensional, class-imbalanced, and real-world datasets.

In this paper, we conducted a systematic robustness analysis of a variational quantum classifier under the combined influence of classical dataset corruption and quantum hardware noise. The results reveal critical vulnerabilities in hybrid quantum–classical learning pipelines and emphasize the necessity of noise-aware evaluation strategies in the Noisy Intermediate-Scale Quantum (NISQ) era~\cite{Preskill2018NISQ}. Looking ahead, future work will investigate the integration of error mitigation techniques, the design of noise-resilient quantum feature maps, and the extension of this framework to larger-scale, real-world learning problems.

While the experimental evaluation is limited to small-scale quantum circuits, the scalability implications of the observed noise interactions are discussed in Section\ref{sec:ResultsAndAnalysis}, providing guidance for extending hybrid quantum–classical learning models to larger NISQ systems

While practical error mitigation methods exist, this work deliberately focuses on unmitigated noise to assess intrinsic robustness under realistic NISQ conditions~\cite{Endo2018PracticalMitigation}. Although this study relies on high-fidelity noise simulation, validating these findings on real quantum hardware remains an important direction for future work.



\newcommand{\FourPanelFigure}[6]{%
\begin{figure*}[!t]
    \centering
    \begin{subfigure}[t]{0.49\linewidth}
        \centering
        \includegraphics[width=\linewidth,trim=4 4 4 4,clip]{#1}
        \caption{Angle-space noise $\sigma = 0$}
        \label{#6:a}
    \end{subfigure}\hfill
    \begin{subfigure}[t]{0.49\linewidth}
        \centering
        \includegraphics[width=\linewidth,trim=4 4 4 4,clip]{#2}
        \caption{Angle-space noise $\sigma = 0.01$}
        \label{#6:b}
    \end{subfigure}

    \medskip

    \begin{subfigure}[t]{0.49\linewidth}
        \centering
        \includegraphics[width=\linewidth,trim=4 4 4 4,clip]{#3}
        \caption{Angle-space noise $\sigma = 0.03$}
        \label{#6:c}
    \end{subfigure}\hfill
    \begin{subfigure}[t]{0.49\linewidth}
        \centering
        \includegraphics[width=\linewidth,trim=4 4 4 4,clip]{#4}
        \caption{Angle-space noise $\sigma = 0.05$}
        \label{#6:d}
    \end{subfigure}

    \caption{#5}
    \label{#6}
\end{figure*}%
}

\begin{appendix}
\section{Complete Loss Curves and Accuracy Plots}
\label{app:all_plots}

This appendix presents the complete set of loss convergence curves and training--testing accuracy bar charts for all evaluated combinations of dataset-level noise, encoding-level (angle-space) noise, and circuit-level quantum noise.

For each dataset noise condition, results are grouped by angle-space noise magnitude ($\sigma \in \{0, 0.01, 0.03, 0.05\}$). Within each subplot, loss curves correspond to all sixteen circuit-level noise configurations, enabling direct comparison of optimization stability and convergence behavior under increasing noise severity.

The accompanying accuracy bar plots summarize the final training and testing accuracies for the same configurations, facilitating analysis of generalization performance and the relative impact of different noise sources.


\FourPanelFigure
{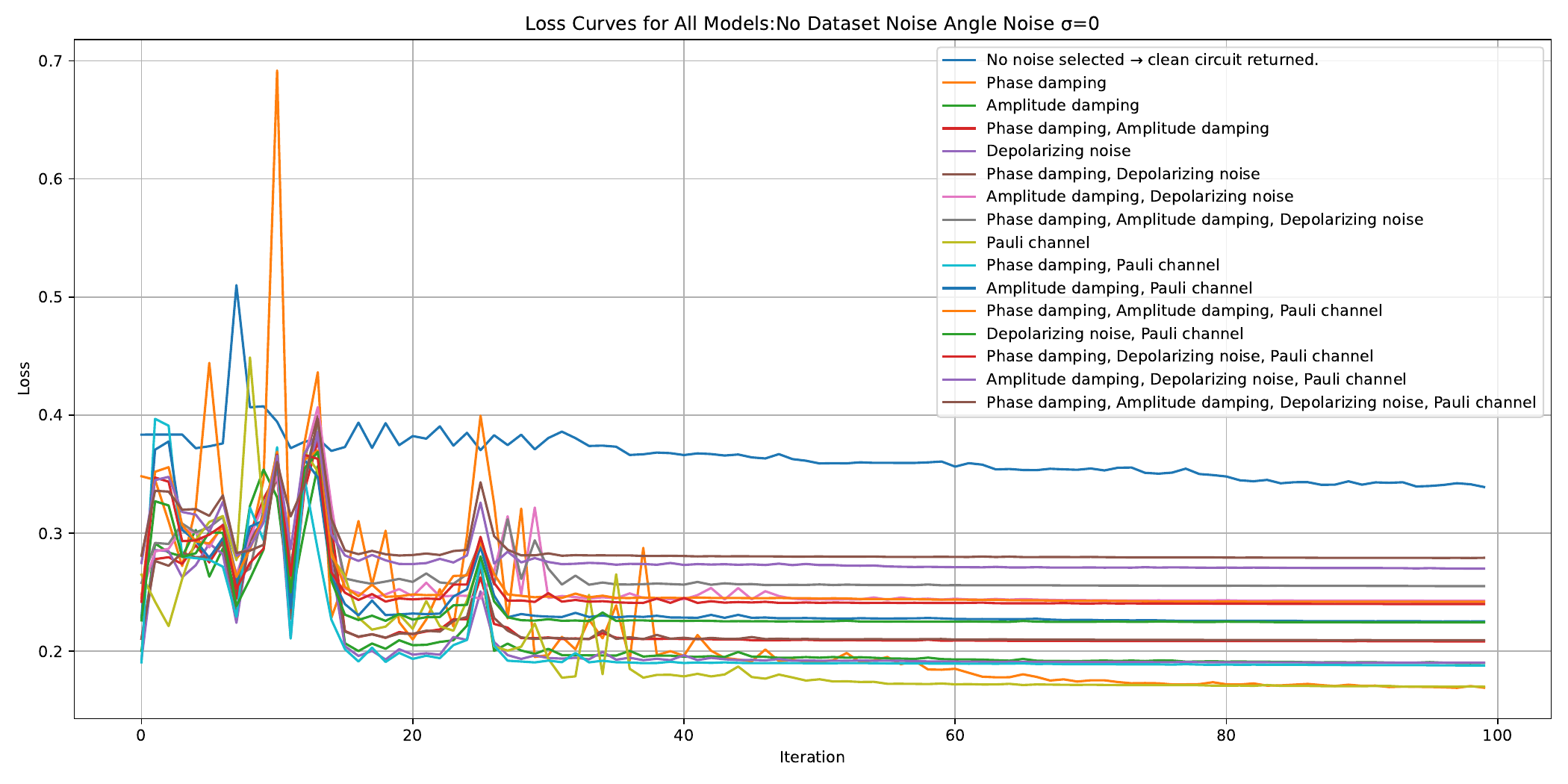}
{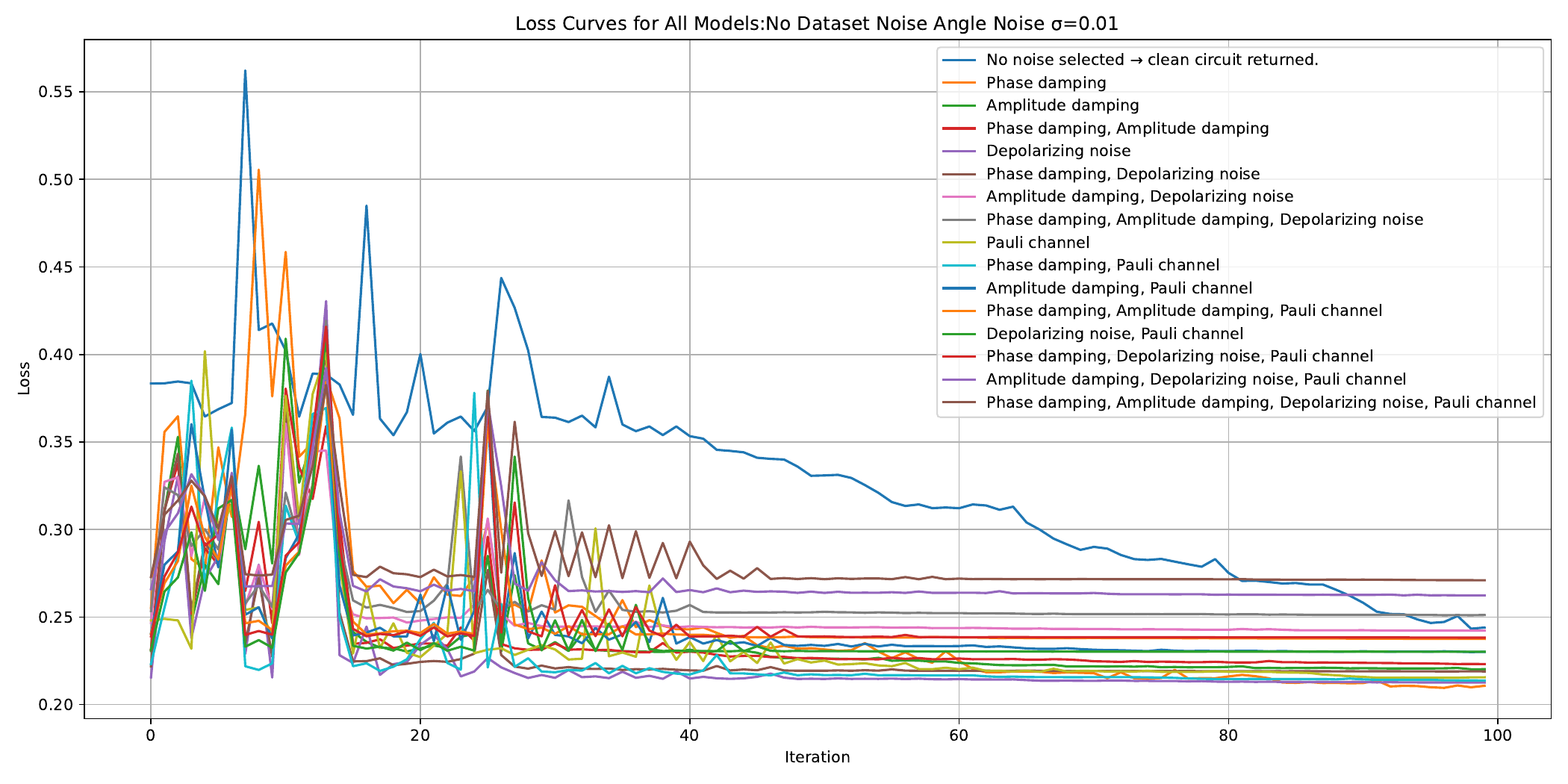}
{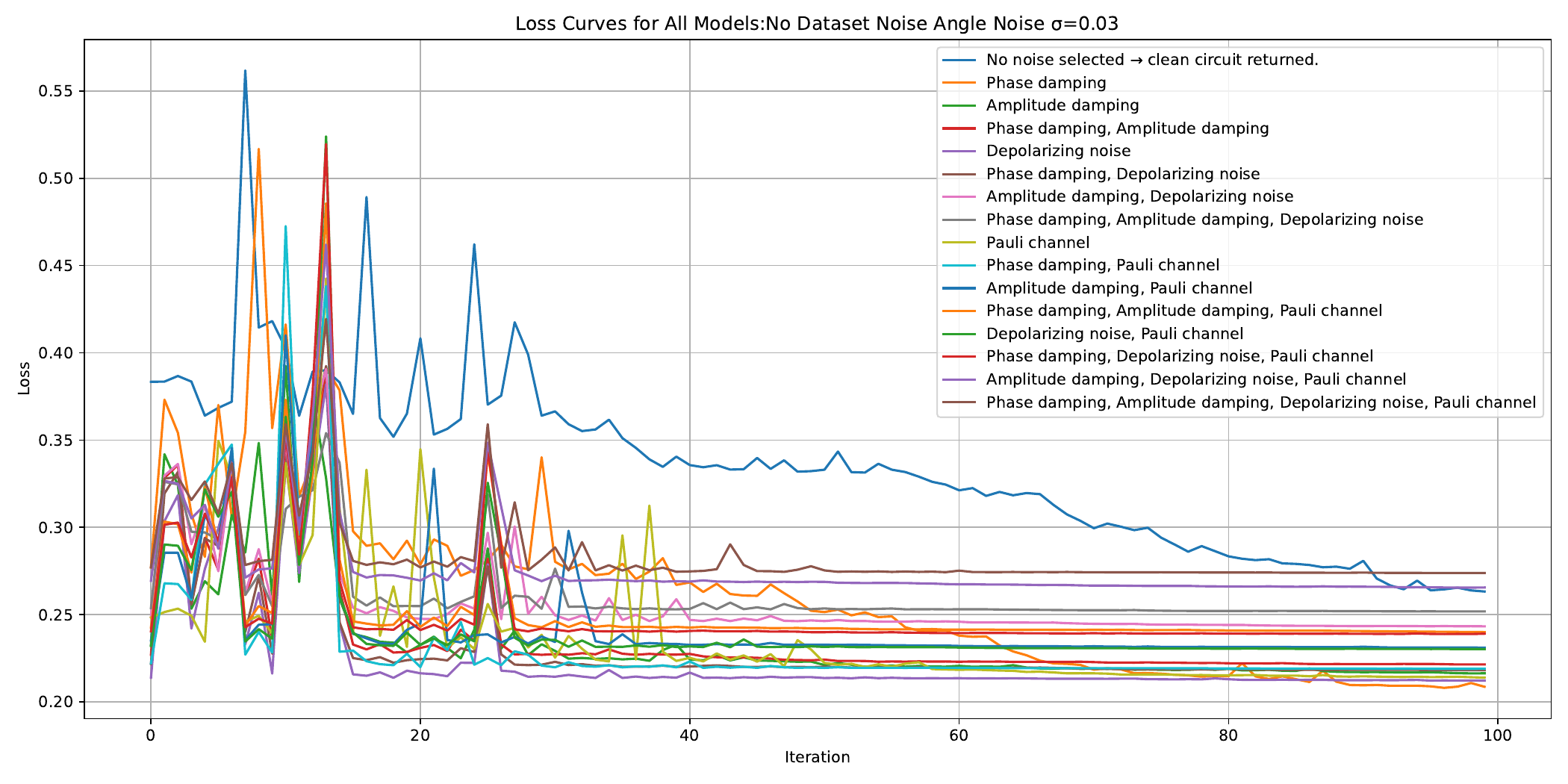}
{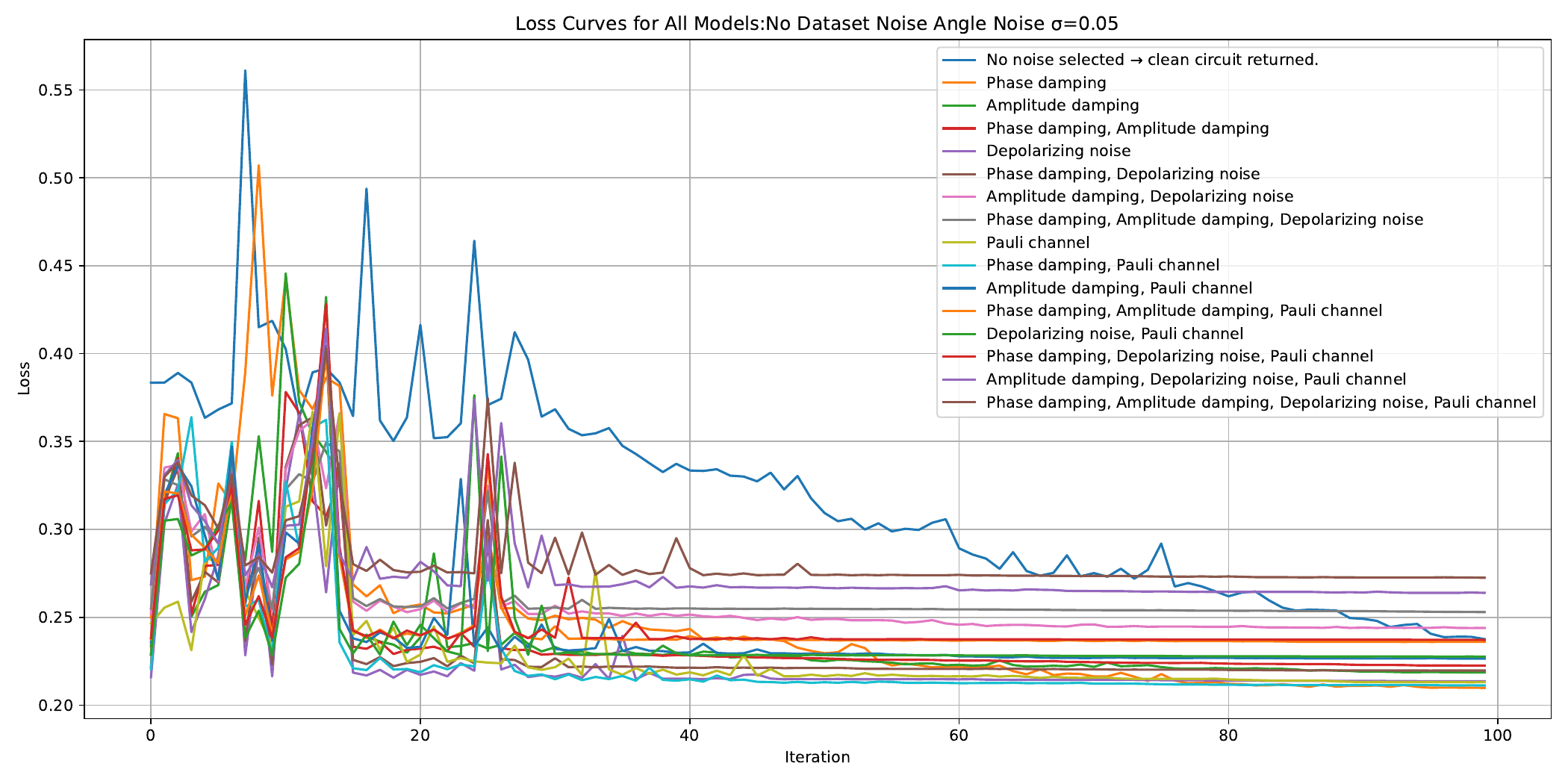}
{Loss convergence curves for all circuit-level noise configurations under \textbf{no dataset noise}, shown for increasing angle-space noise magnitudes.}
{fig:loss_nods_grouped}

\FourPanelFigure
{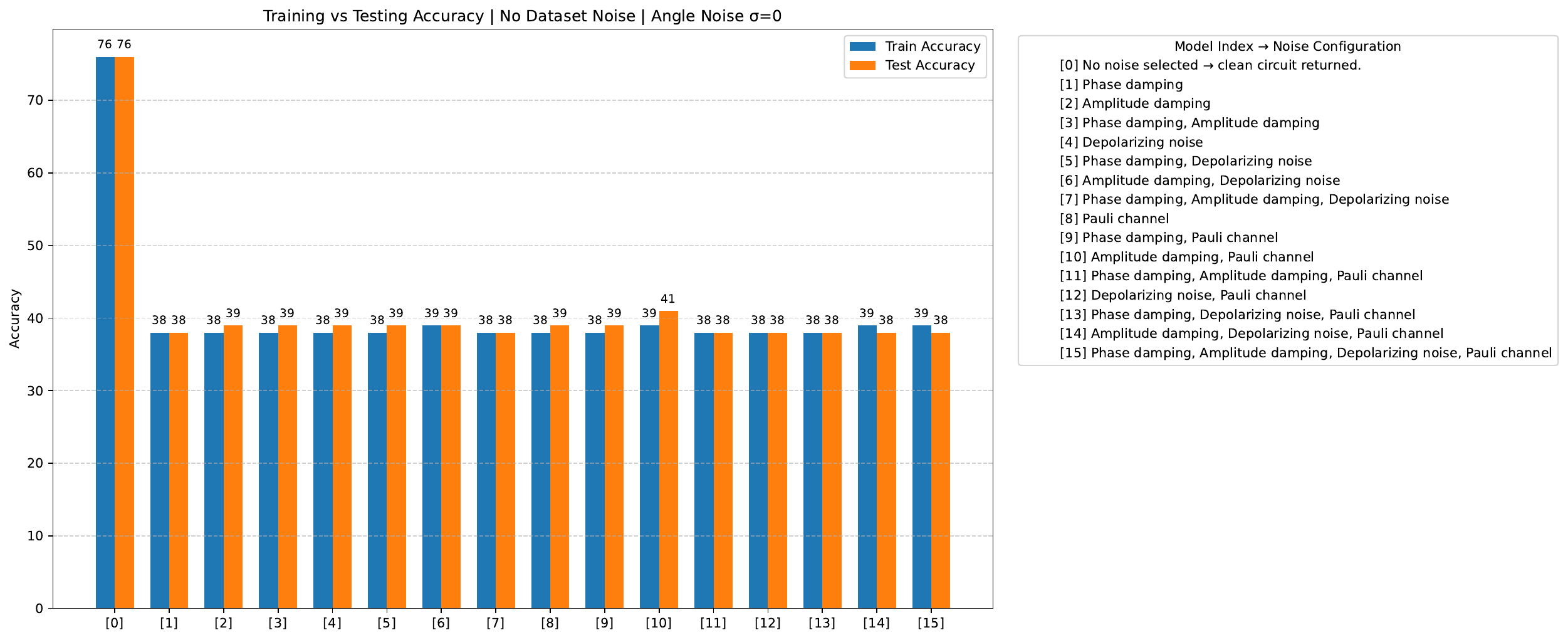}
{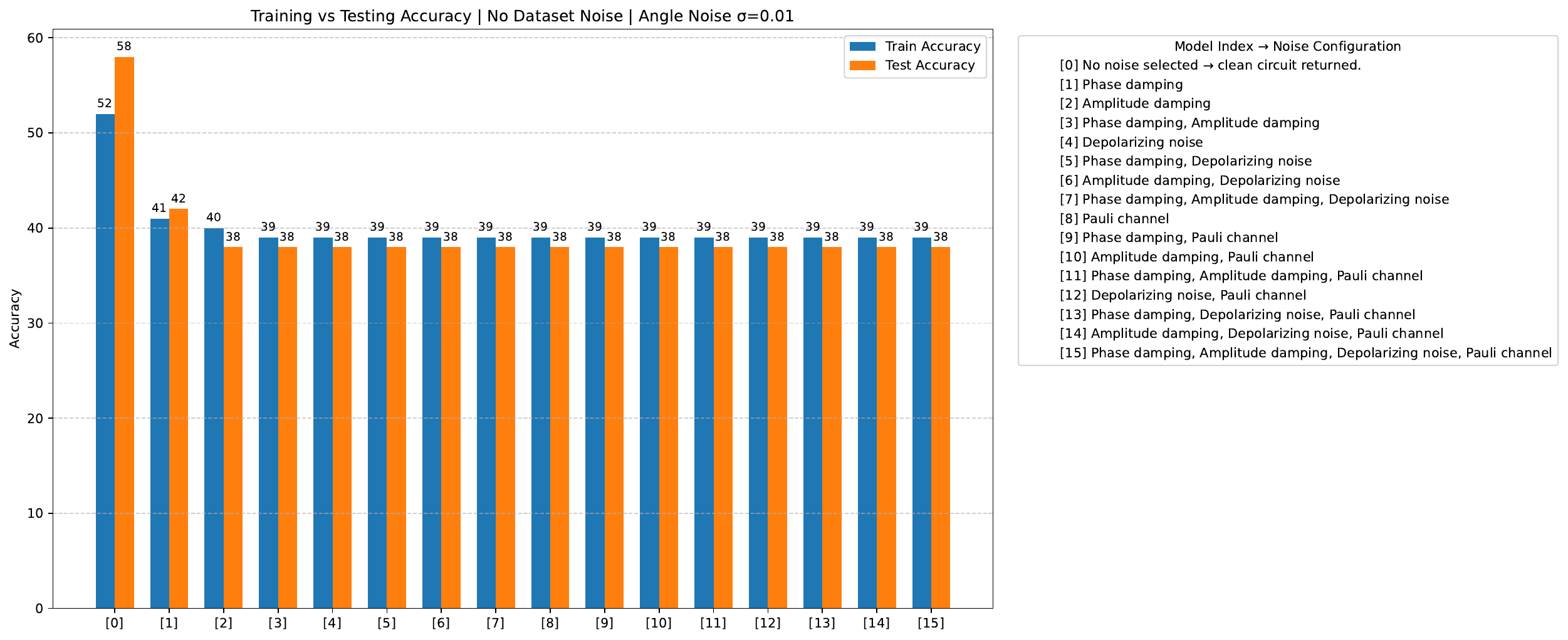}
{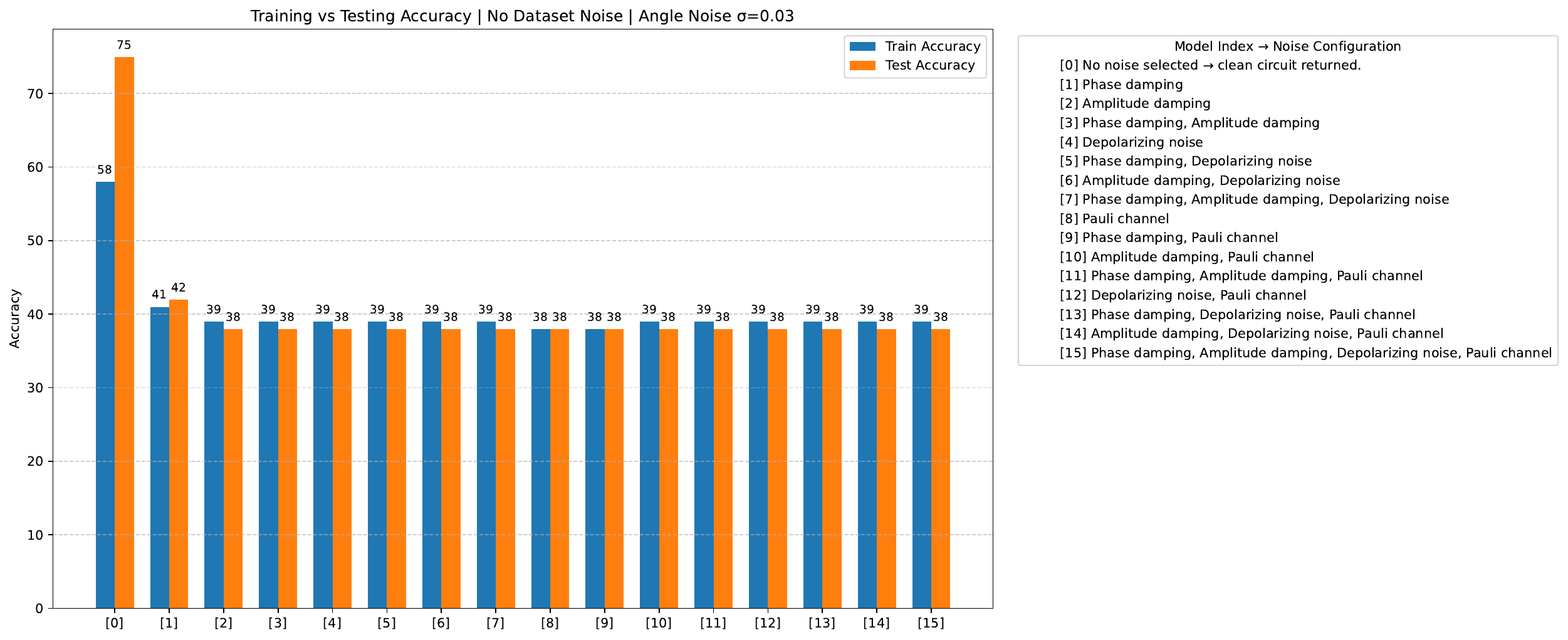}
{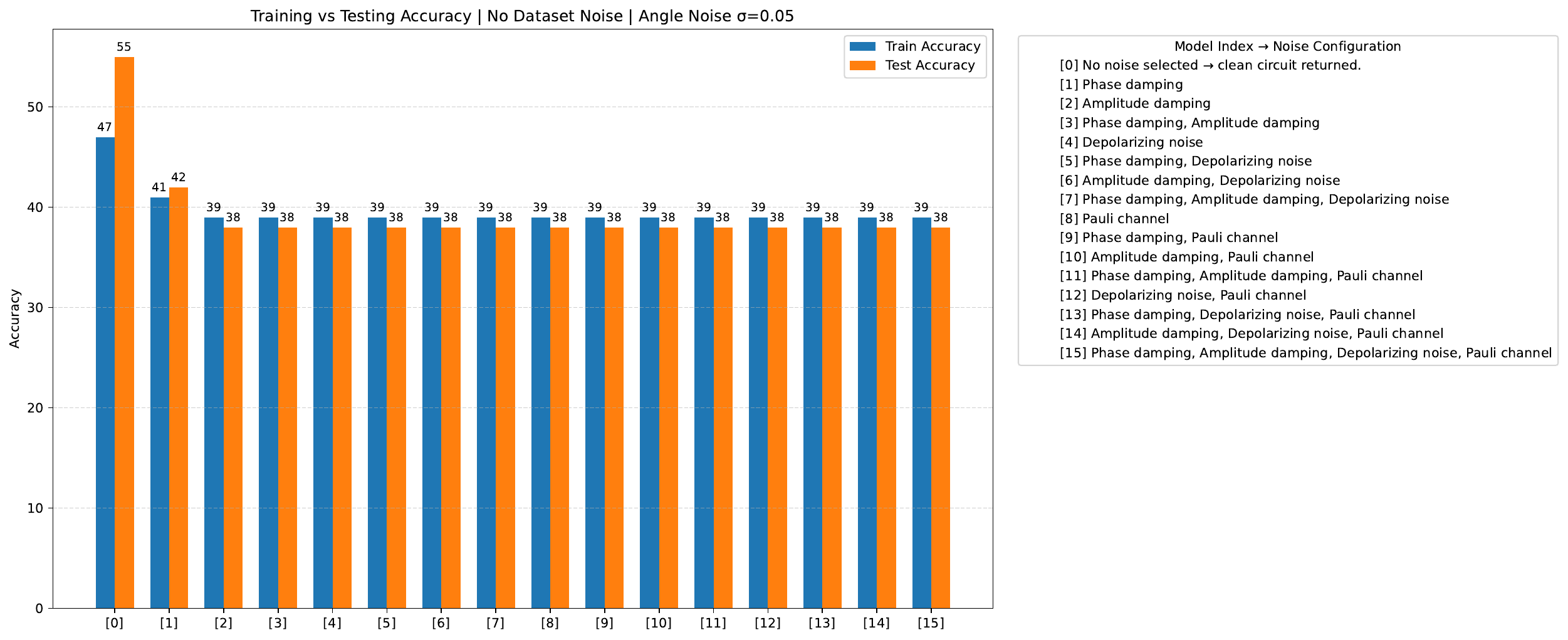}
{Training and testing accuracy comparison under \textbf{no dataset noise}, shown for increasing angle-space perturbations.}
{fig:acc_nods_grouped}


\FourPanelFigure
{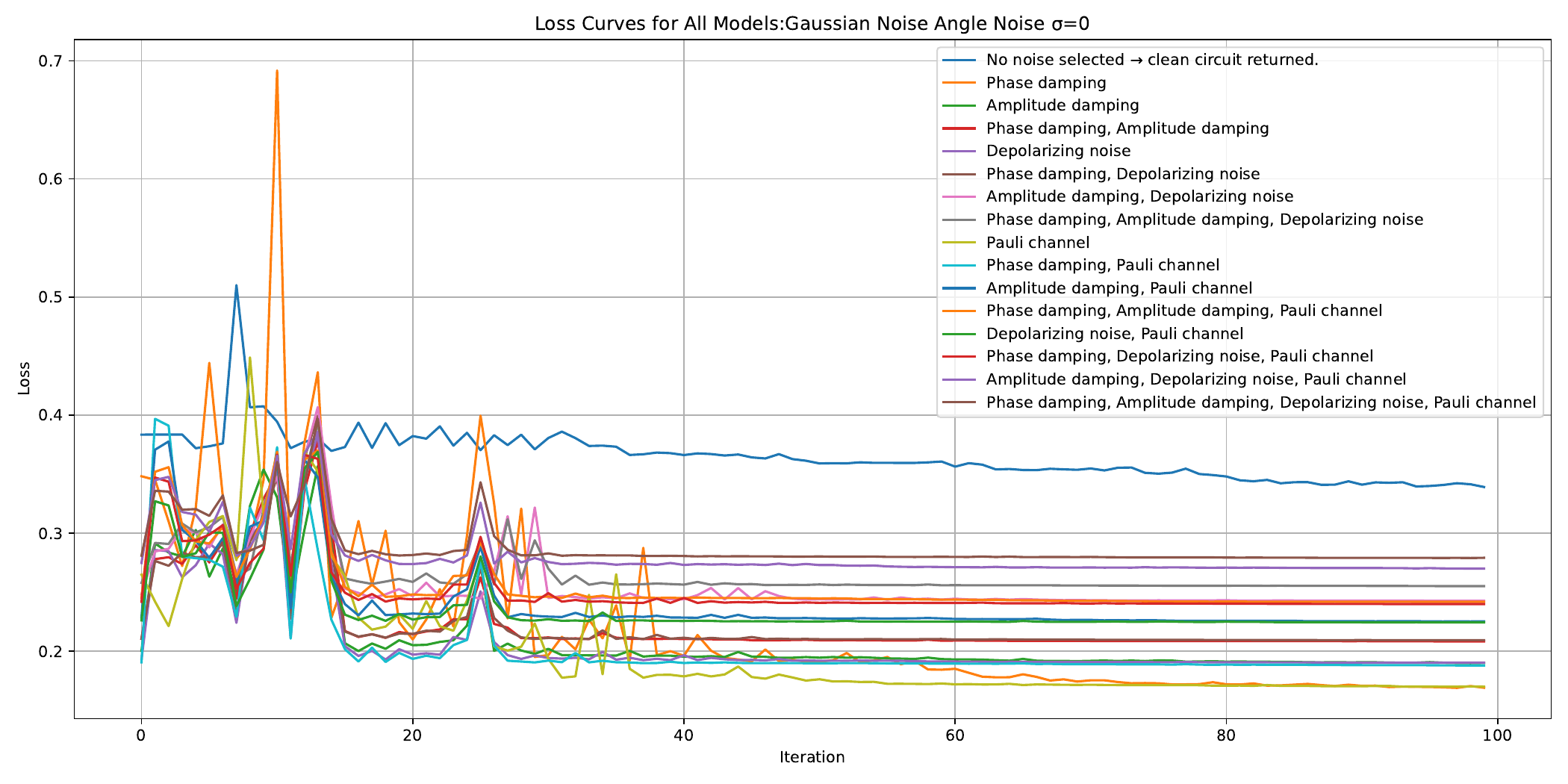}
{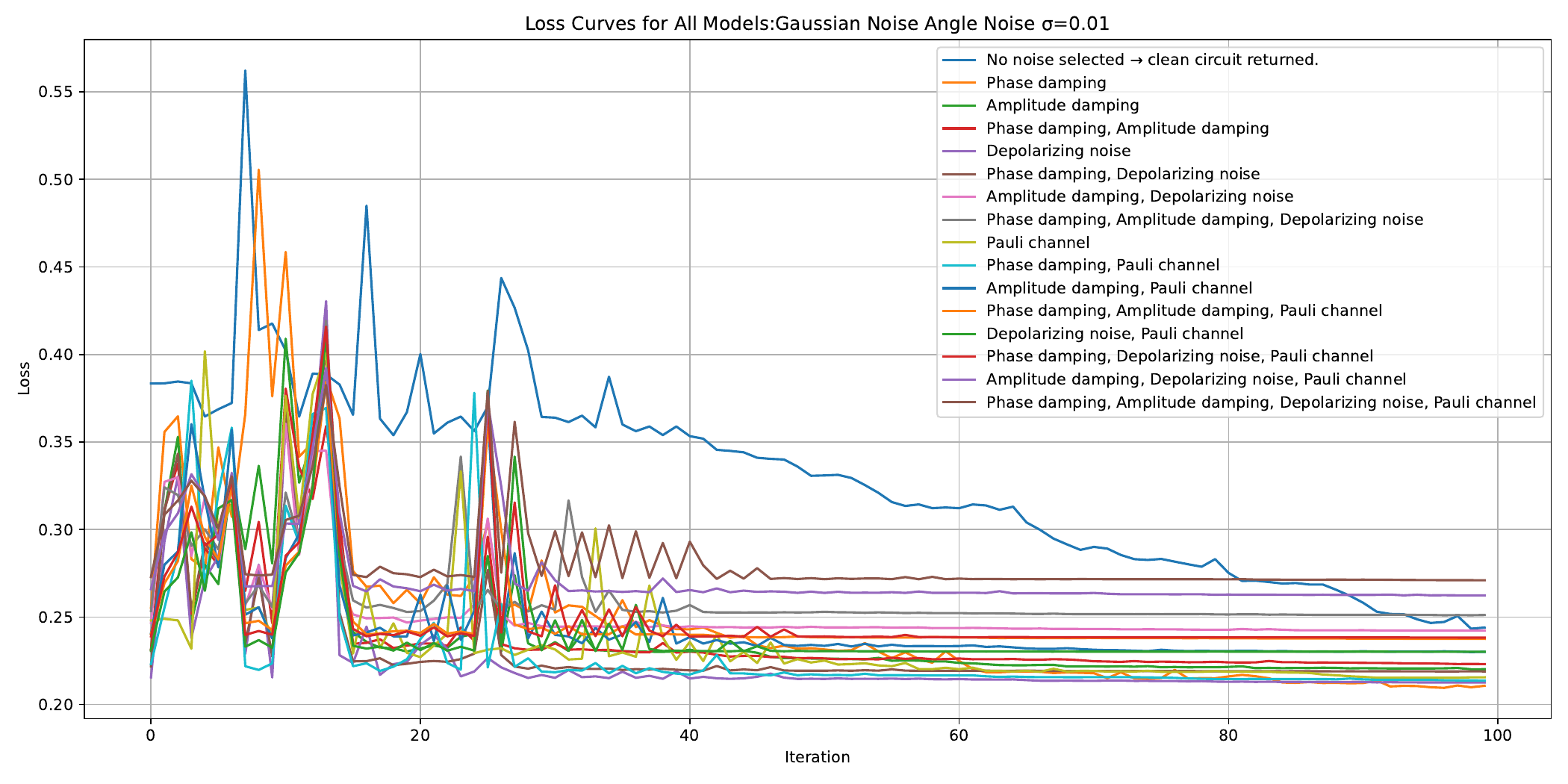}
{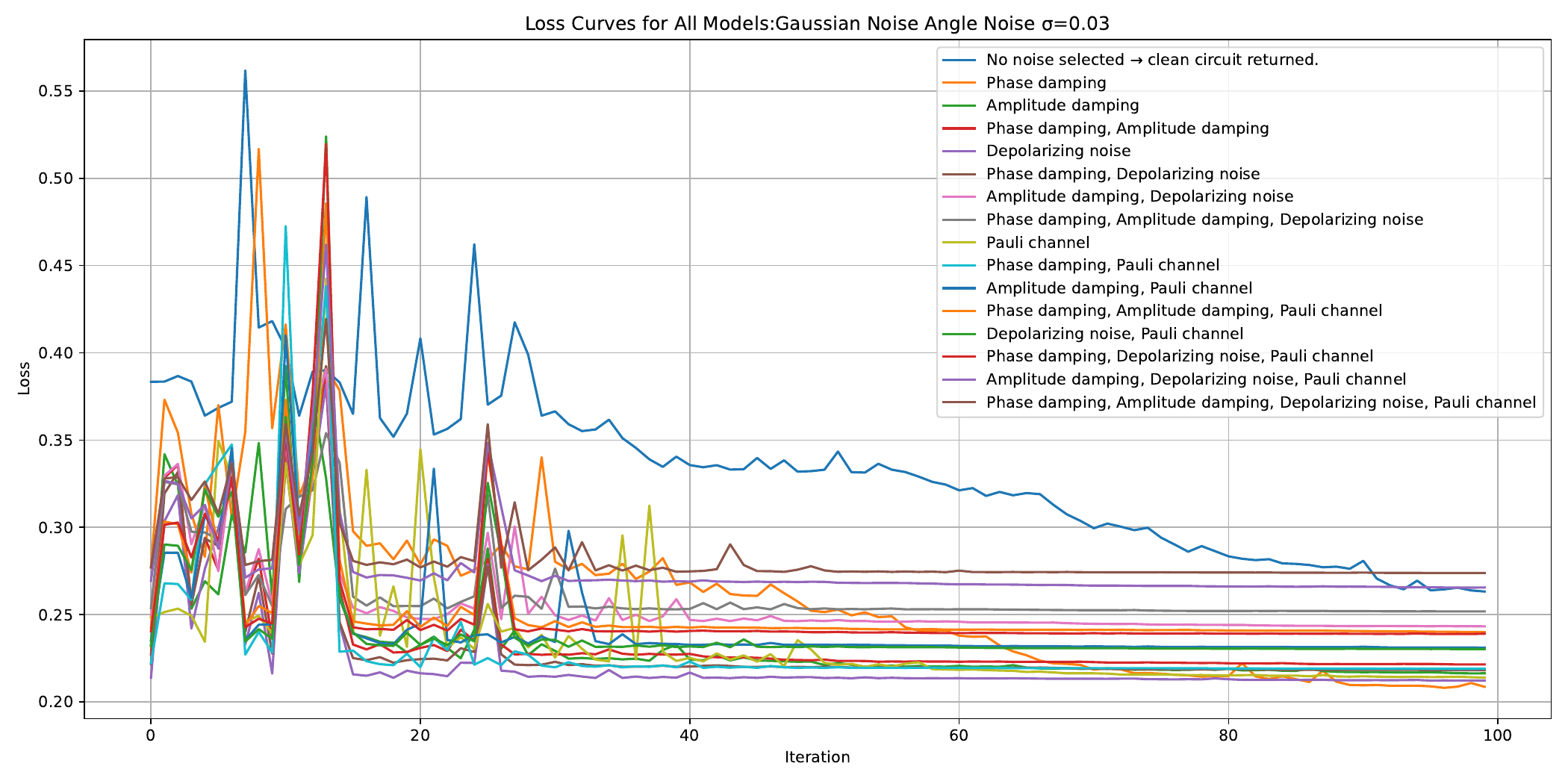}
{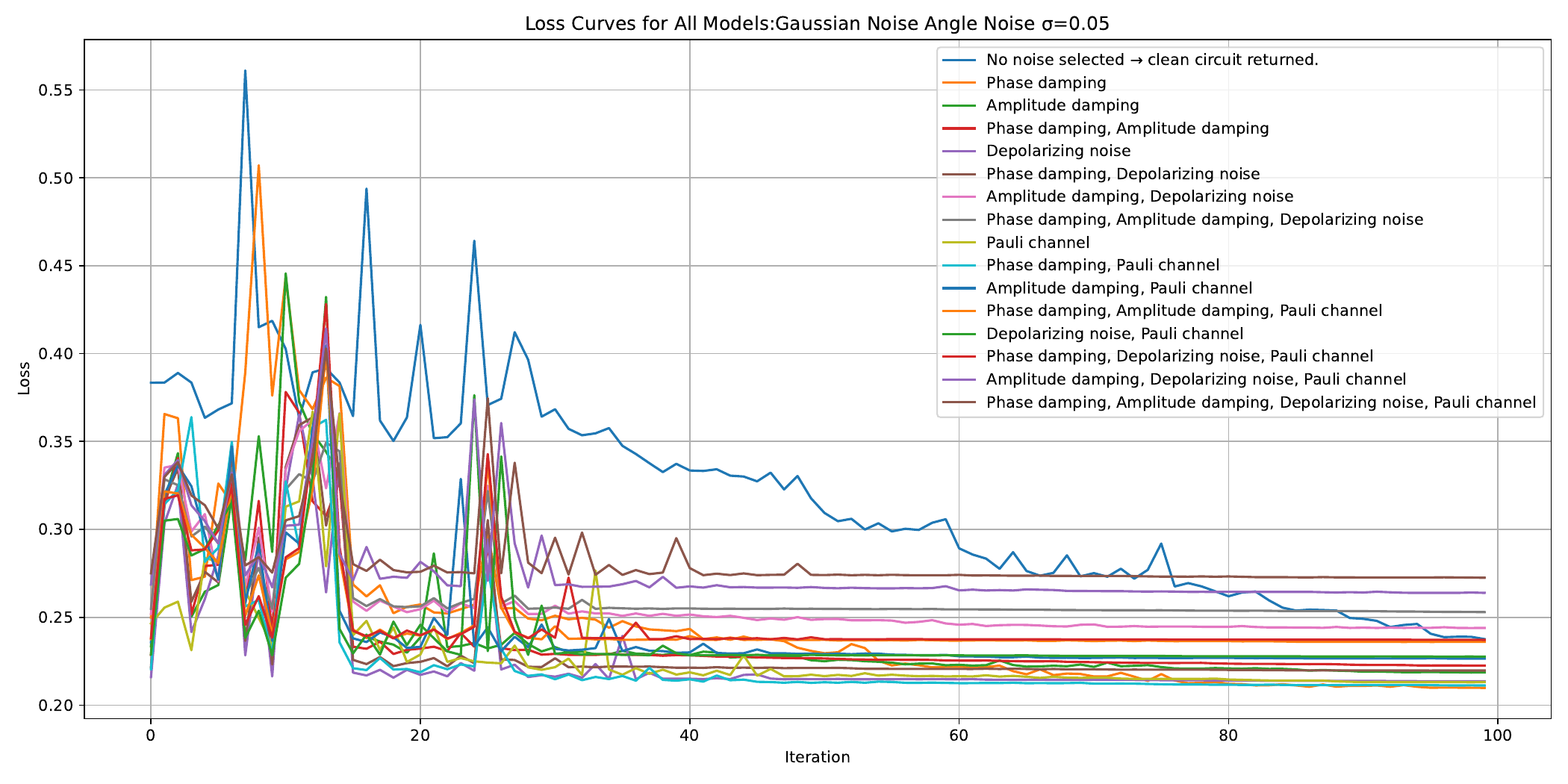}
{Loss convergence curves under \textbf{Gaussian dataset noise}.}
{fig:loss_gauss_grouped}

\FourPanelFigure
{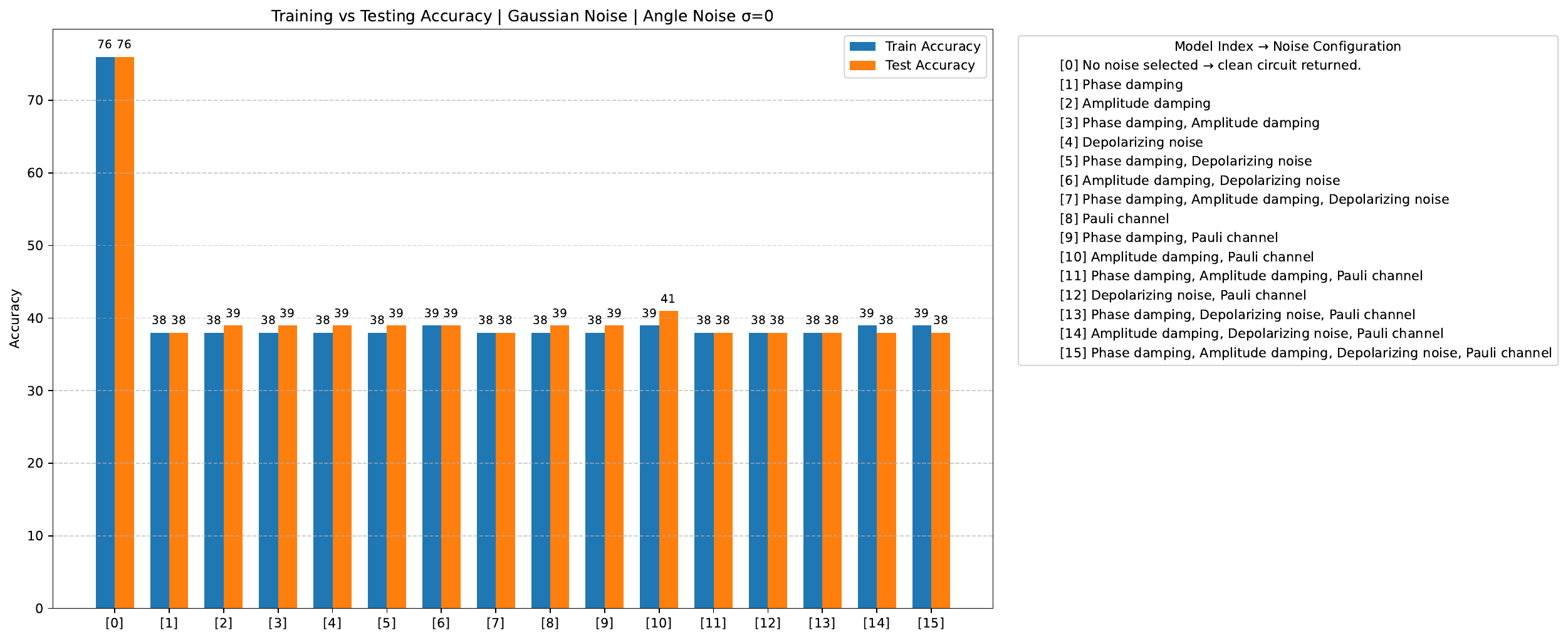}
{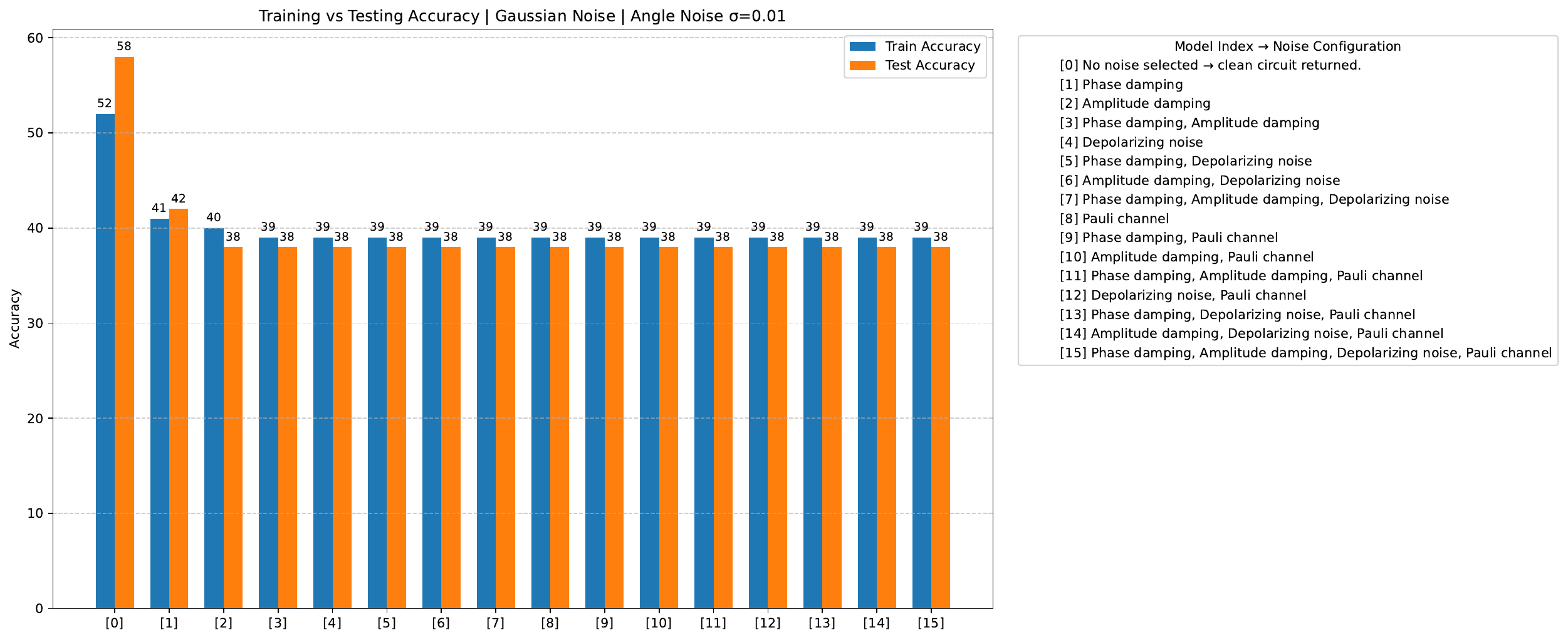}
{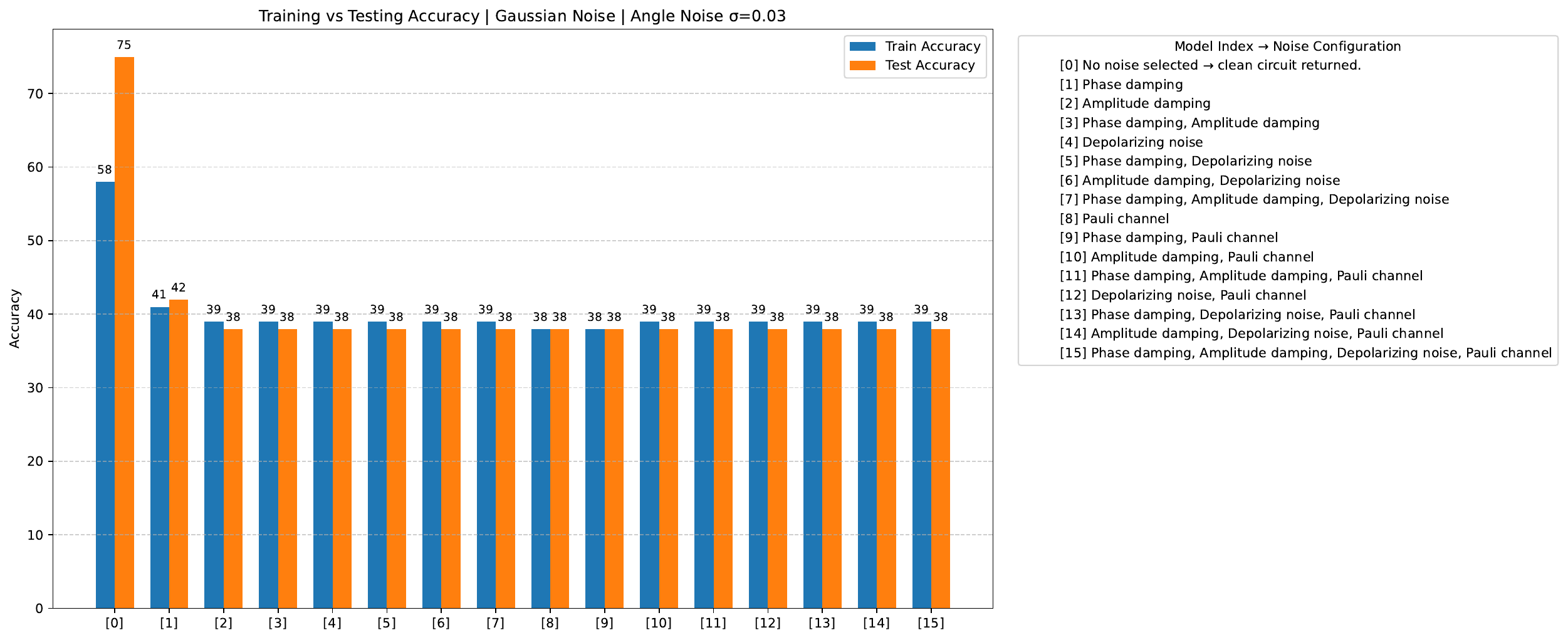}
{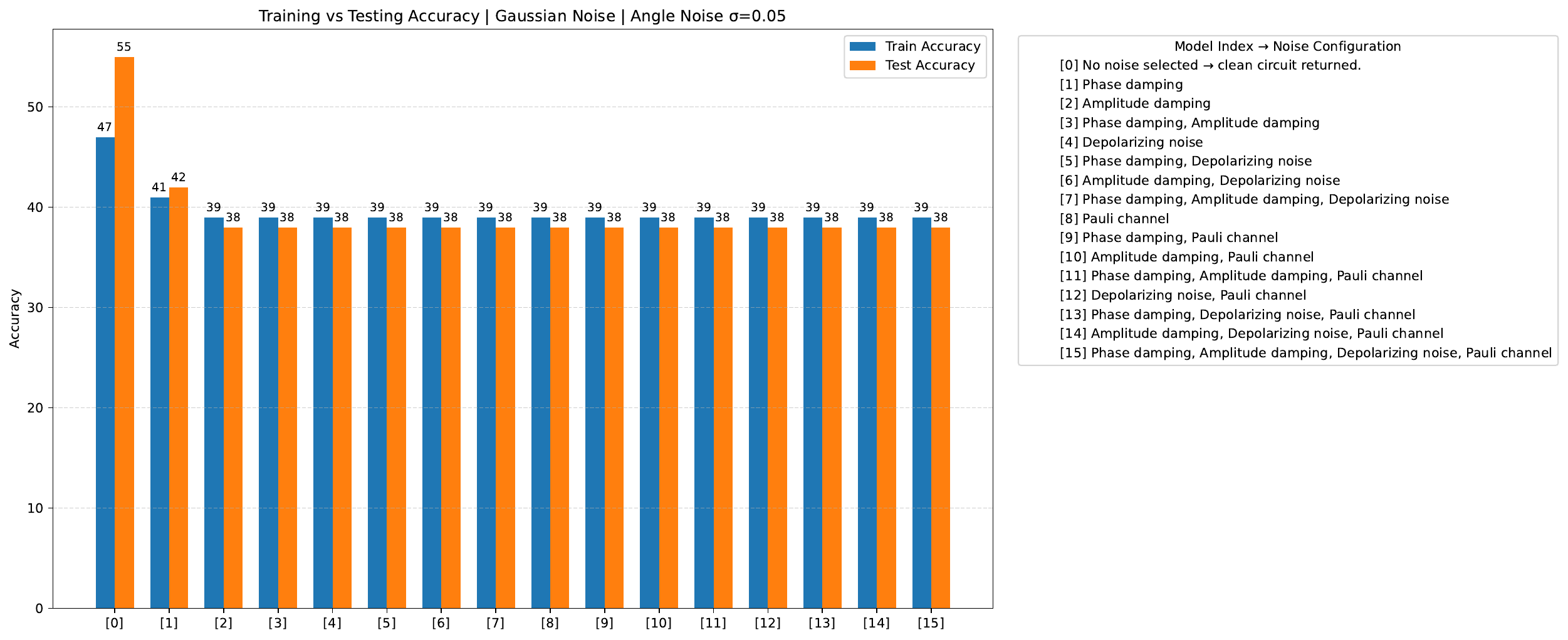}
{Accuracy comparison under \textbf{Gaussian dataset noise}.}
{fig:acc_gauss_grouped}


\FourPanelFigure
{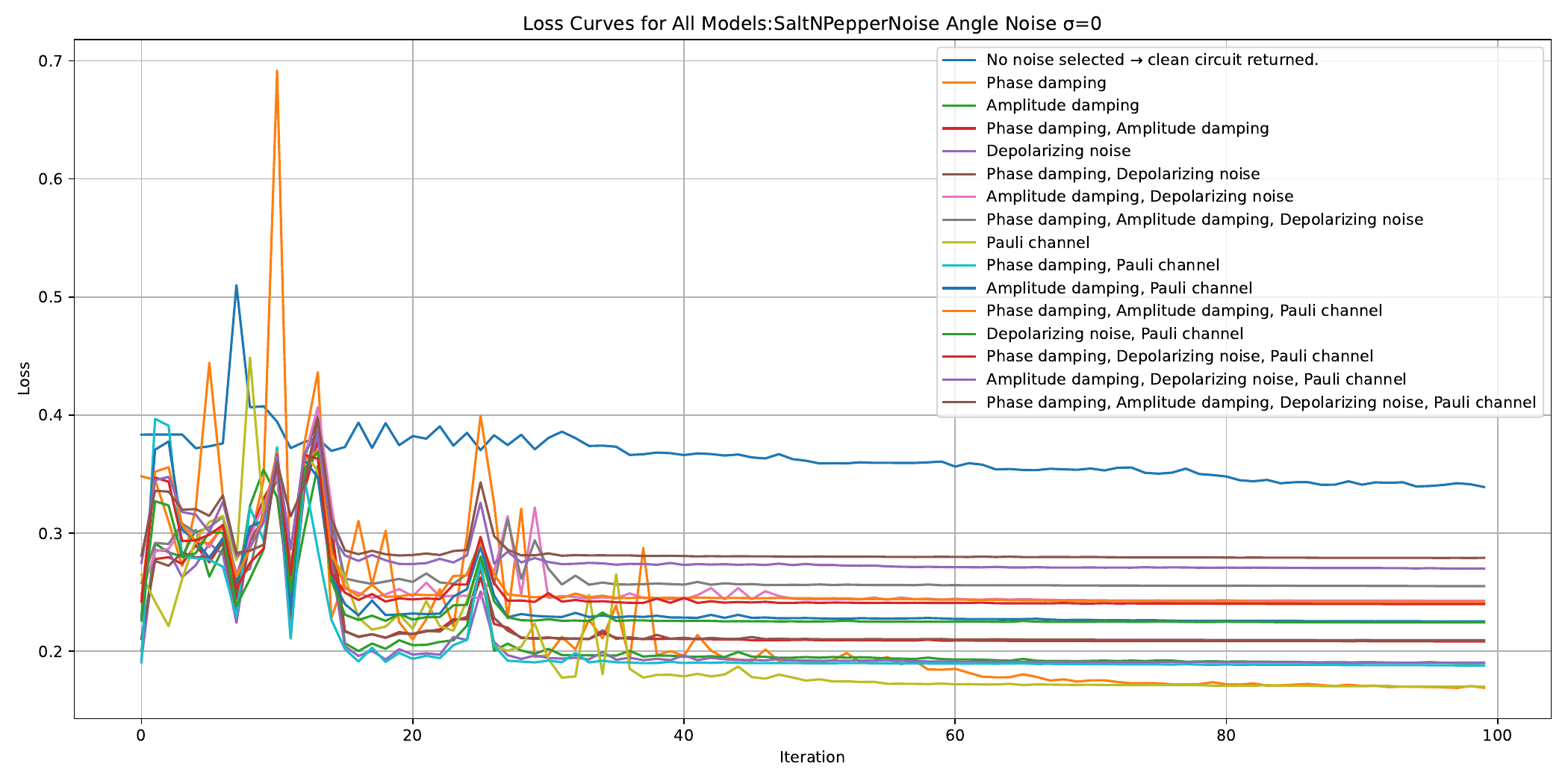}
{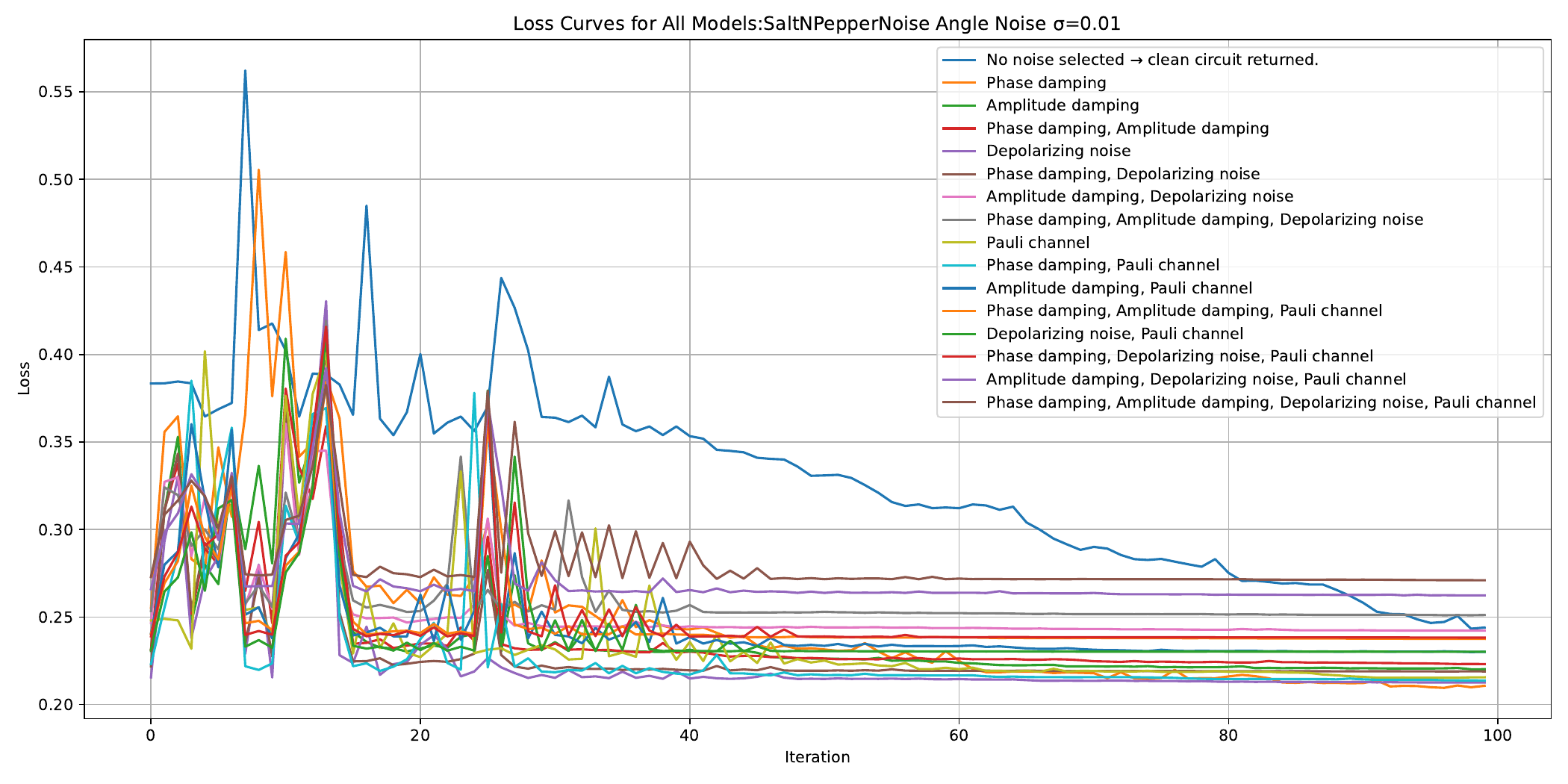}
{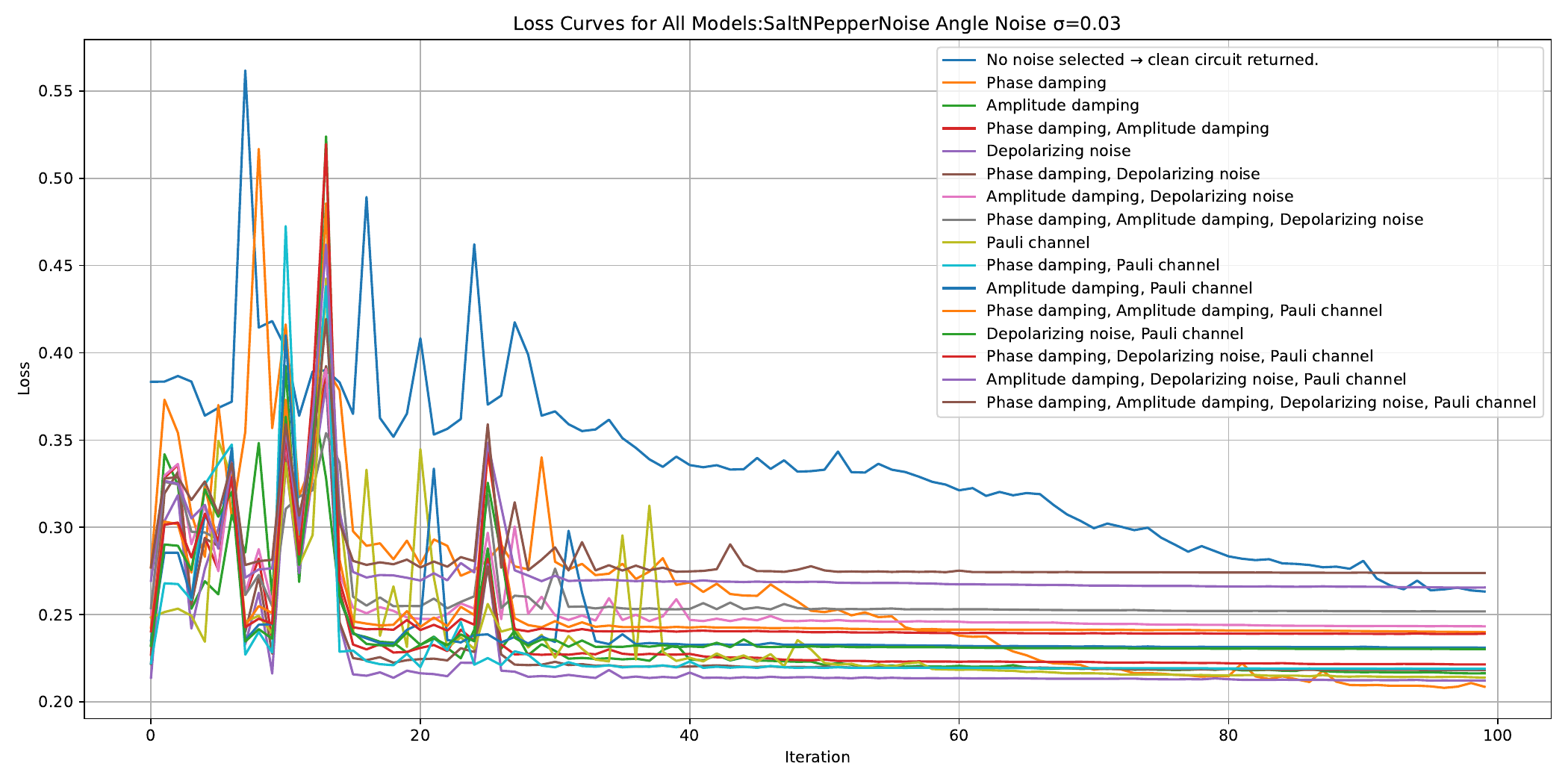}
{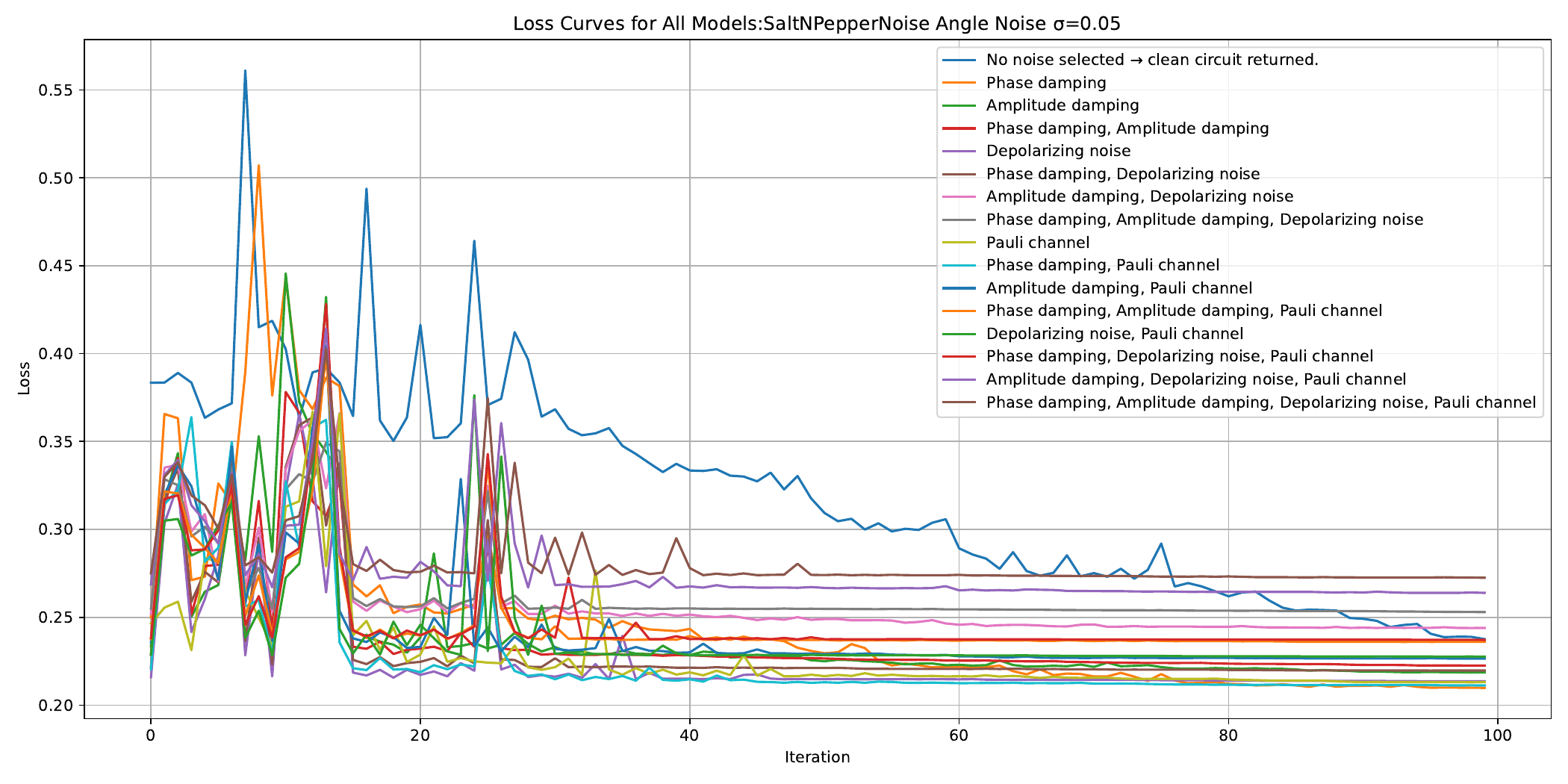}
{Loss convergence curves under \textbf{salt-and-pepper noise}.}
{fig:loss_snp_grouped}

\FourPanelFigure
{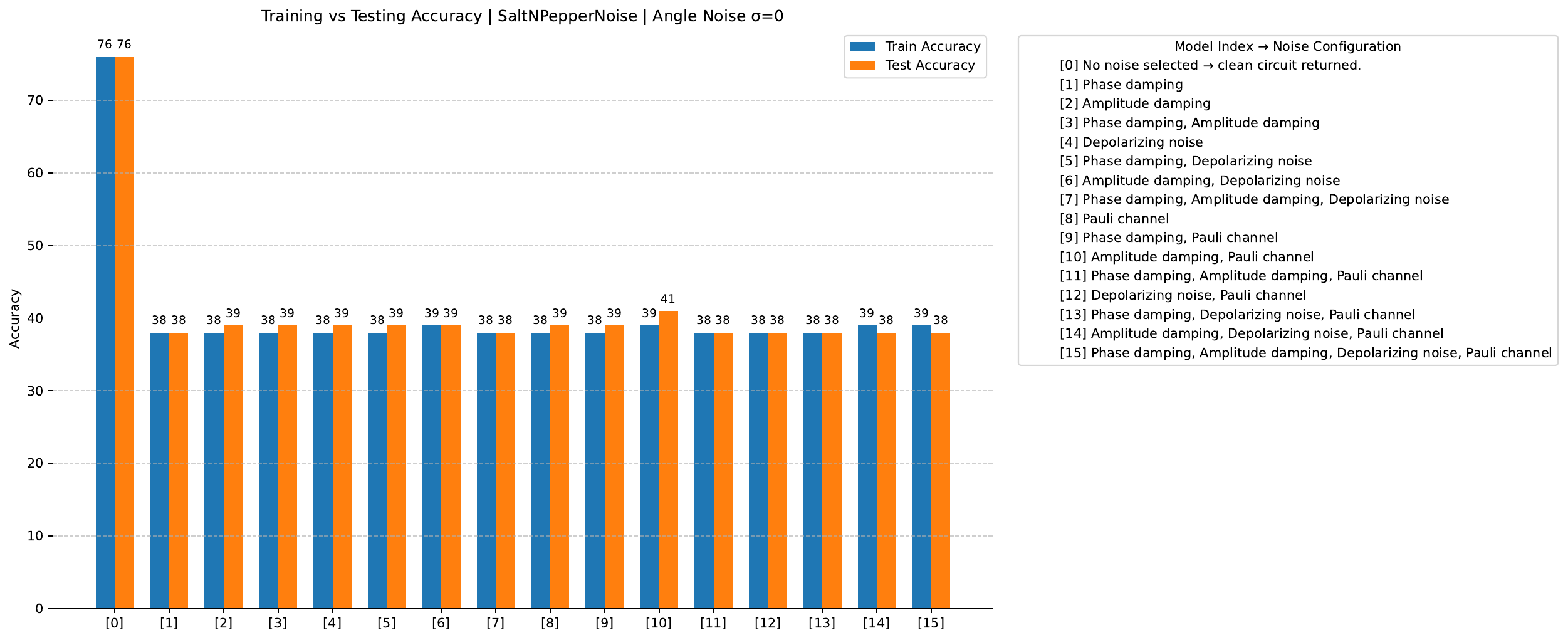}
{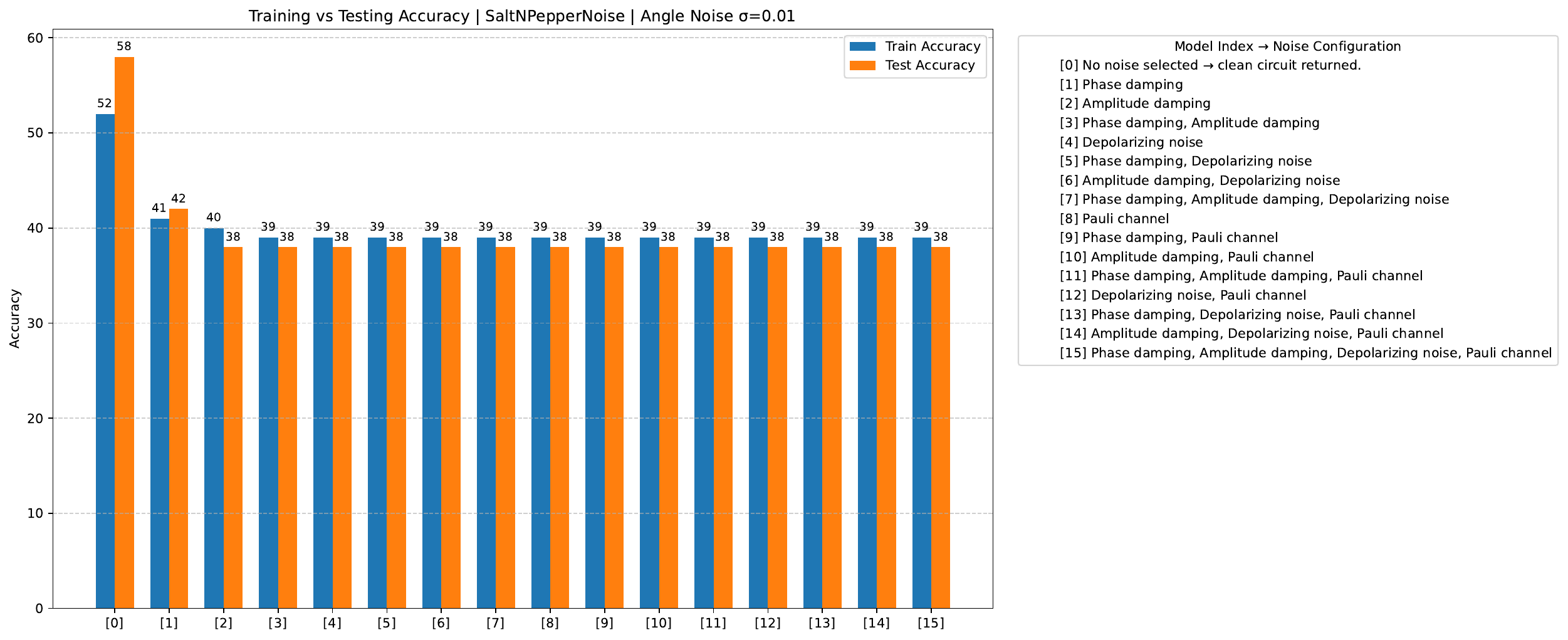}
{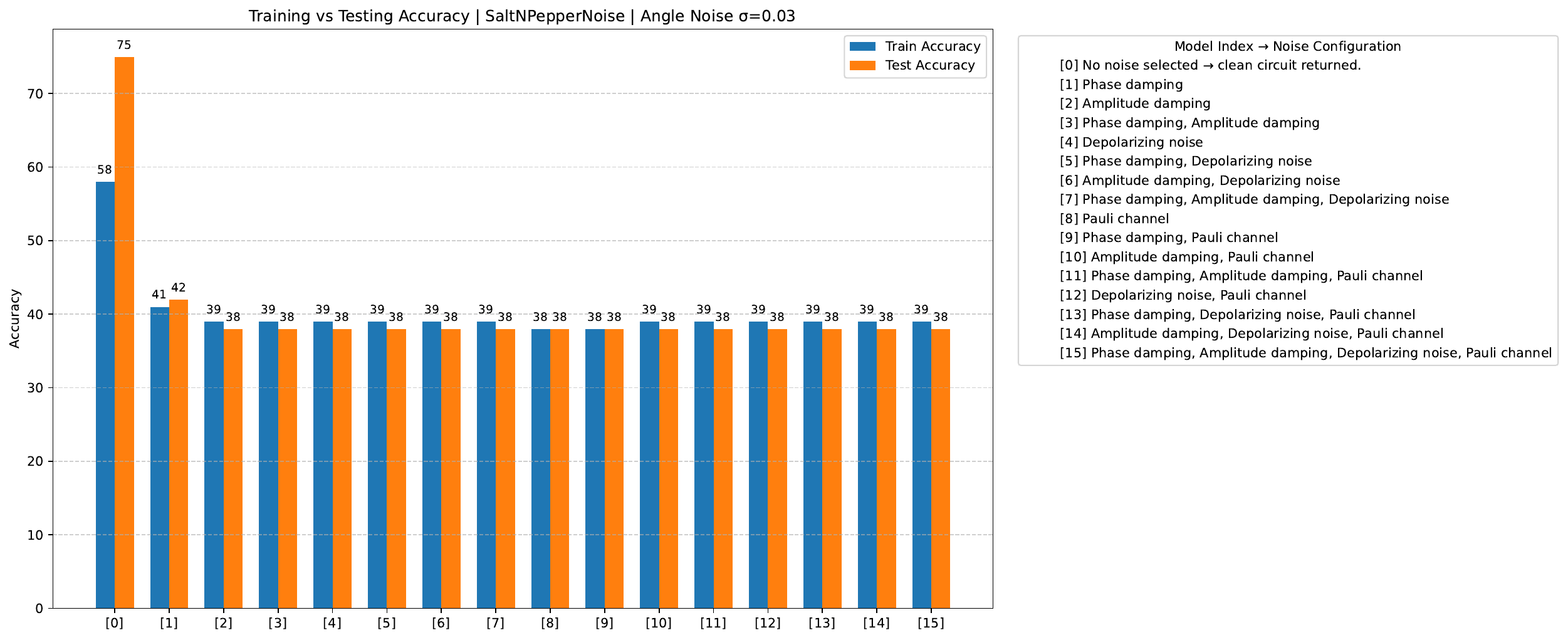}
{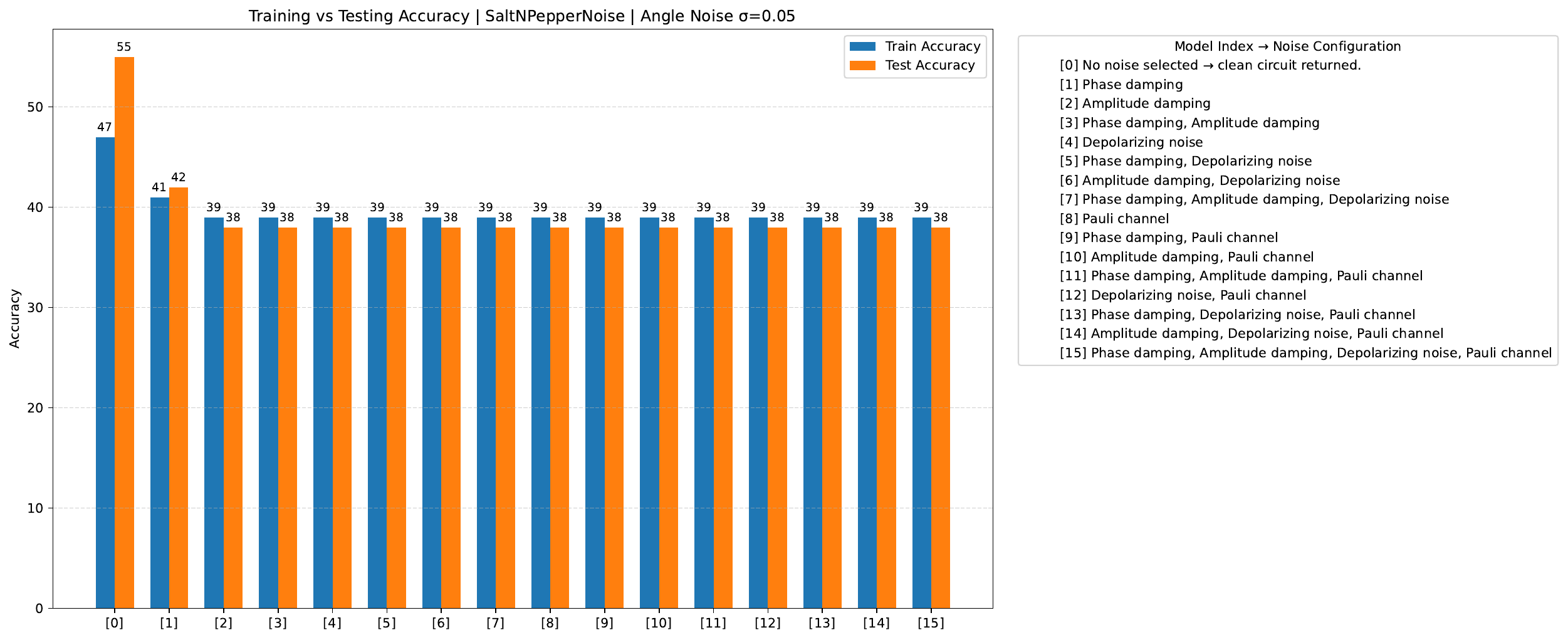}
{Accuracy comparison under \textbf{salt-and-pepper noise}.}
{fig:acc_snp_grouped}


\FourPanelFigure
{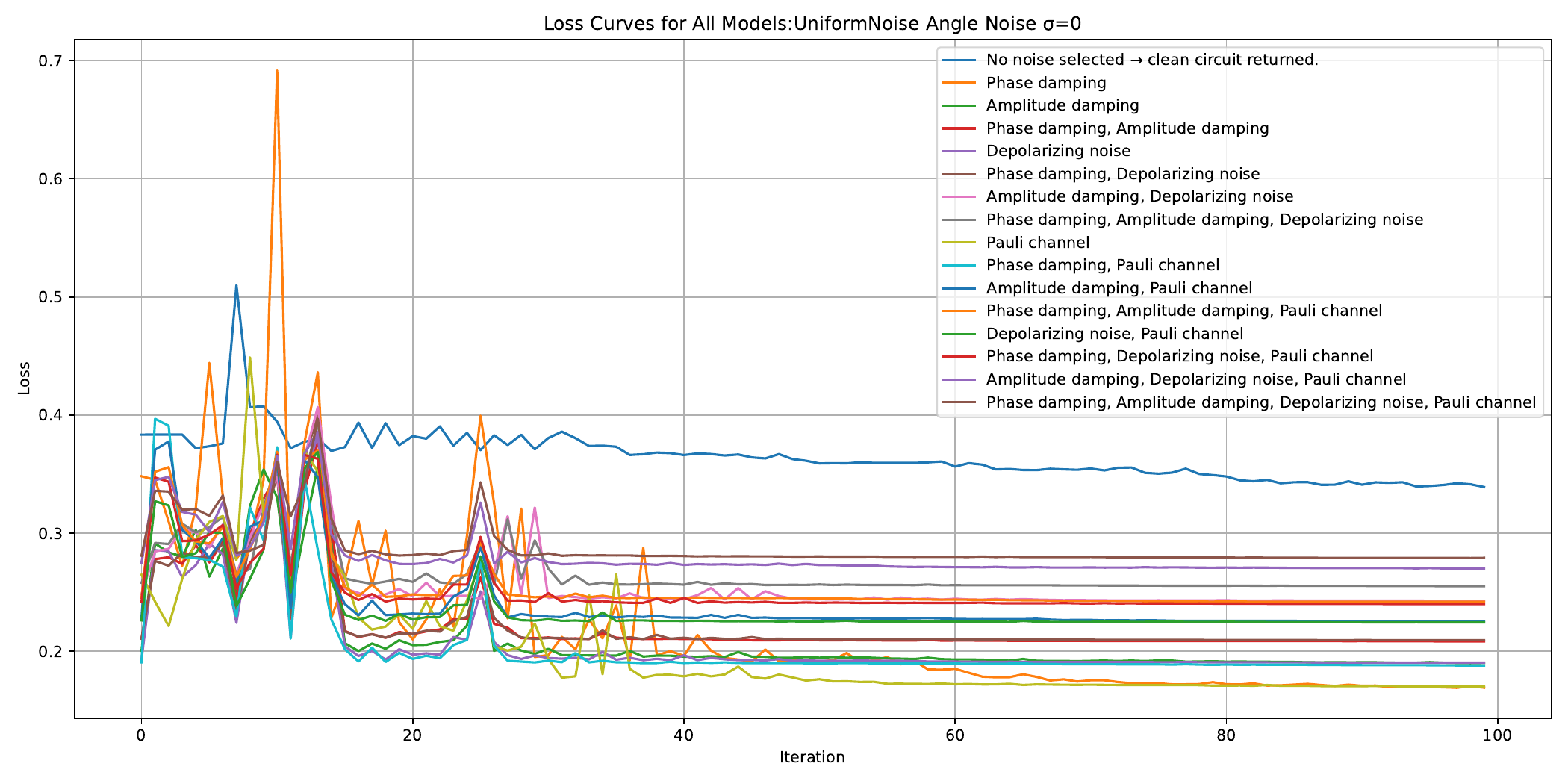}
{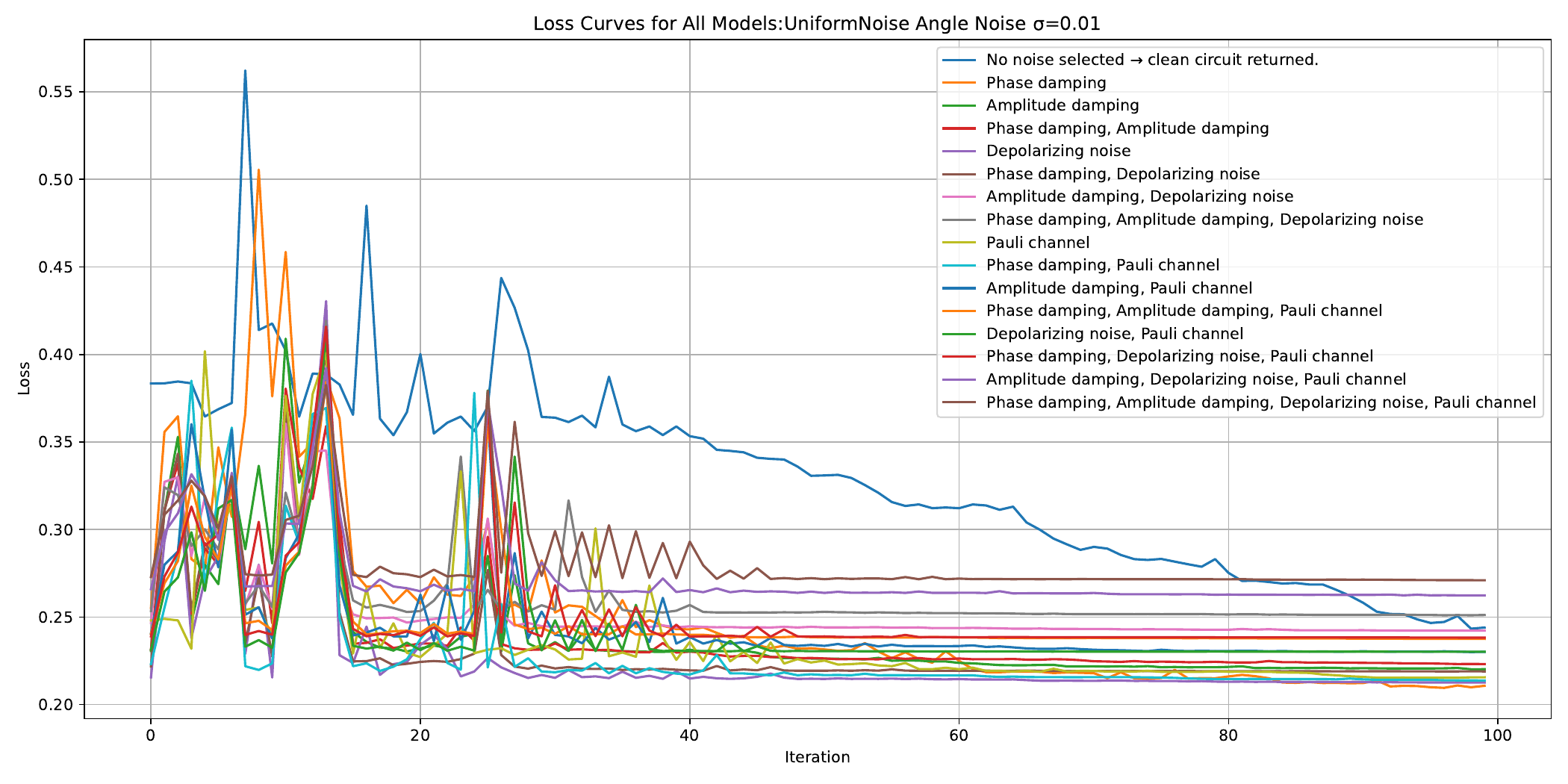}
{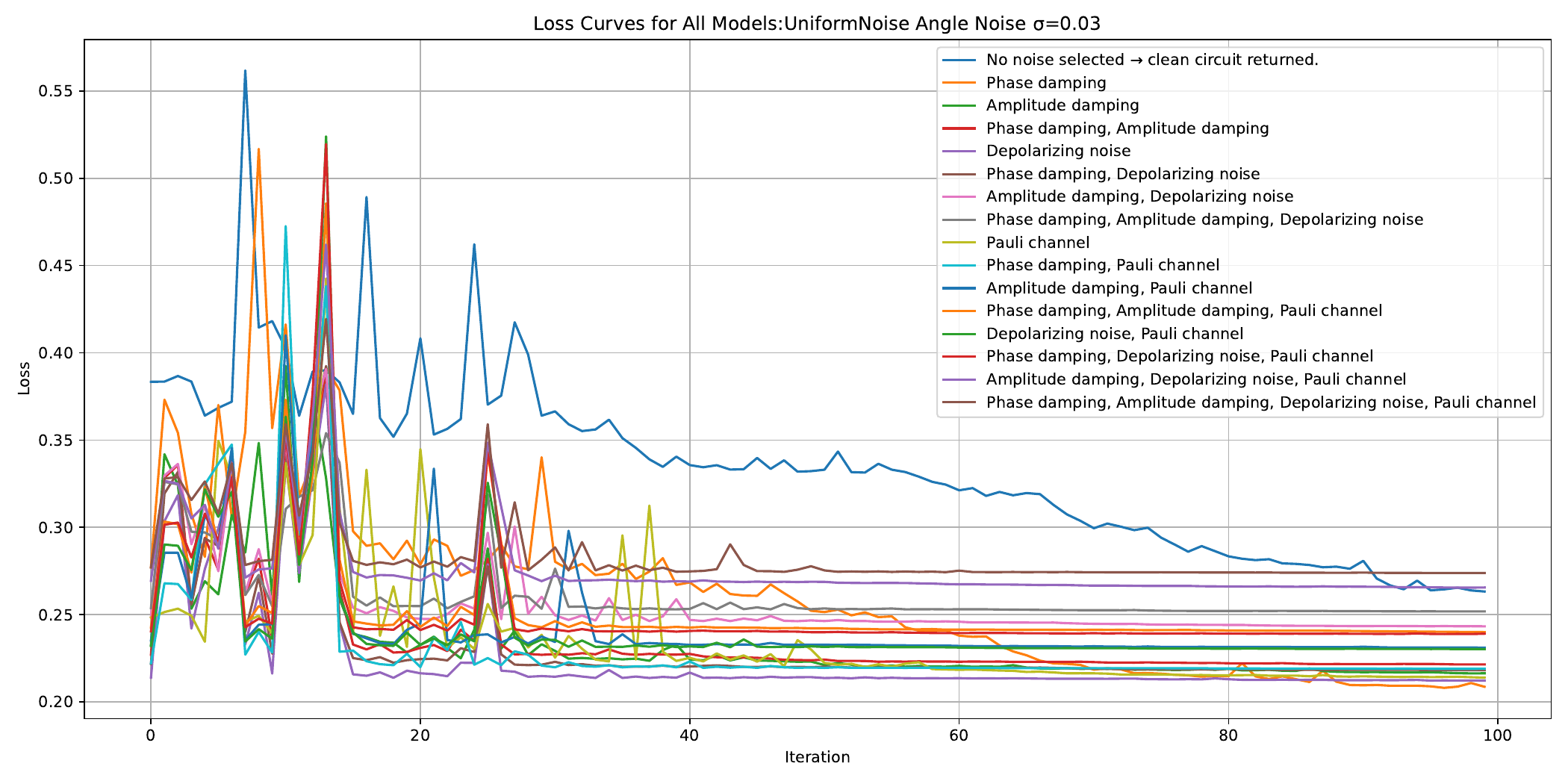}
{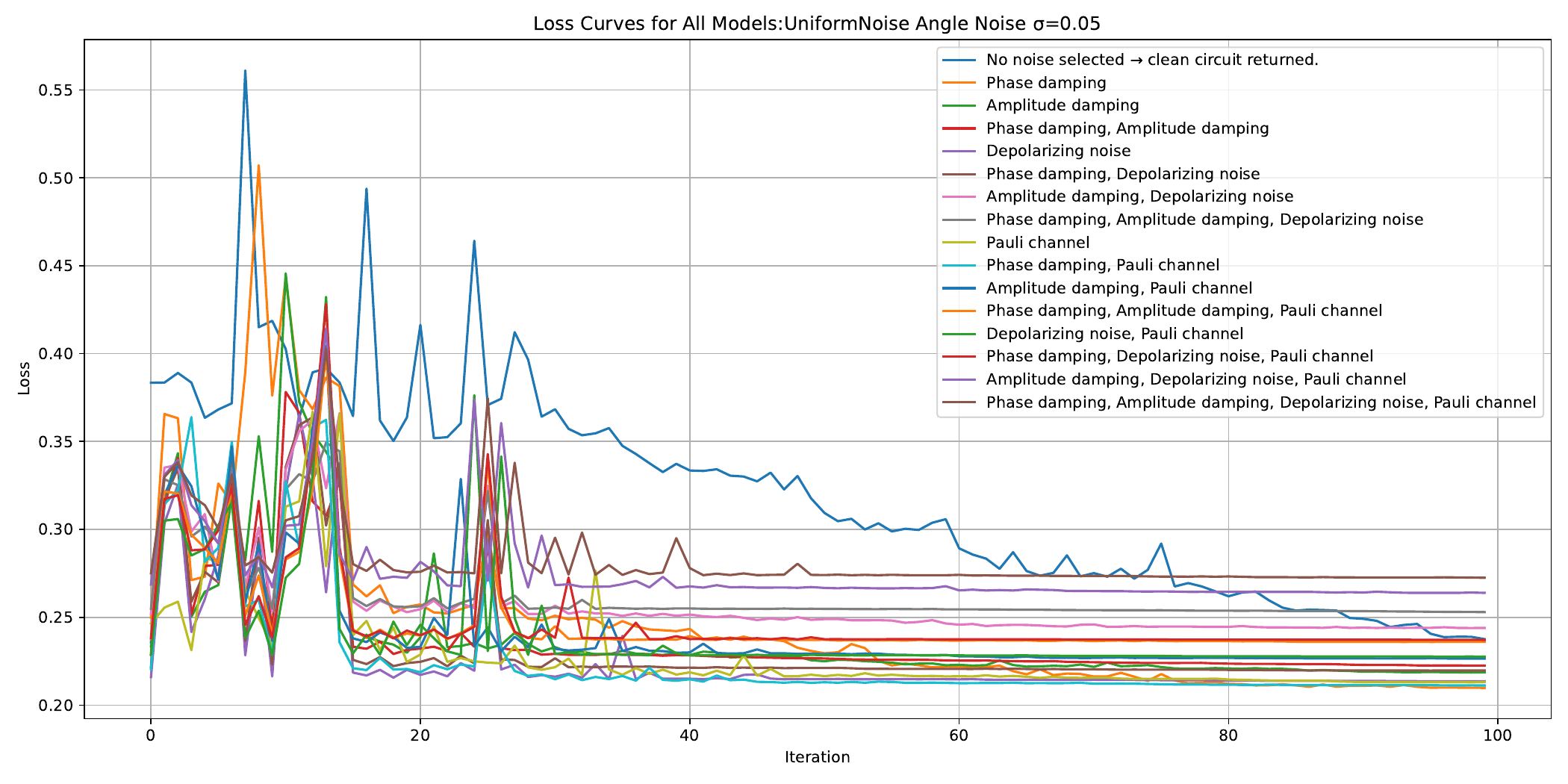}
{Loss convergence curves under \textbf{uniform noise}.}
{fig:loss_unif_grouped}

\FourPanelFigure
{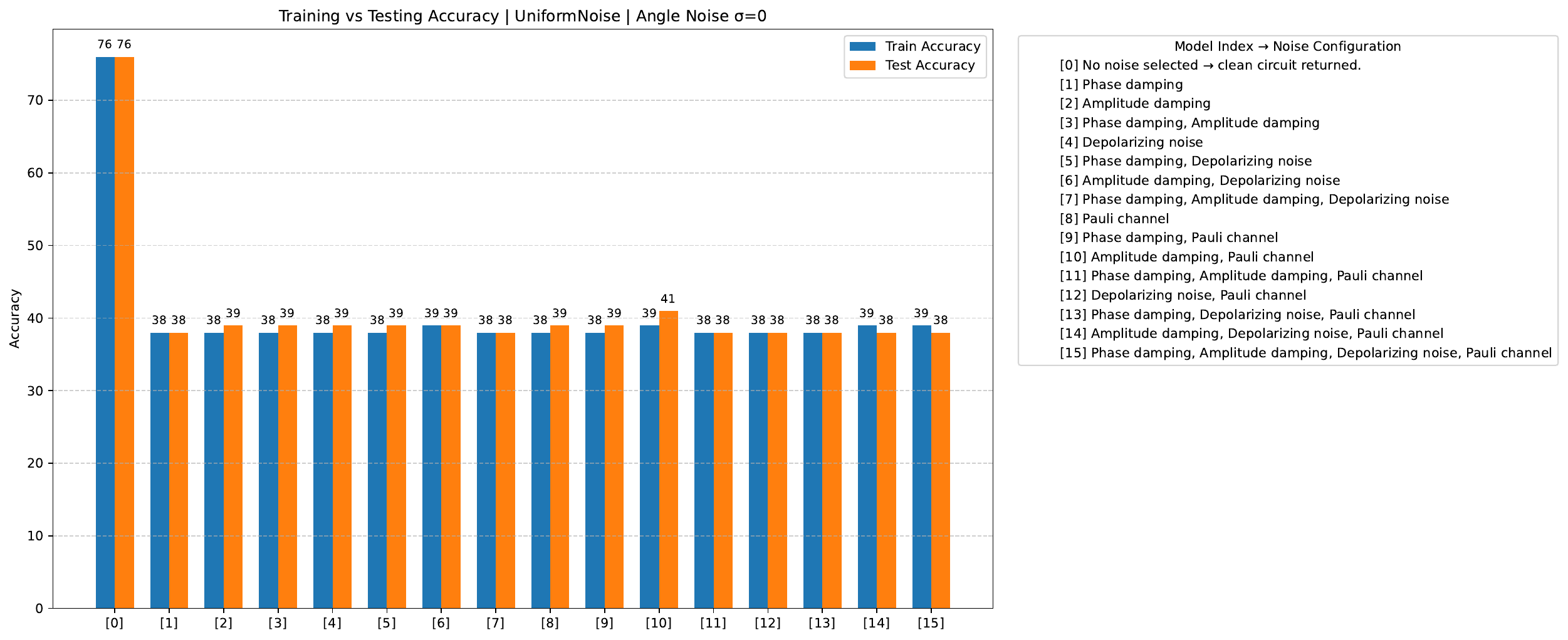}
{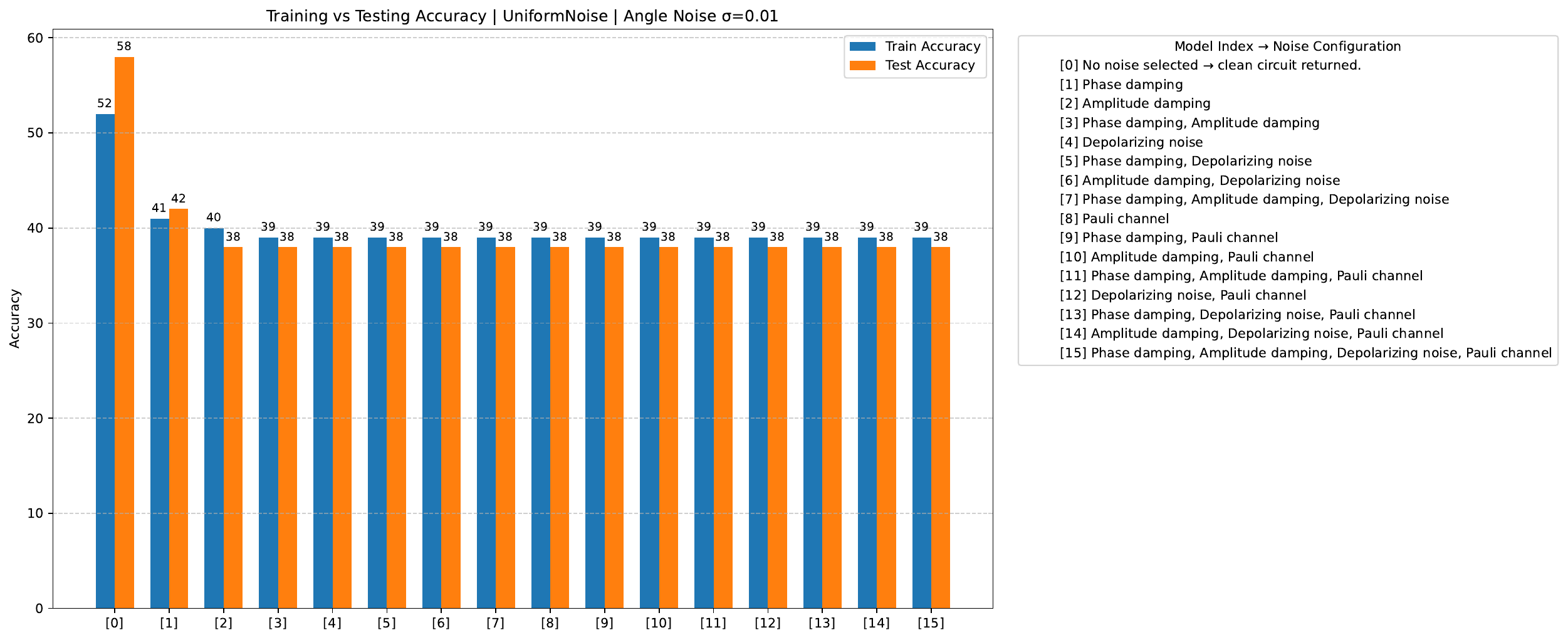}
{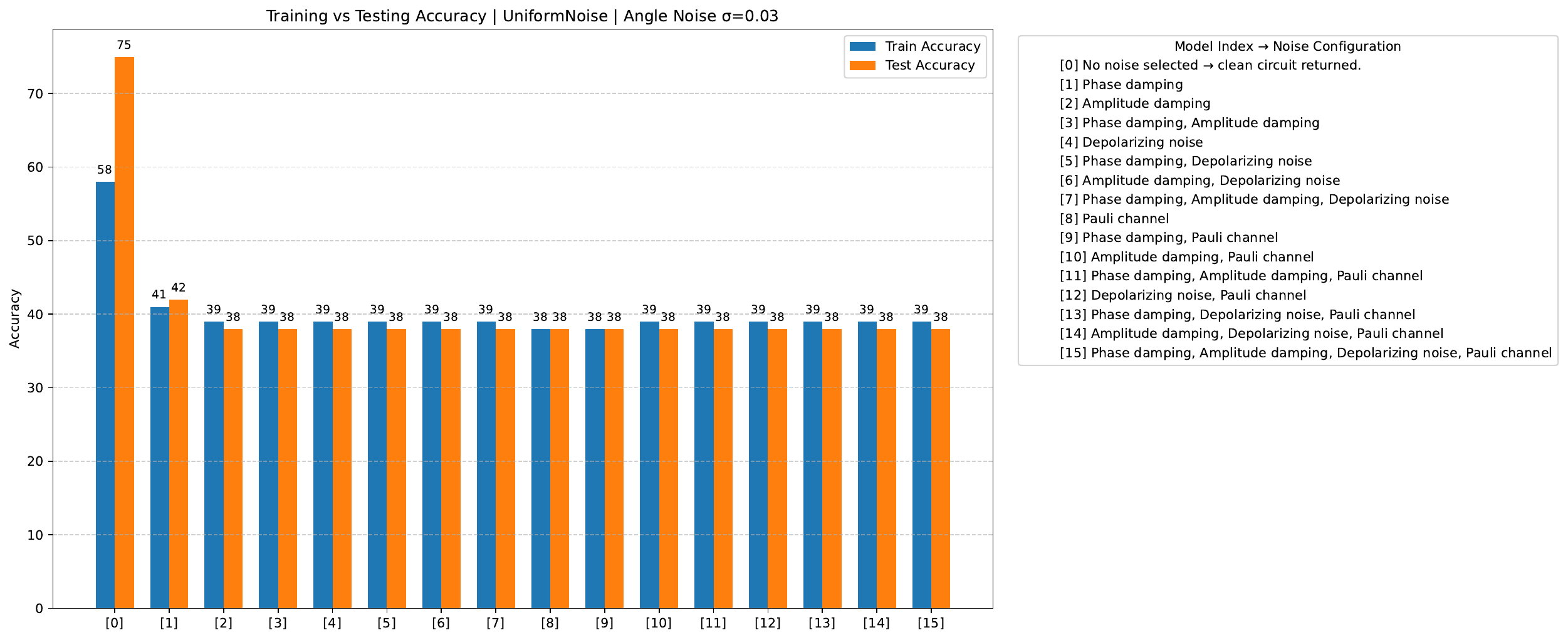}
{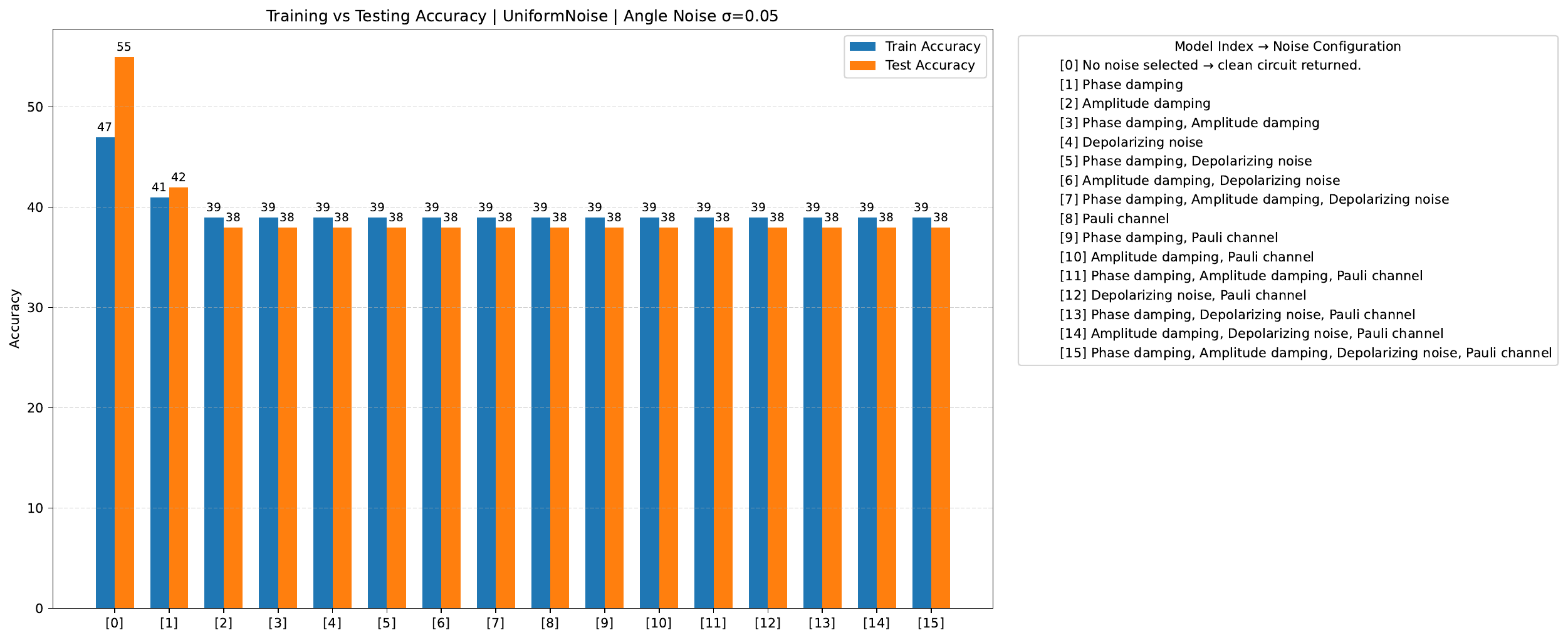}
{Accuracy comparison under \textbf{uniform noise}.}
{fig:acc_unif_grouped}


\FourPanelFigure
{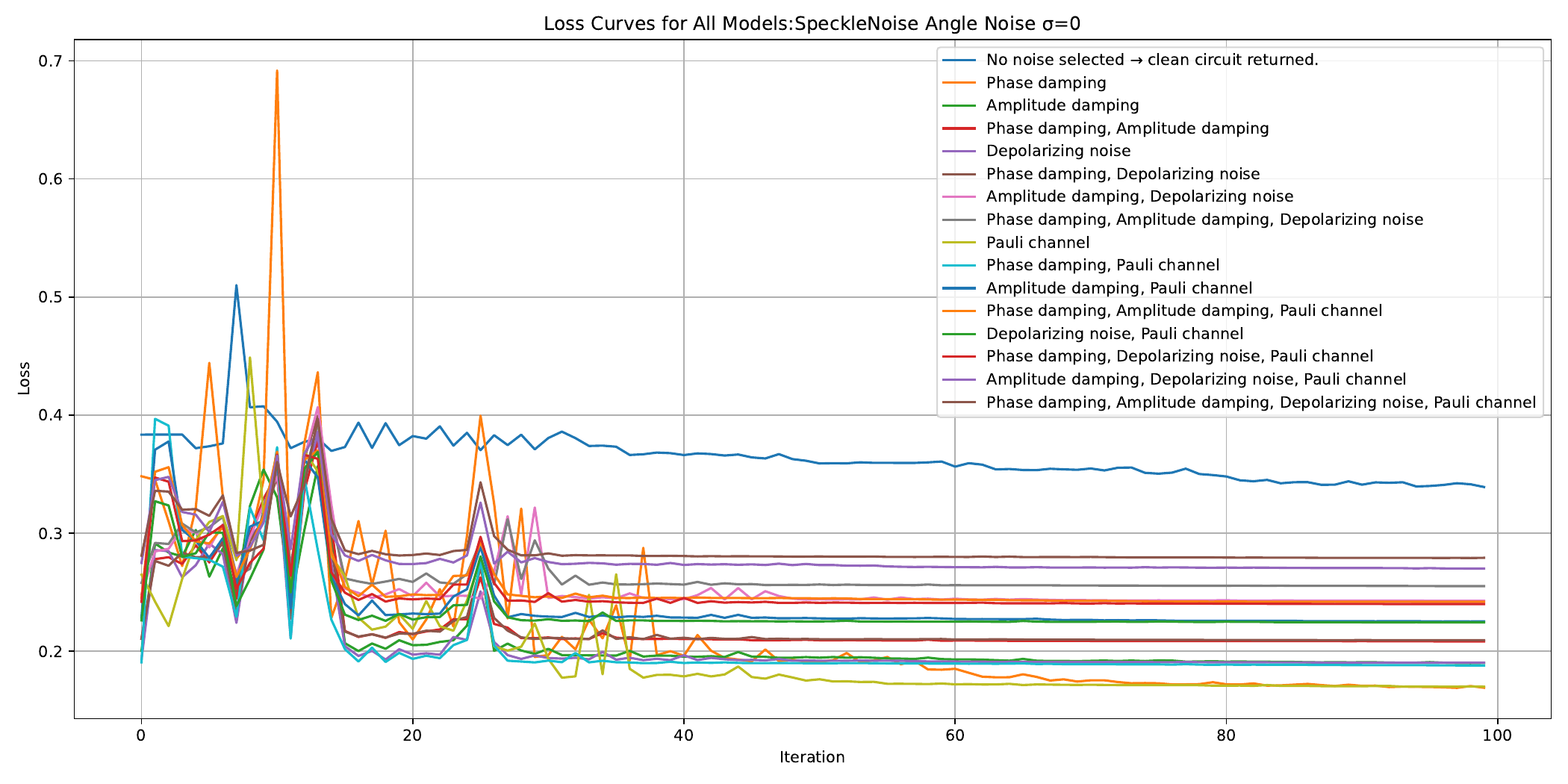}
{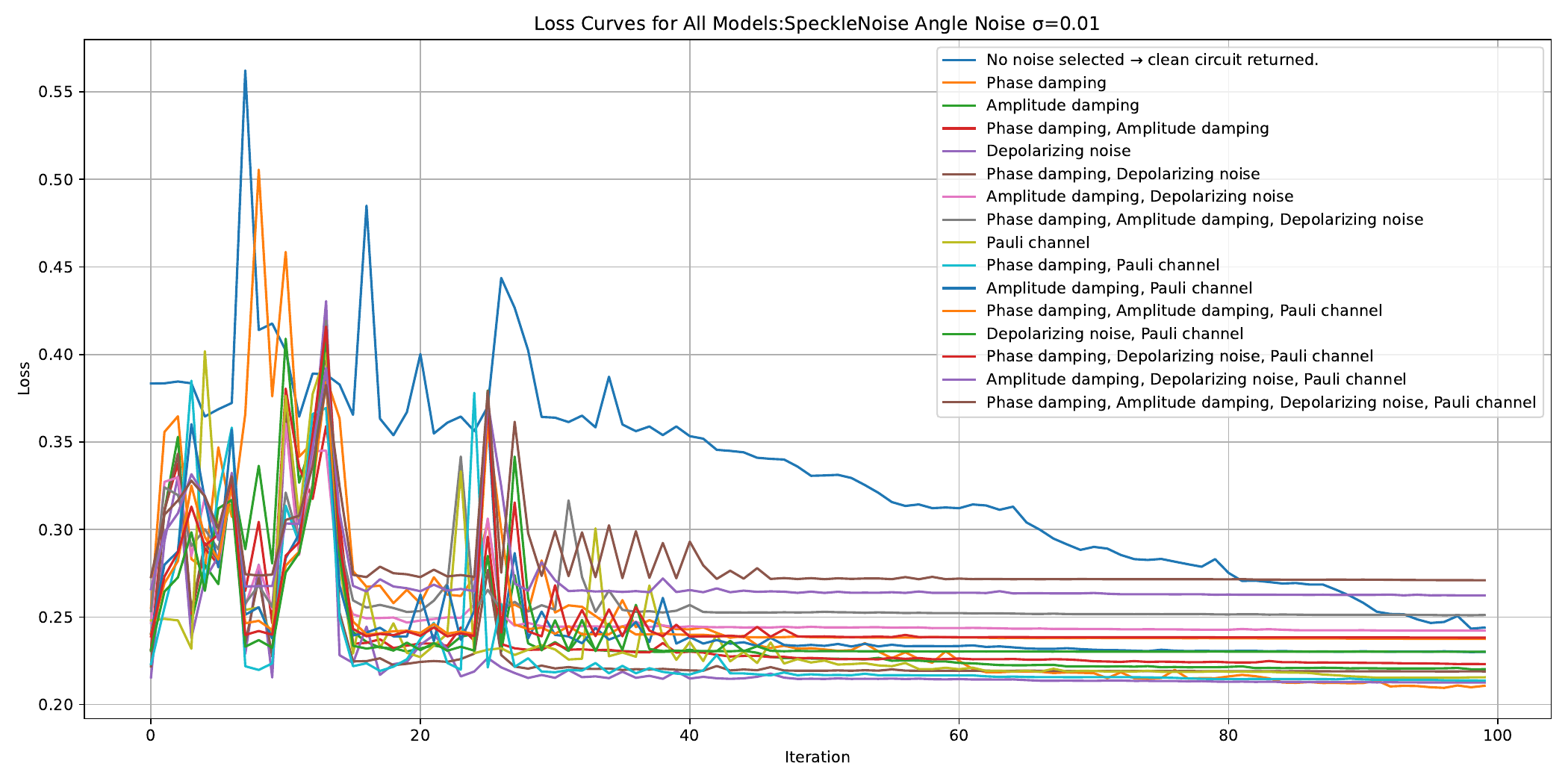}
{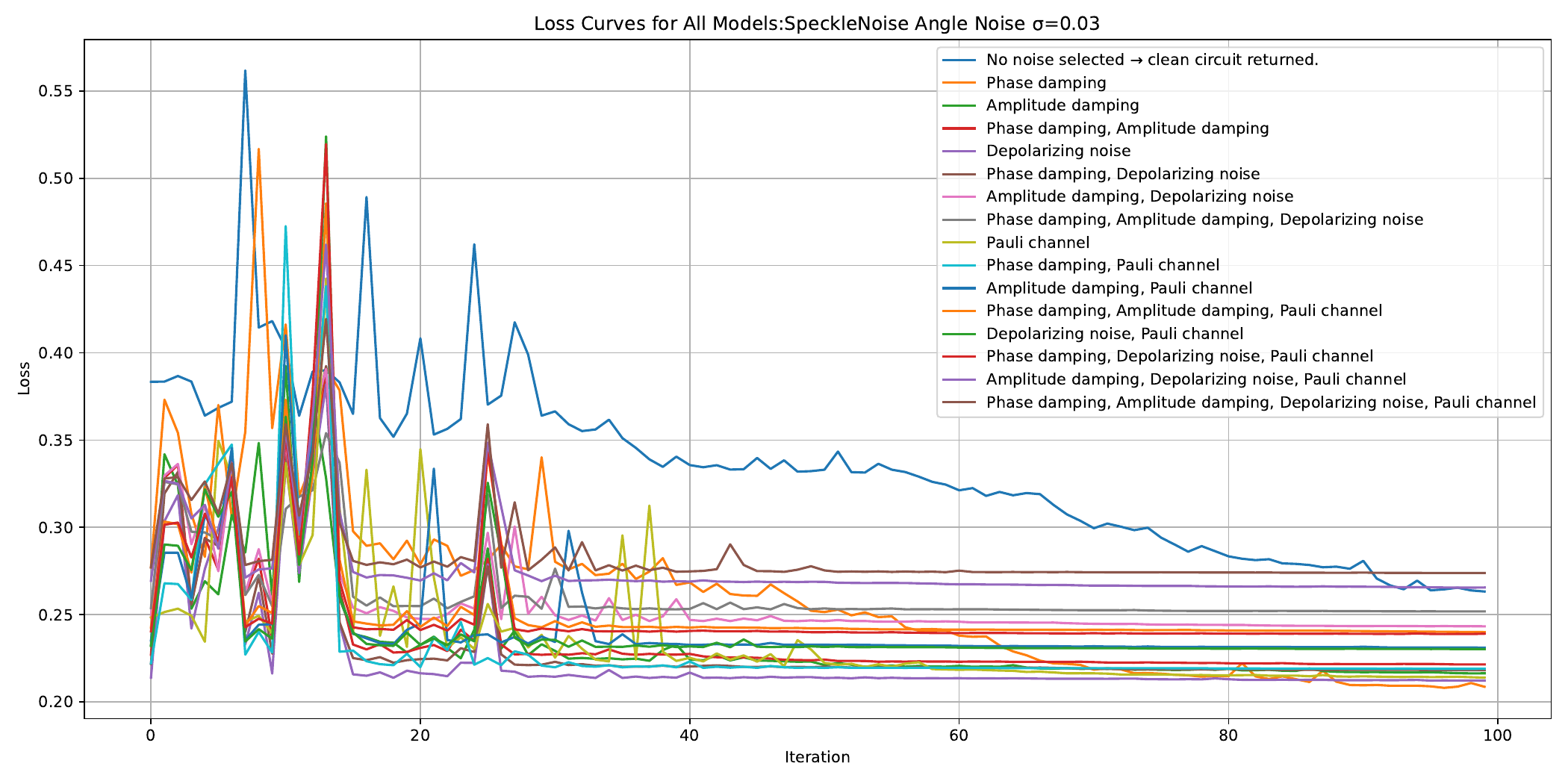}
{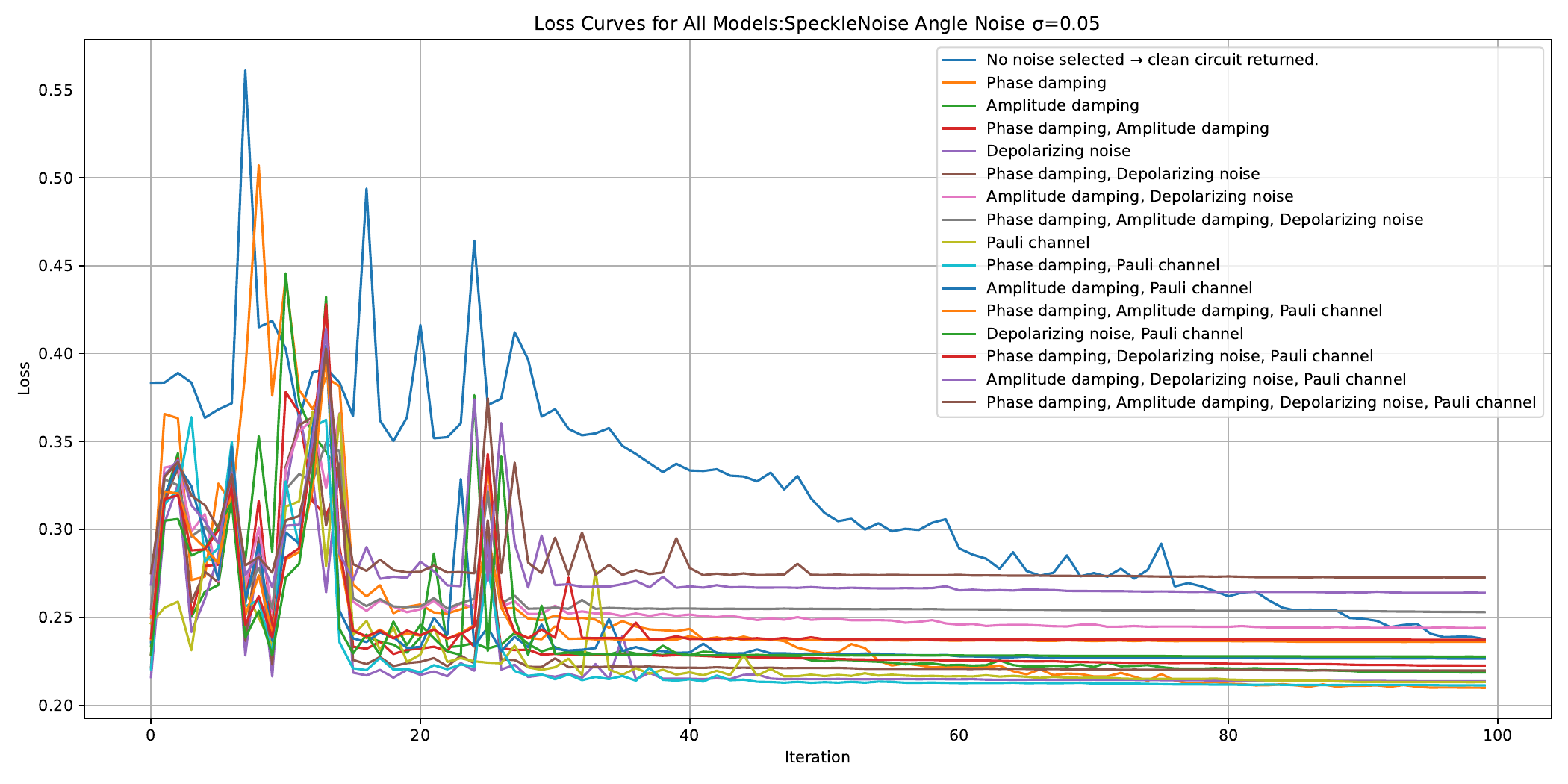}
{Loss convergence curves under \textbf{speckle noise}.}
{fig:loss_spec_grouped}

\FourPanelFigure
{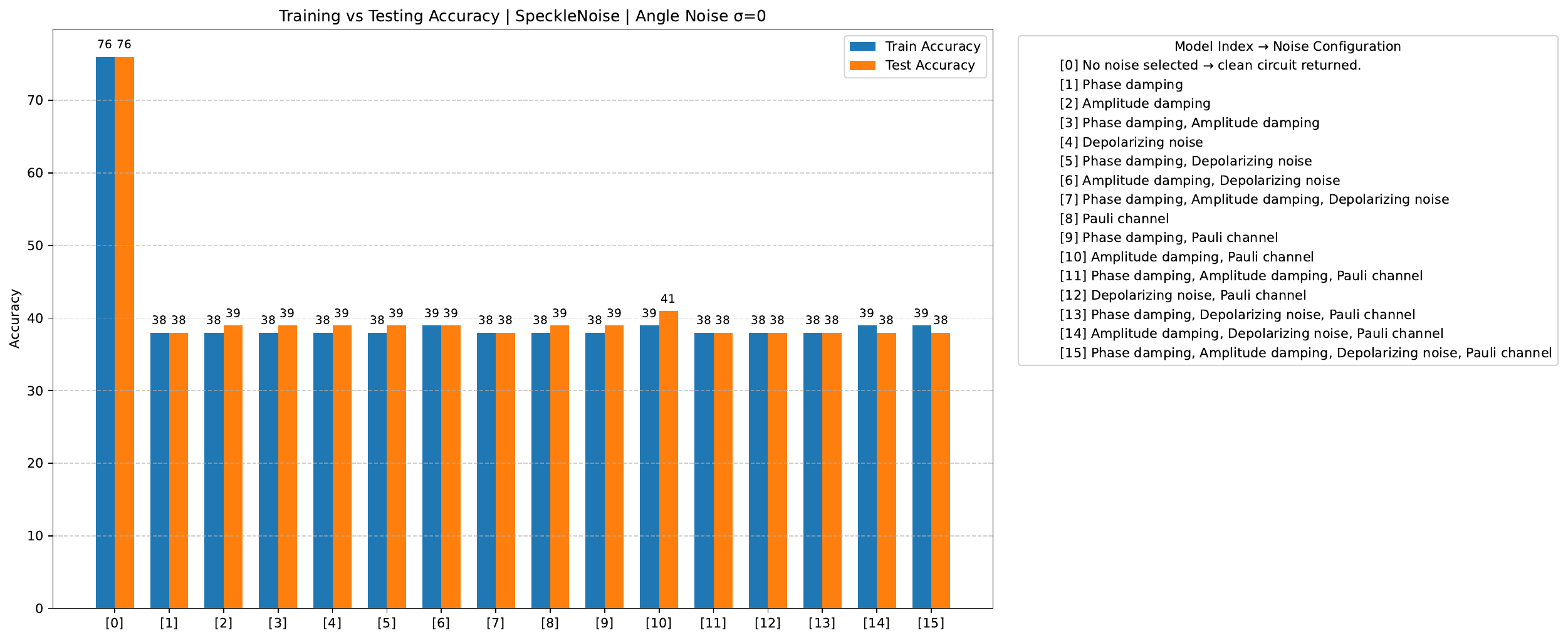}
{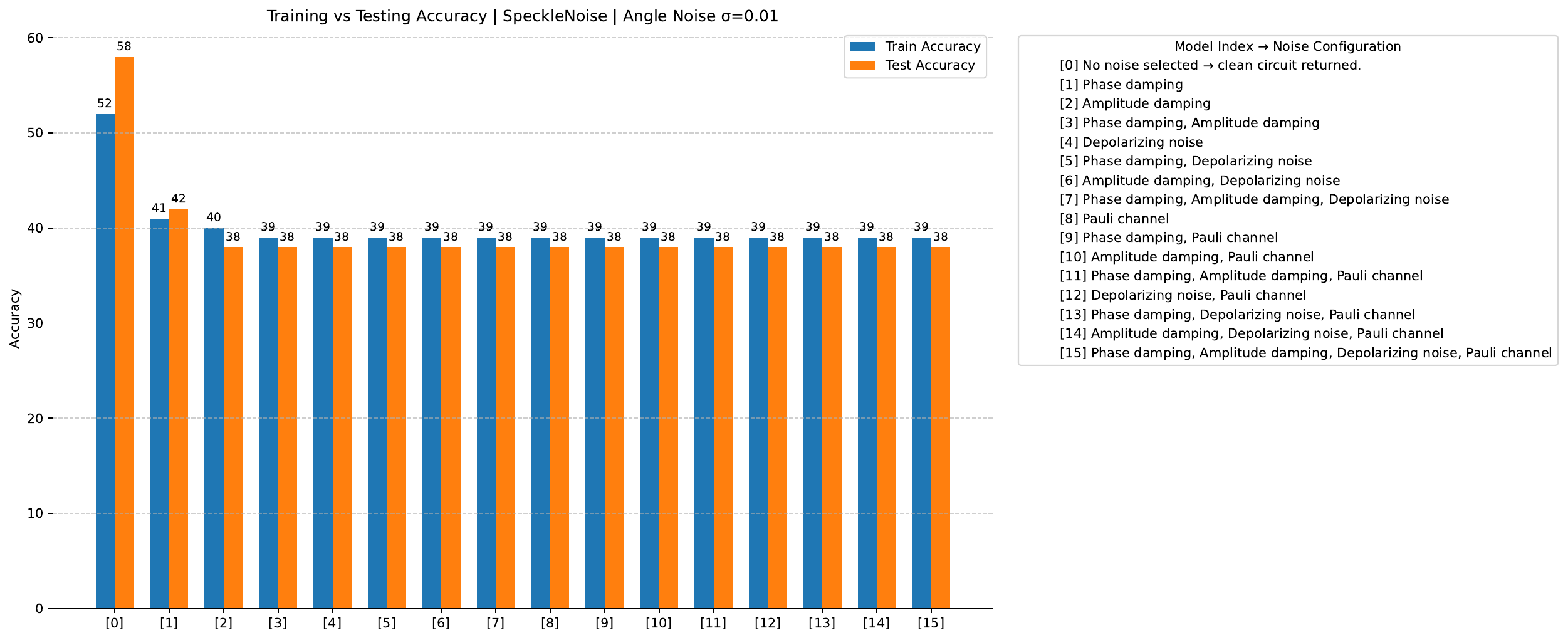}
{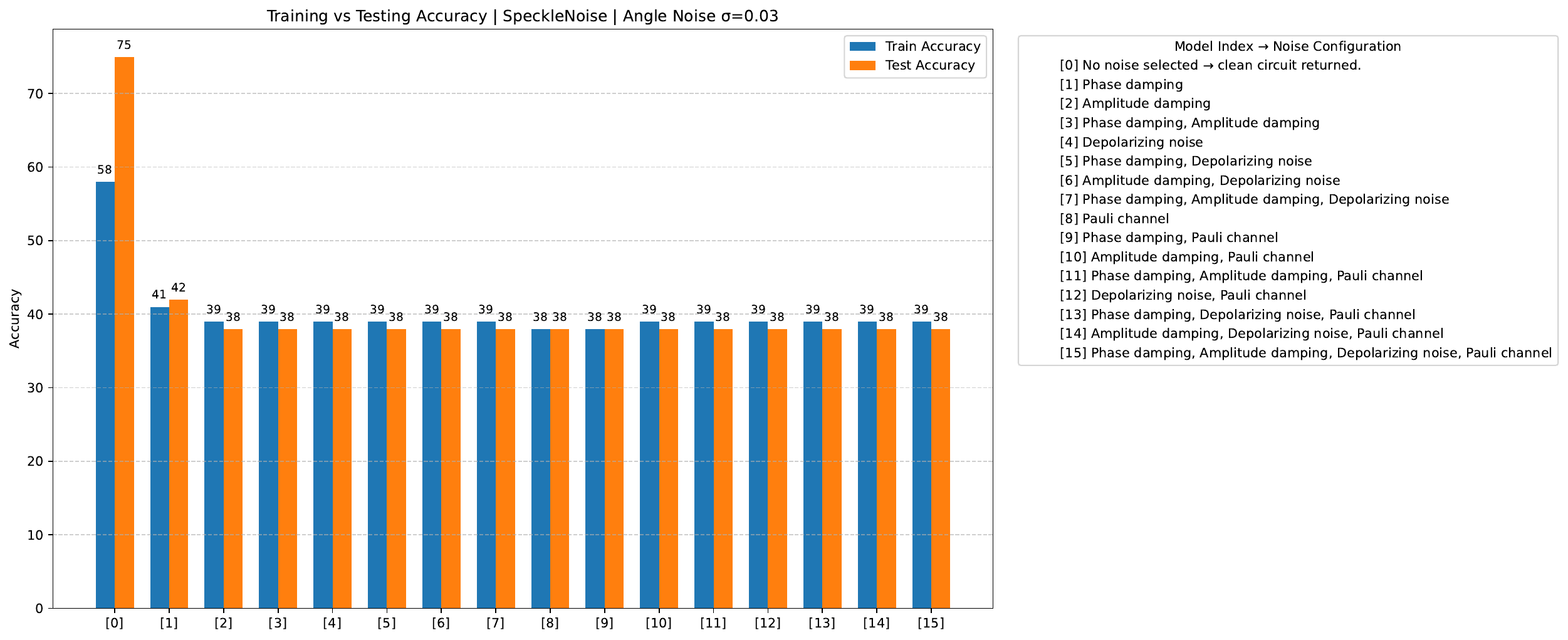}
{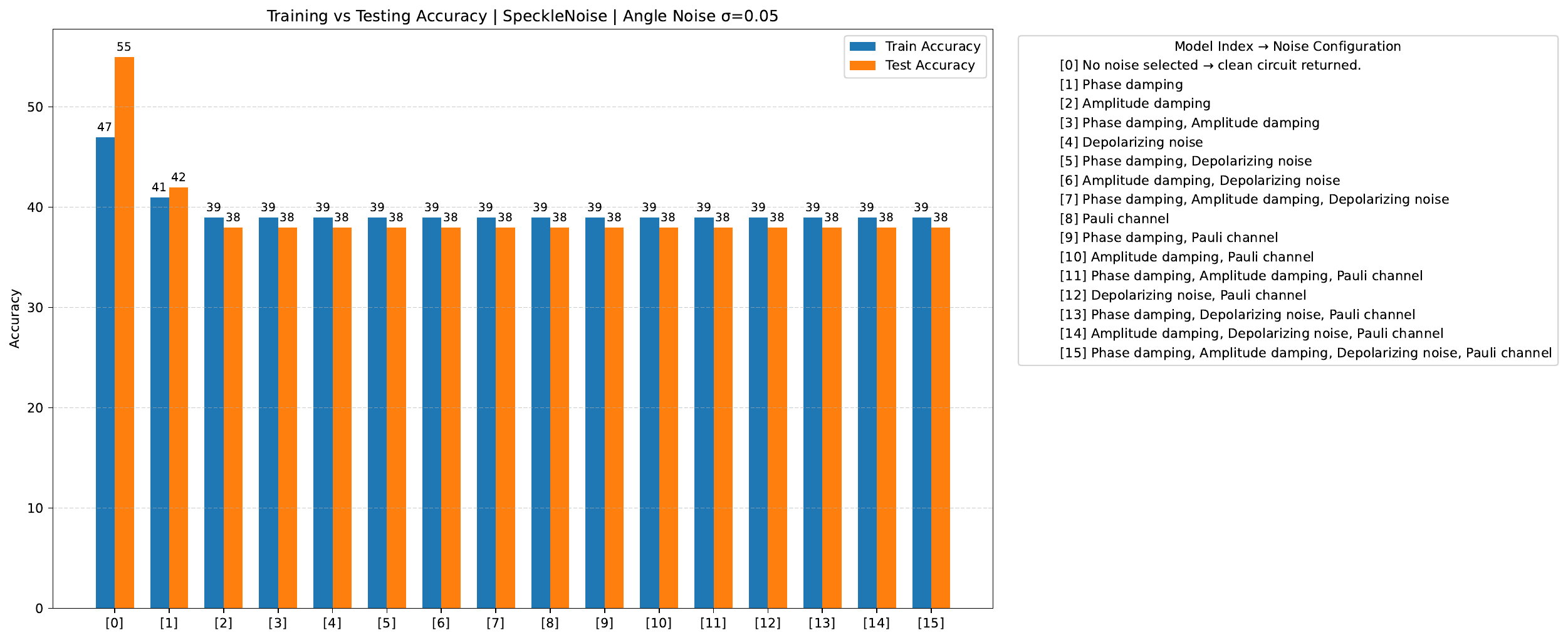}
{Accuracy comparison under \textbf{speckle noise}.}
{fig:acc_spec_grouped}


\FourPanelFigure
{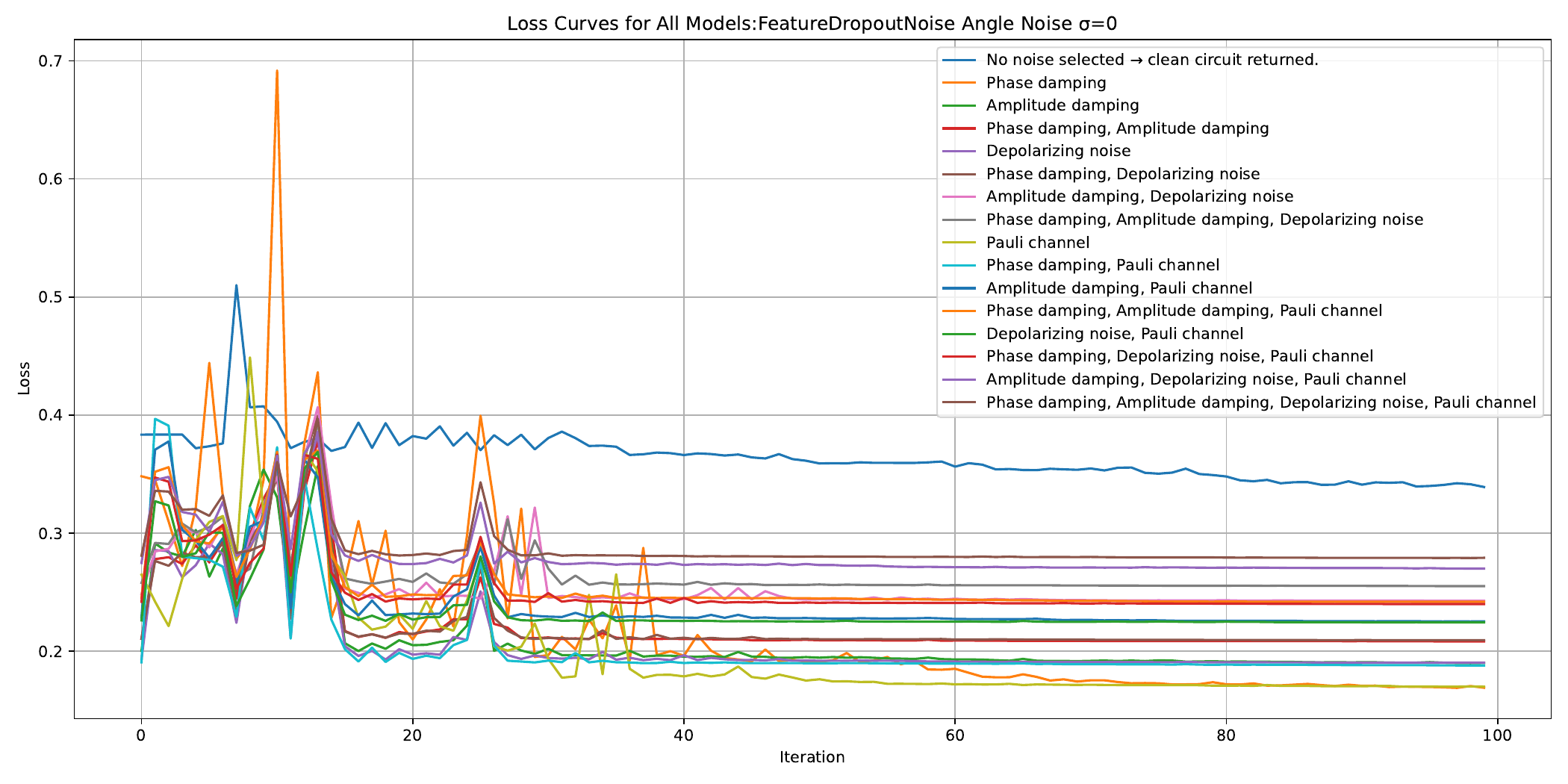}
{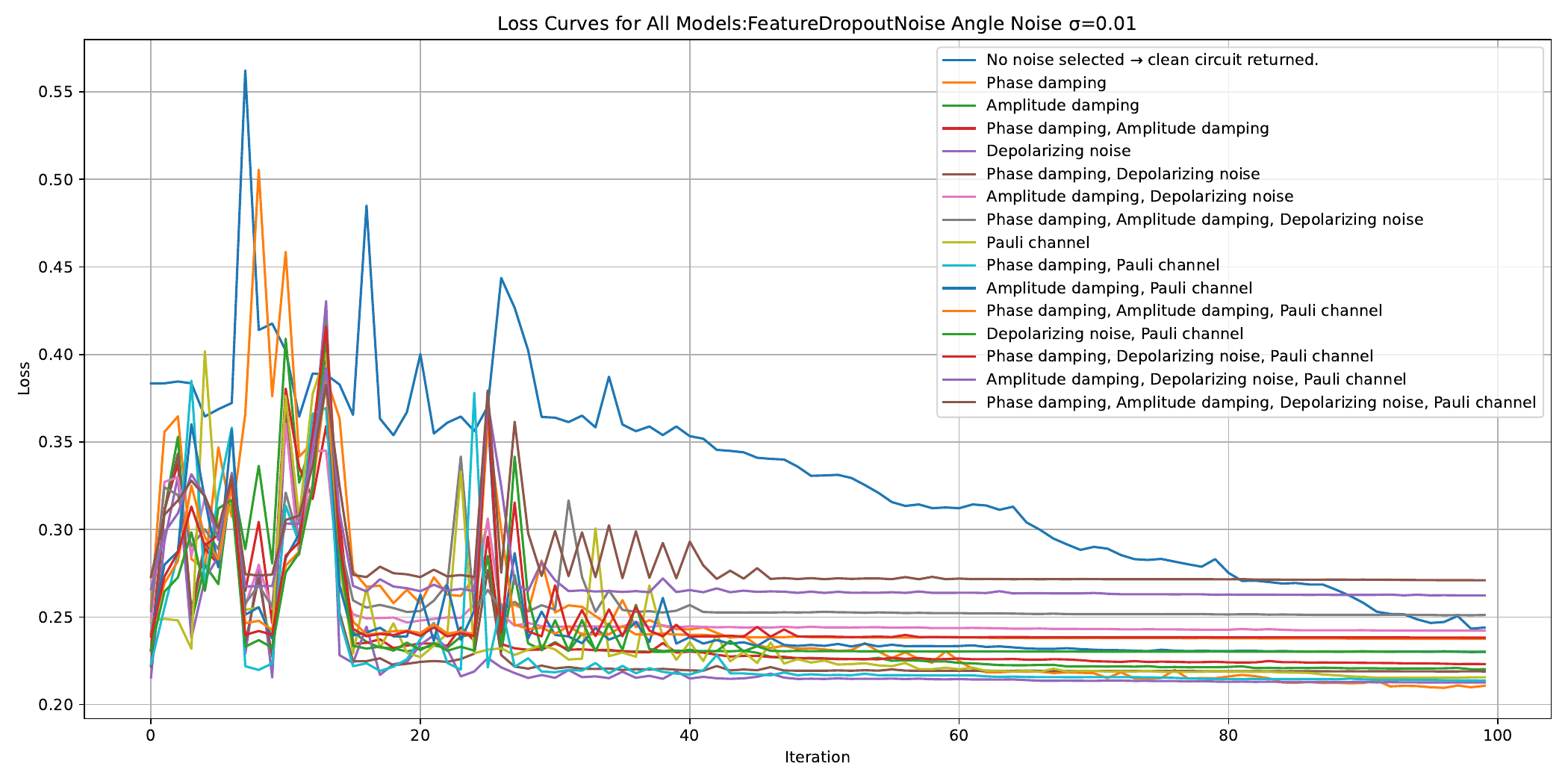}
{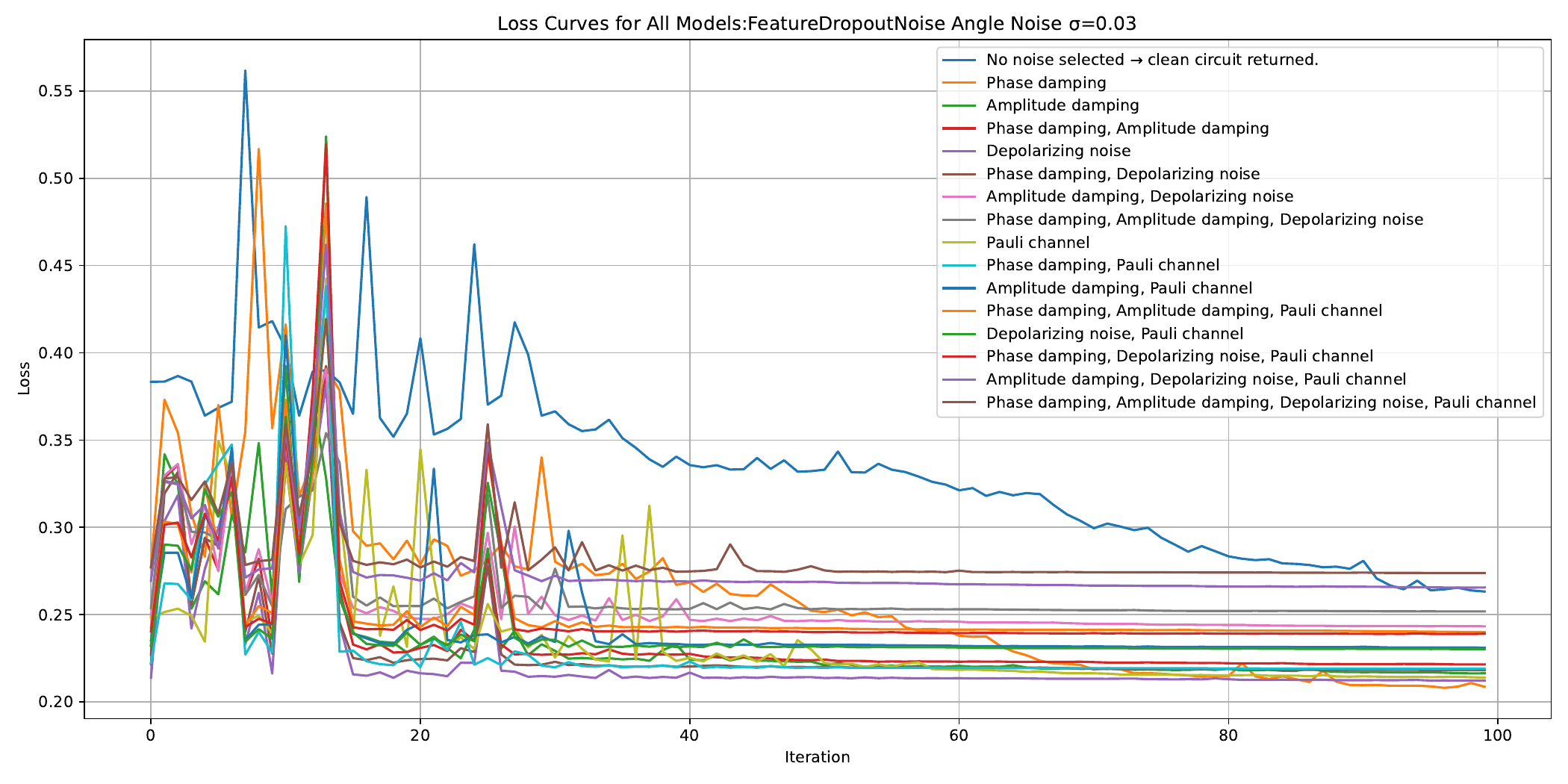}
{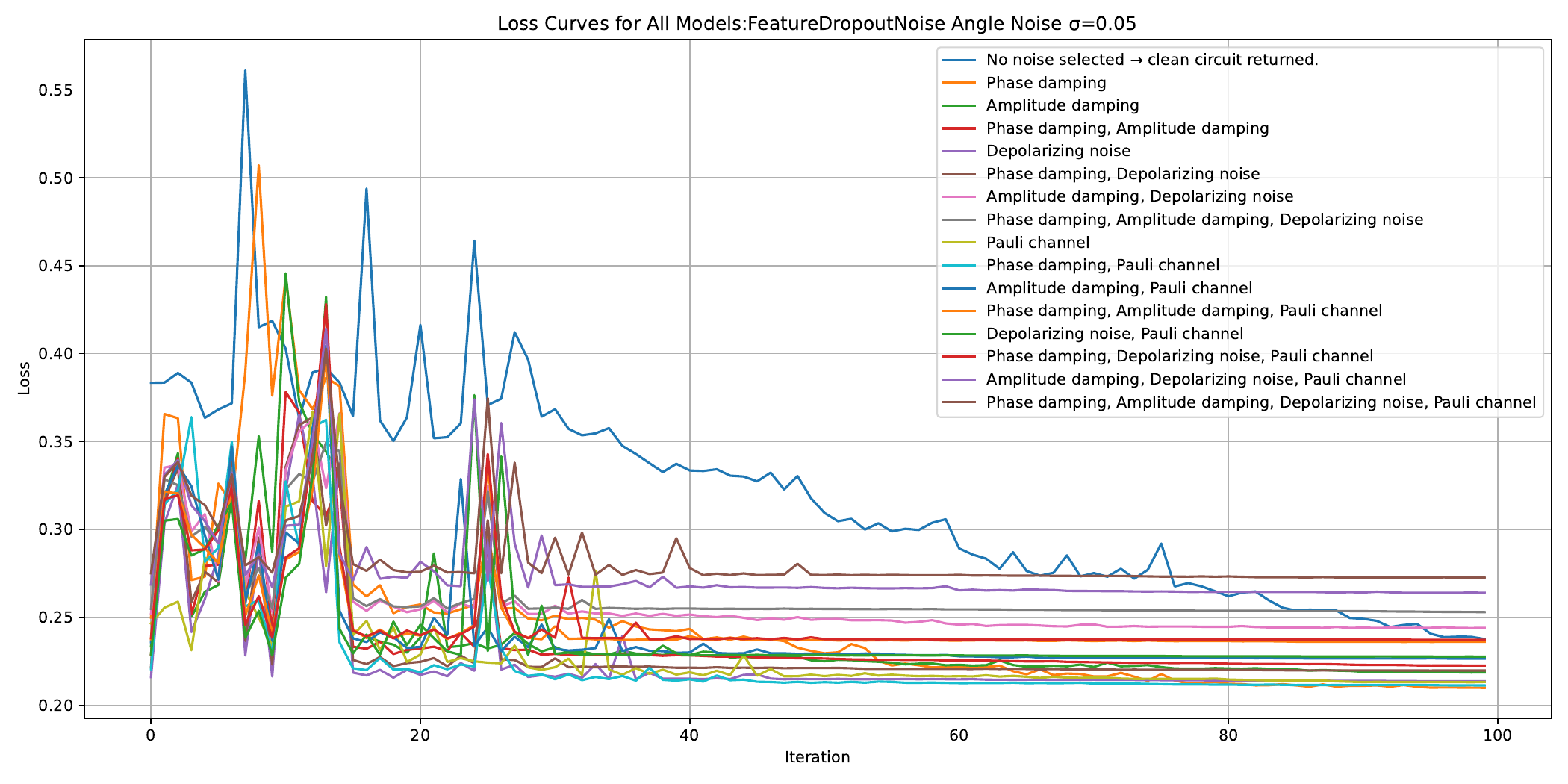}
{Loss convergence curves under \textbf{feature dropout noise}.}
{fig:loss_FD_grouped}

\FourPanelFigure
{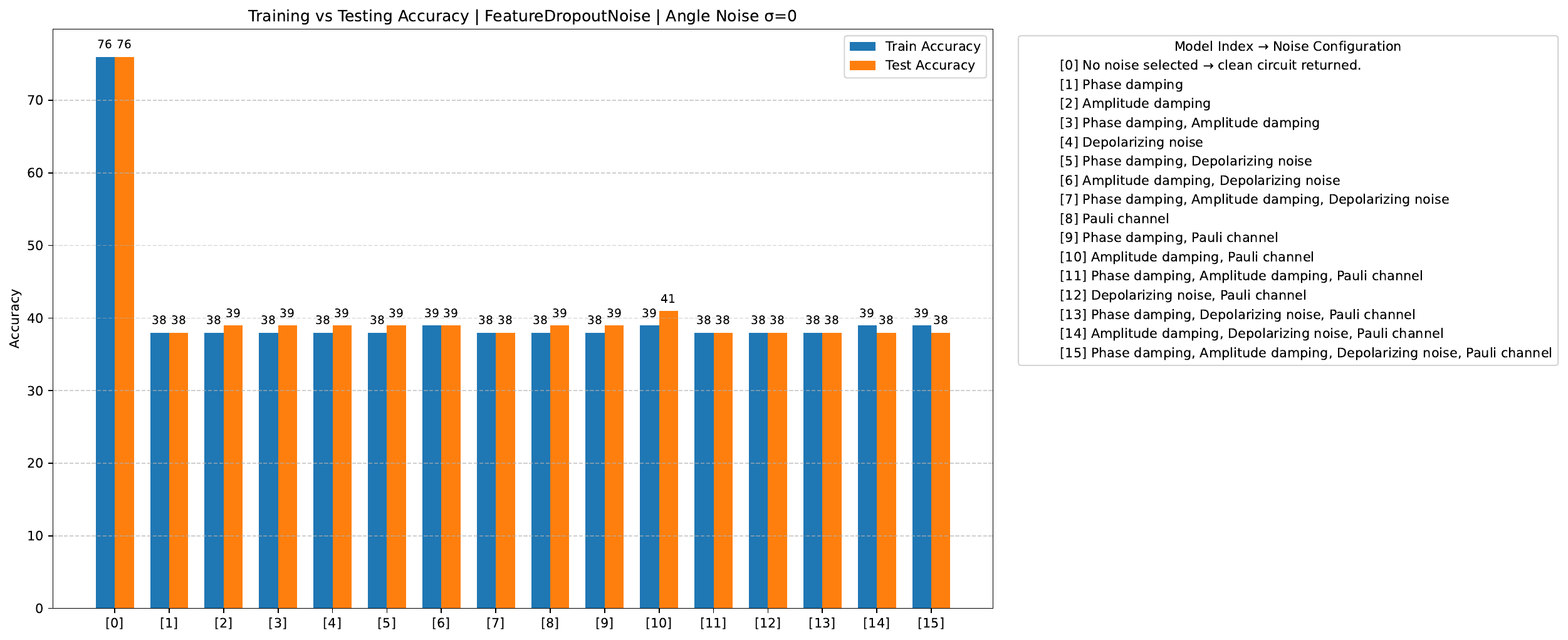}
{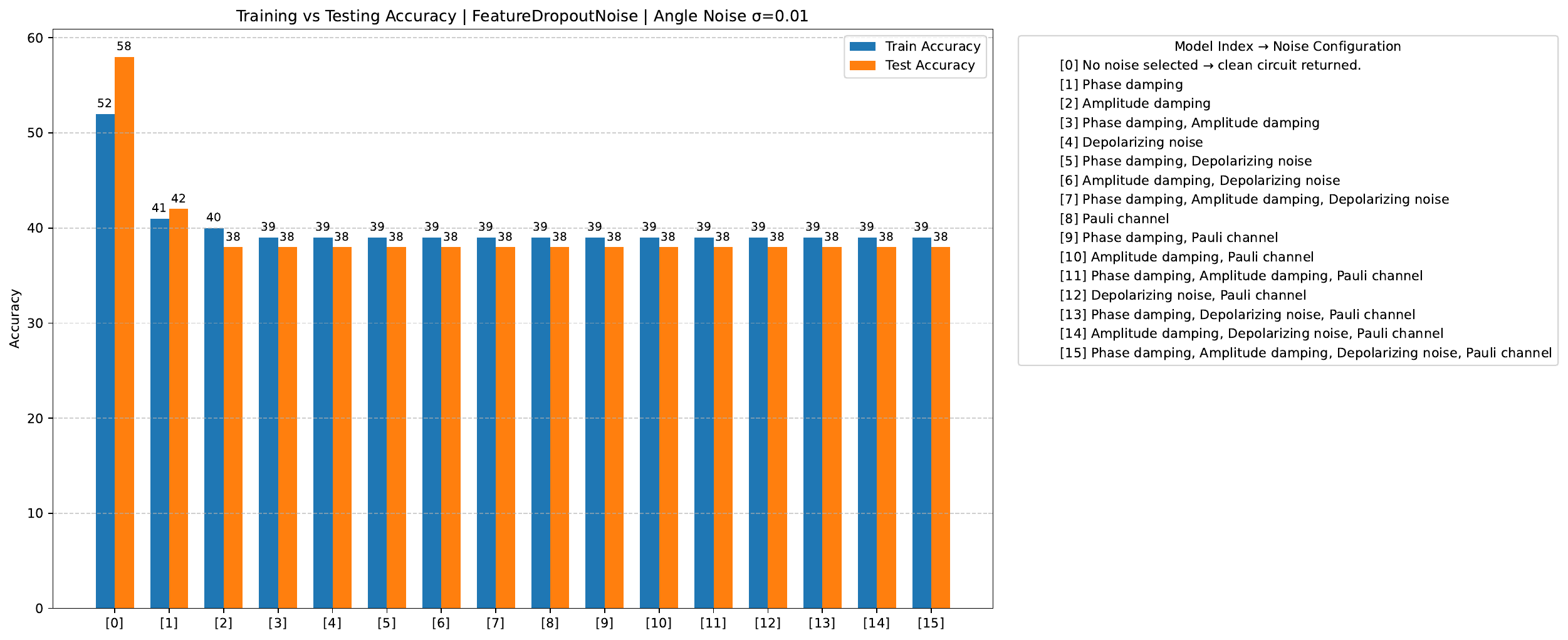}
{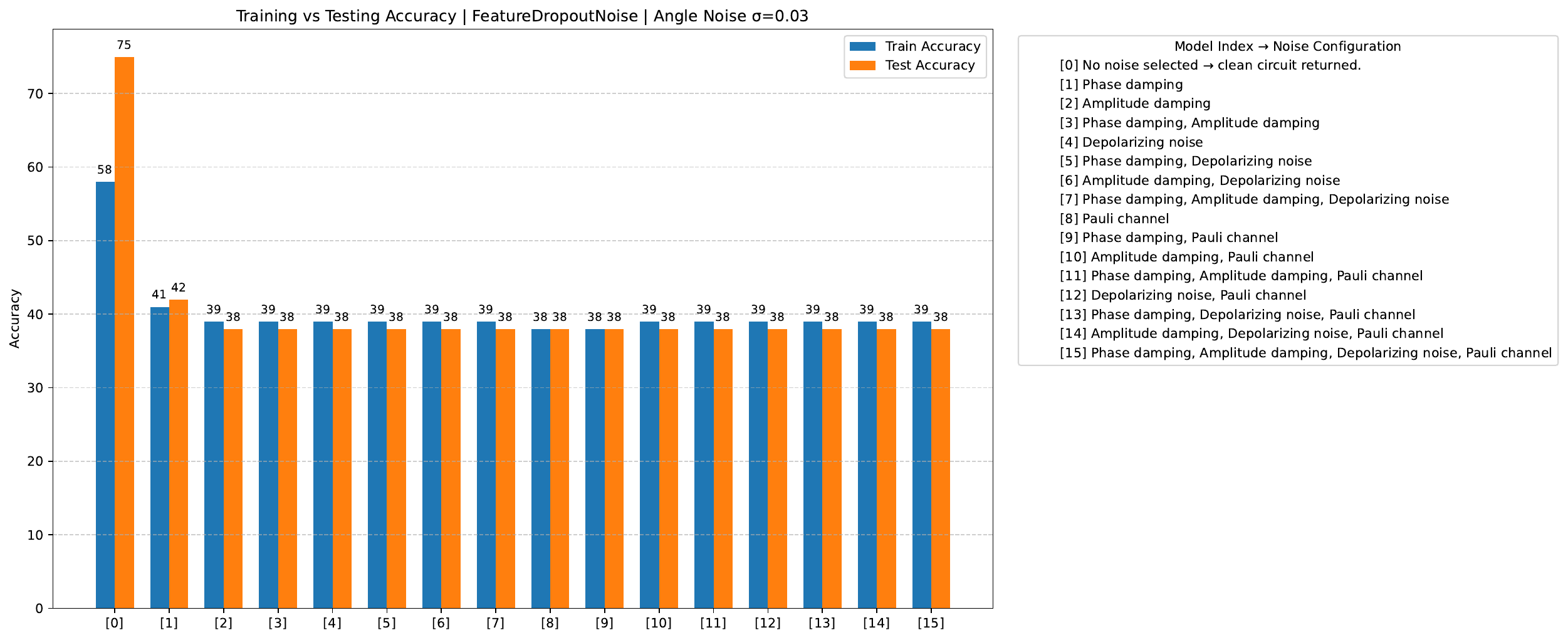}
{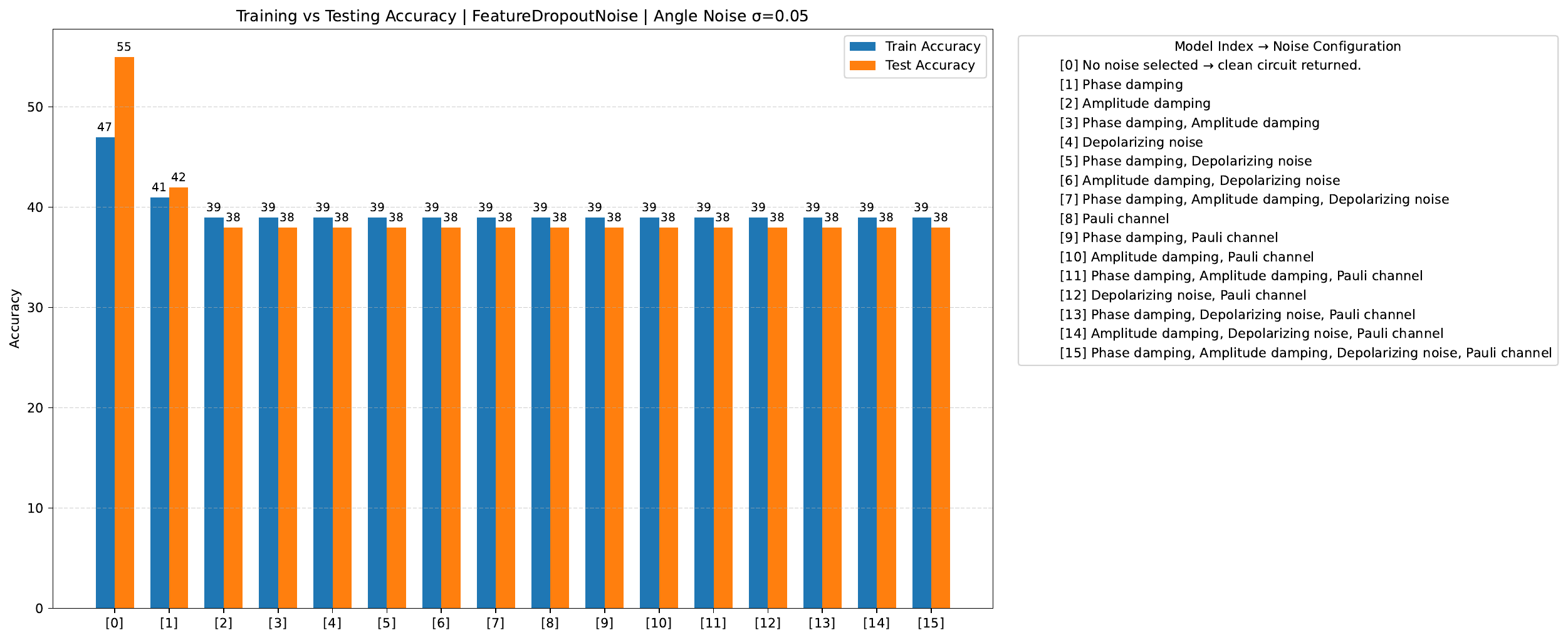}
{Accuracy comparison under \textbf{feature dropout noise}.}
{fig:acc_FD_grouped}


\FourPanelFigure
{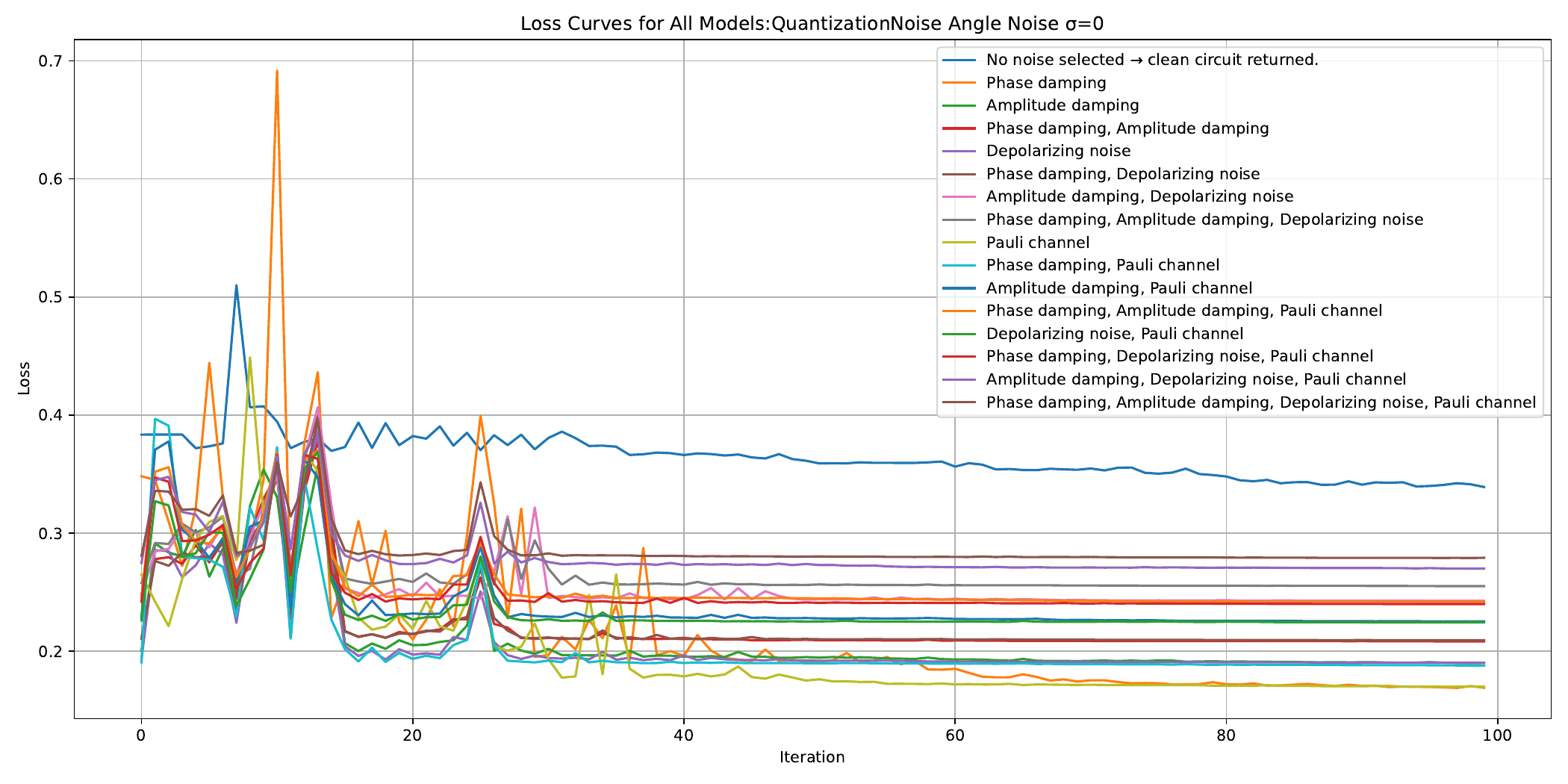}
{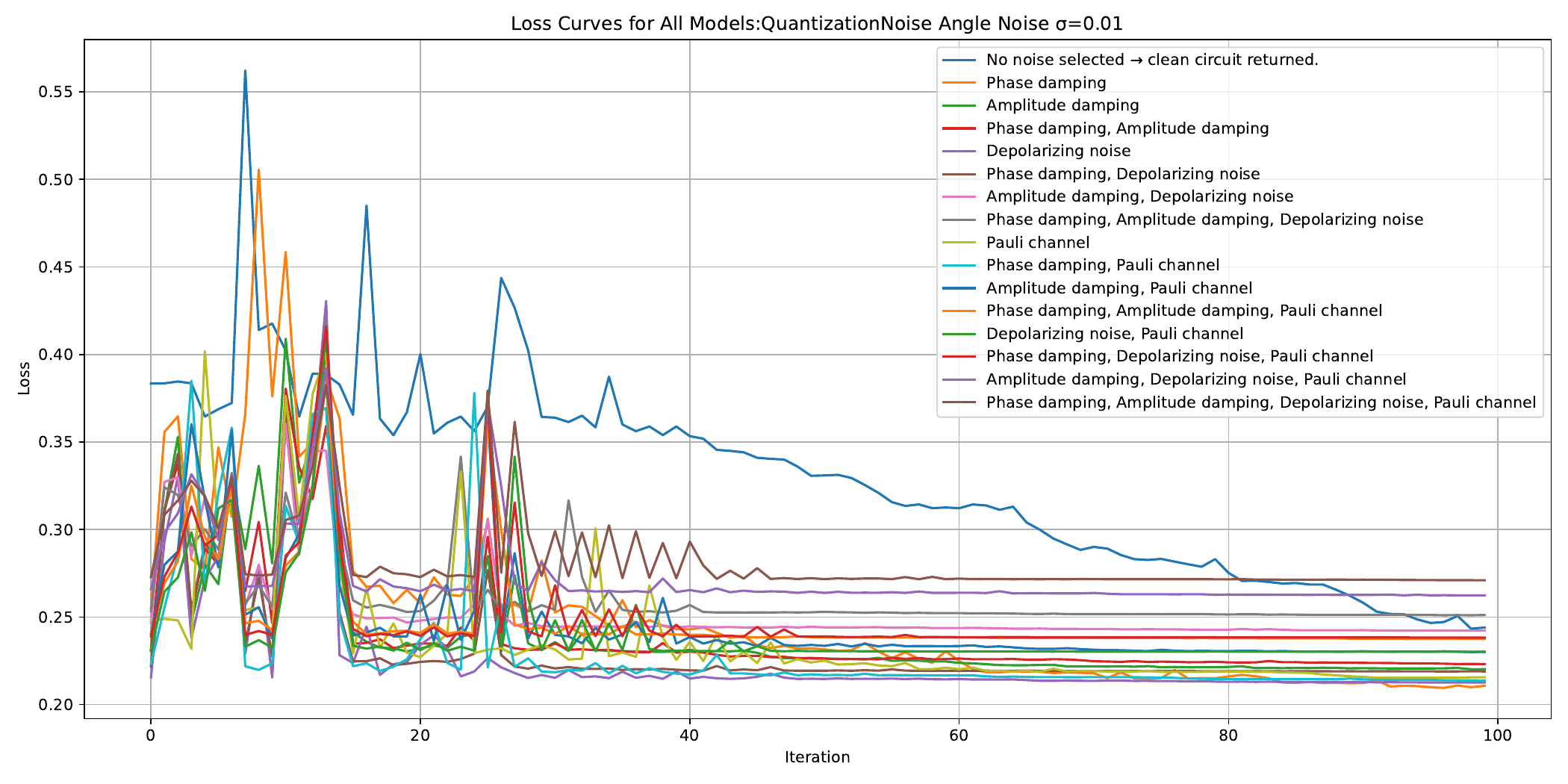}
{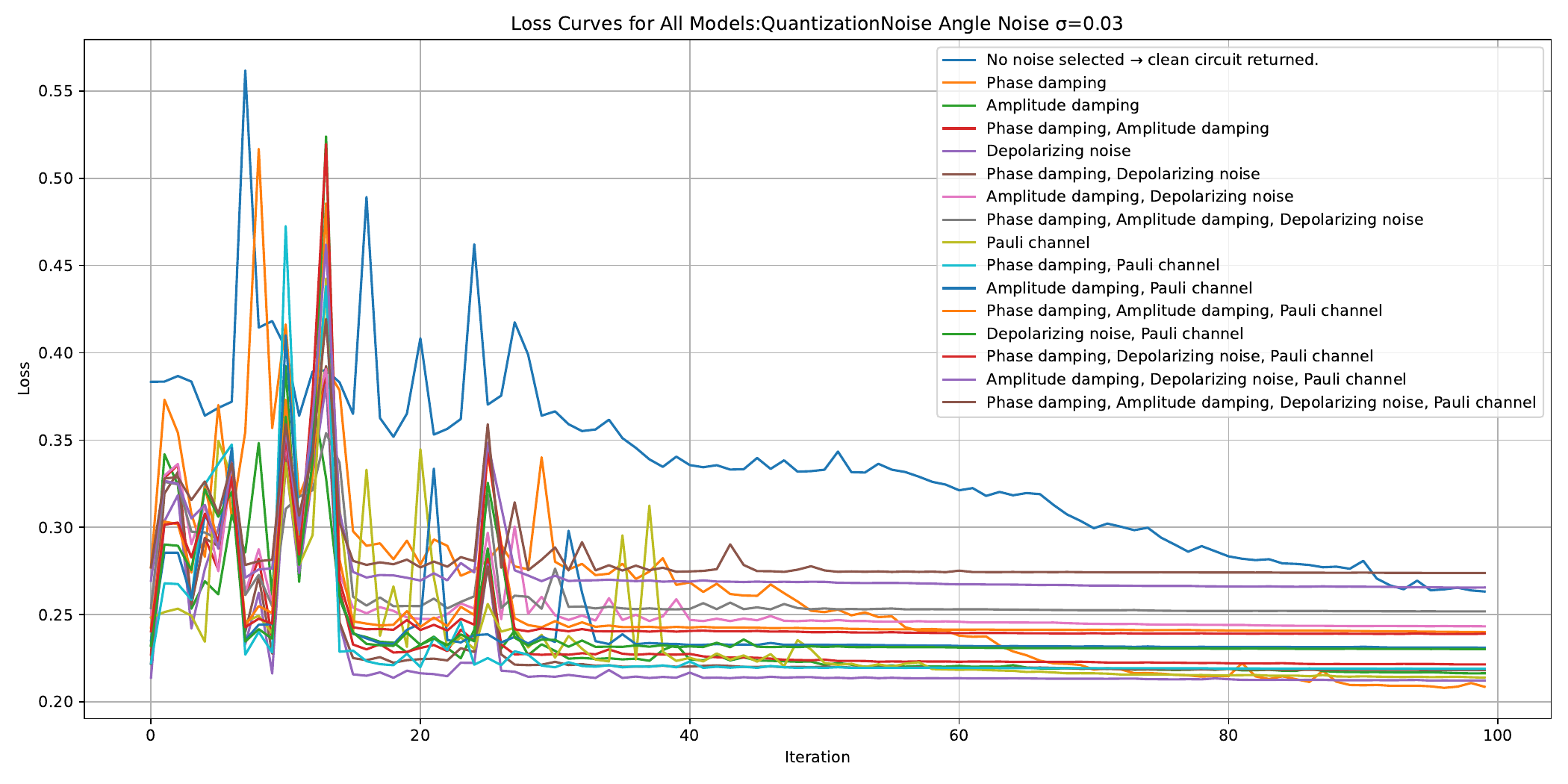}
{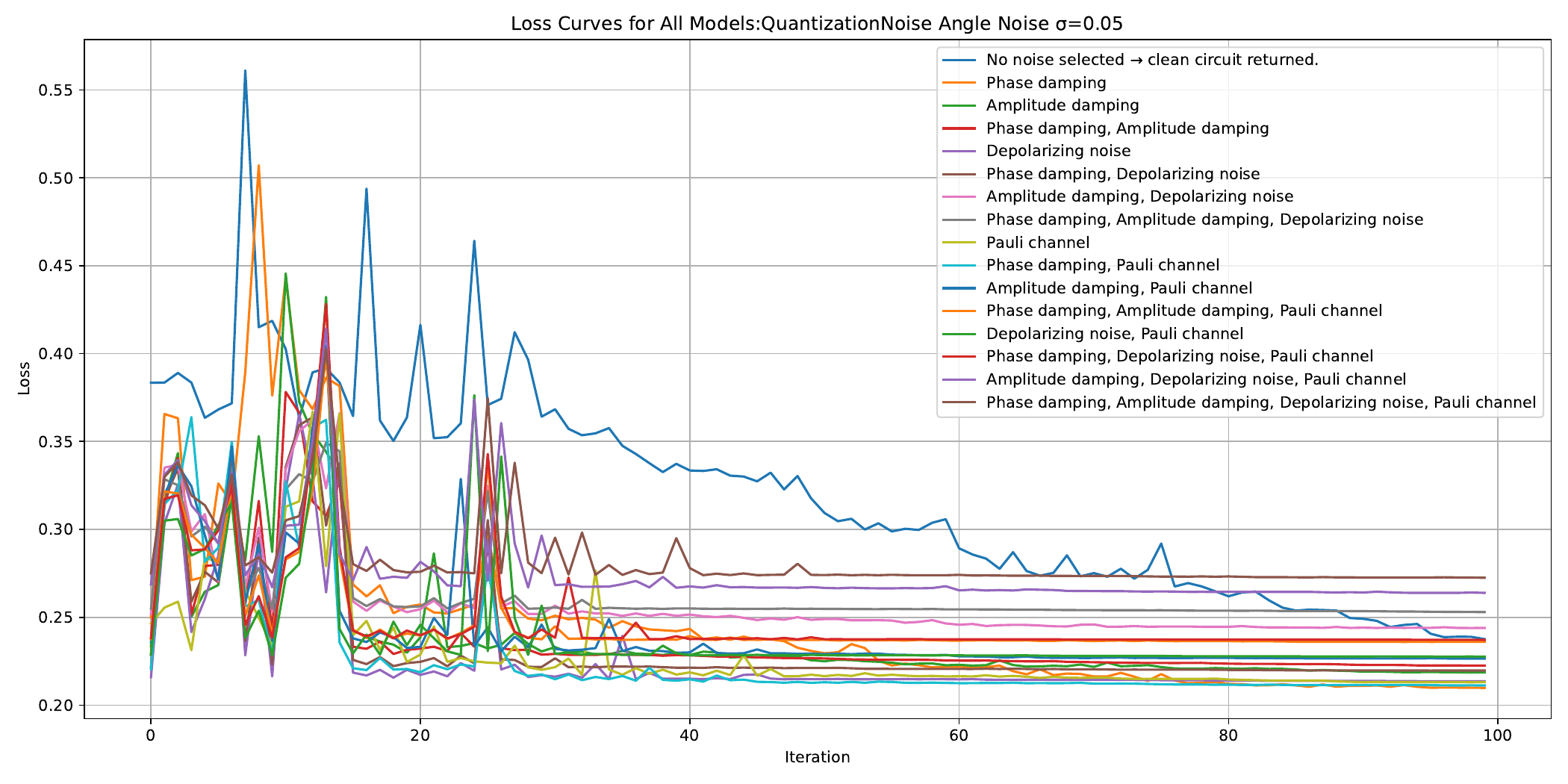}
{Loss convergence curves under \textbf{quantization noise}.}
{fig:loss_Quant_grouped}

\FourPanelFigure
{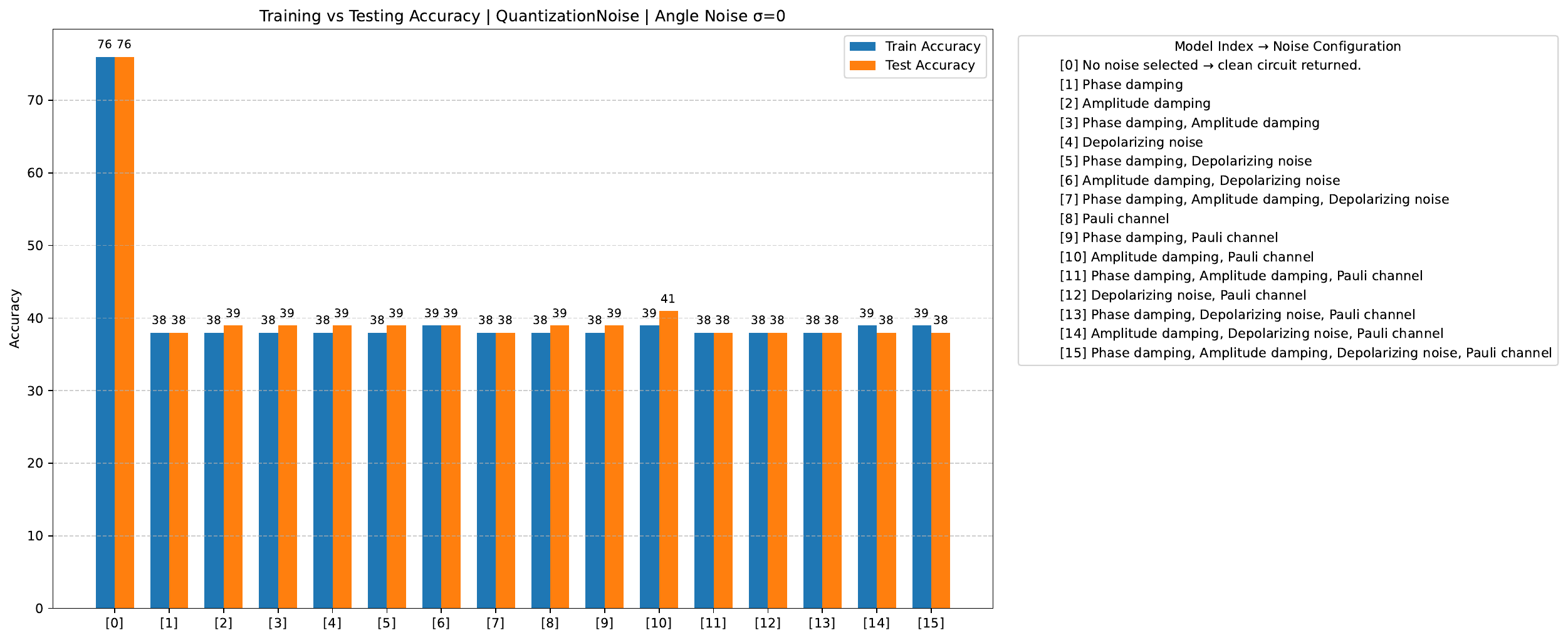}
{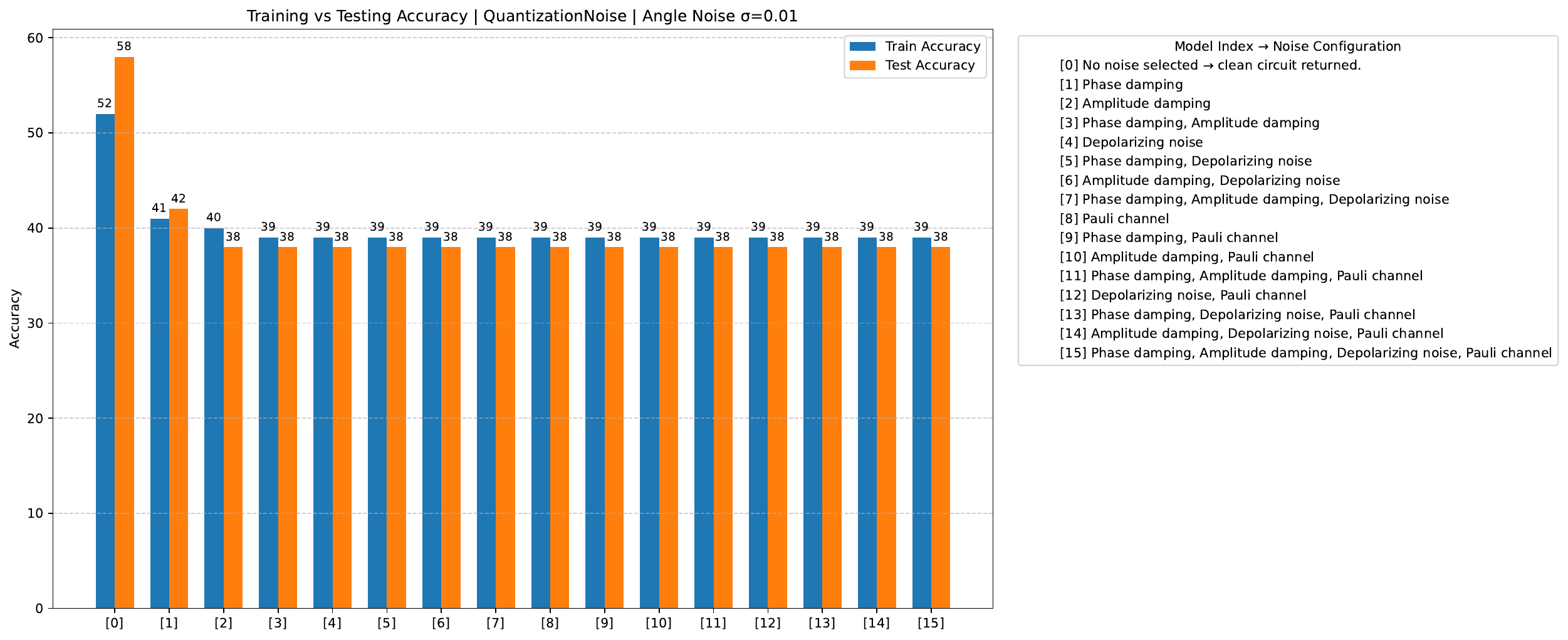}
{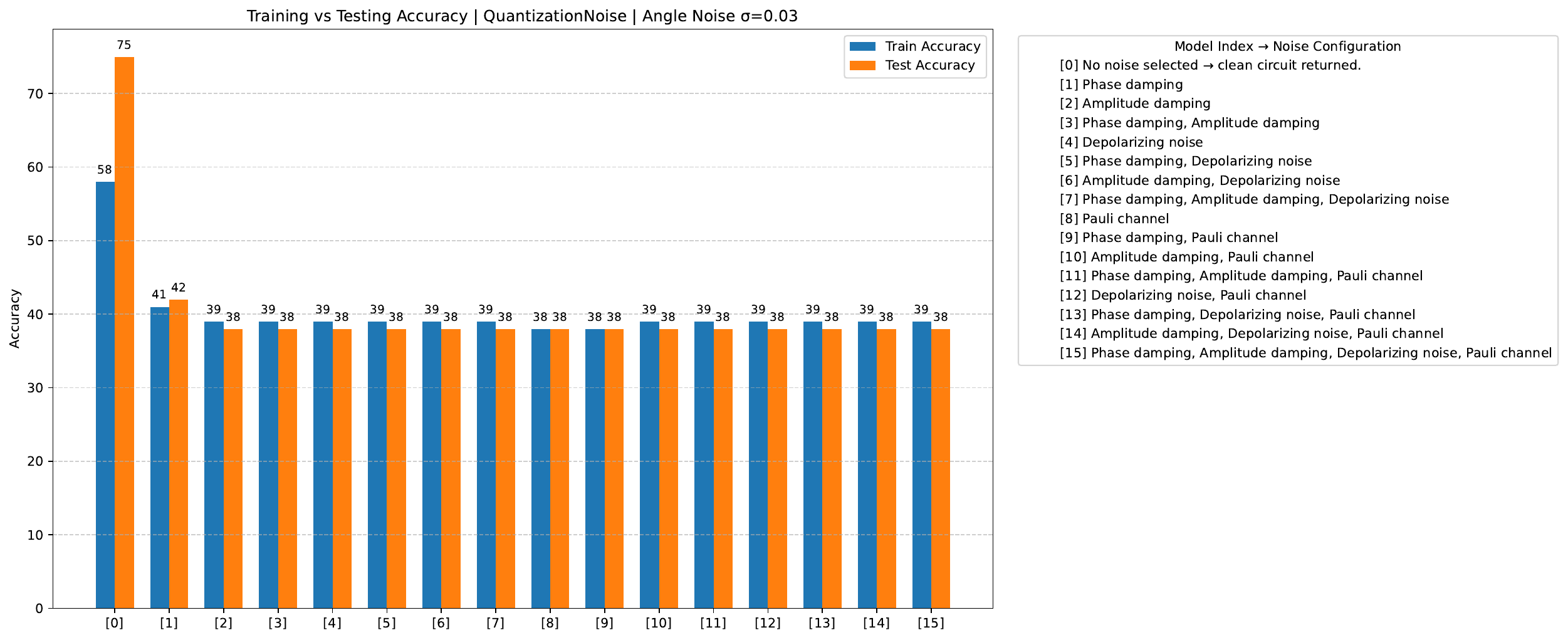}
{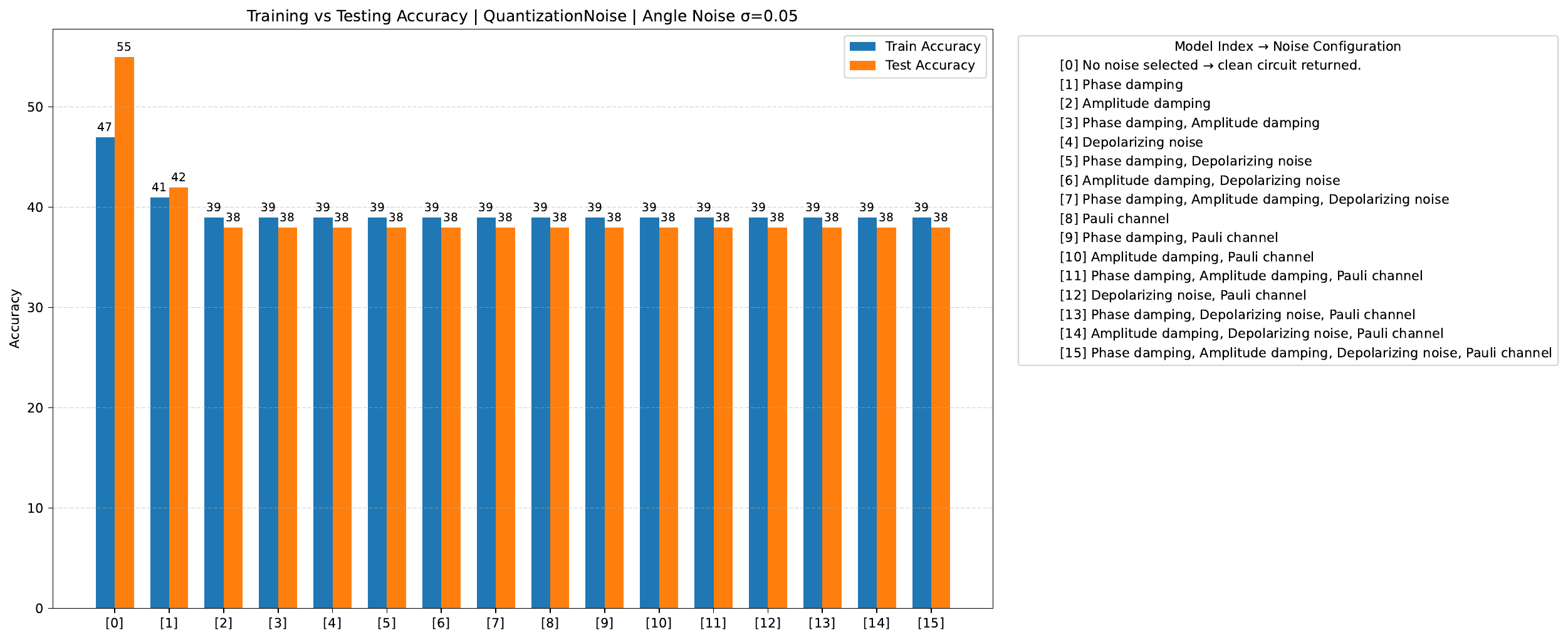}
{Accuracy comparison under \textbf{quantization noise}.}
{fig:acc_Quant_grouped}


\FourPanelFigure
{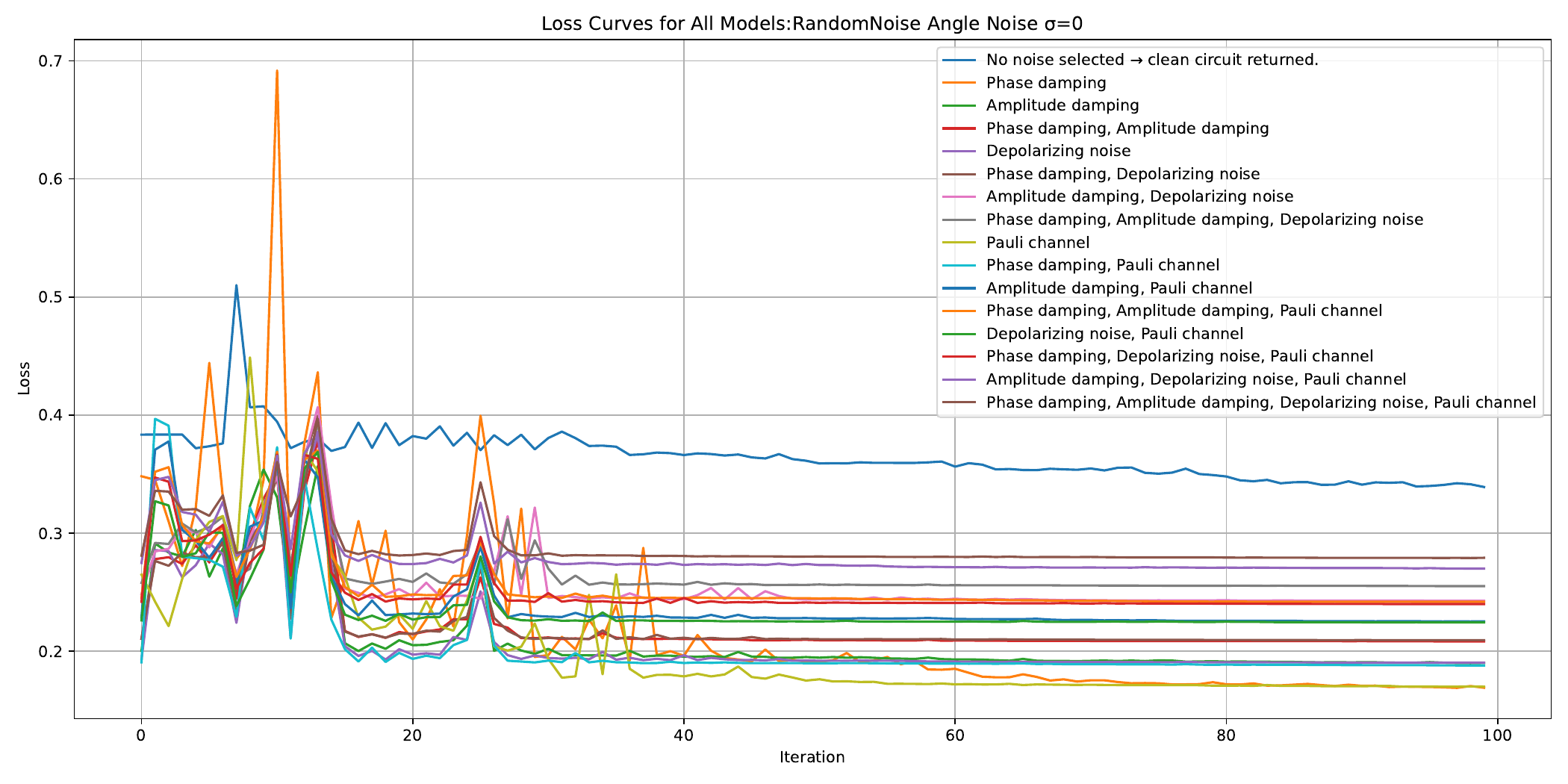}
{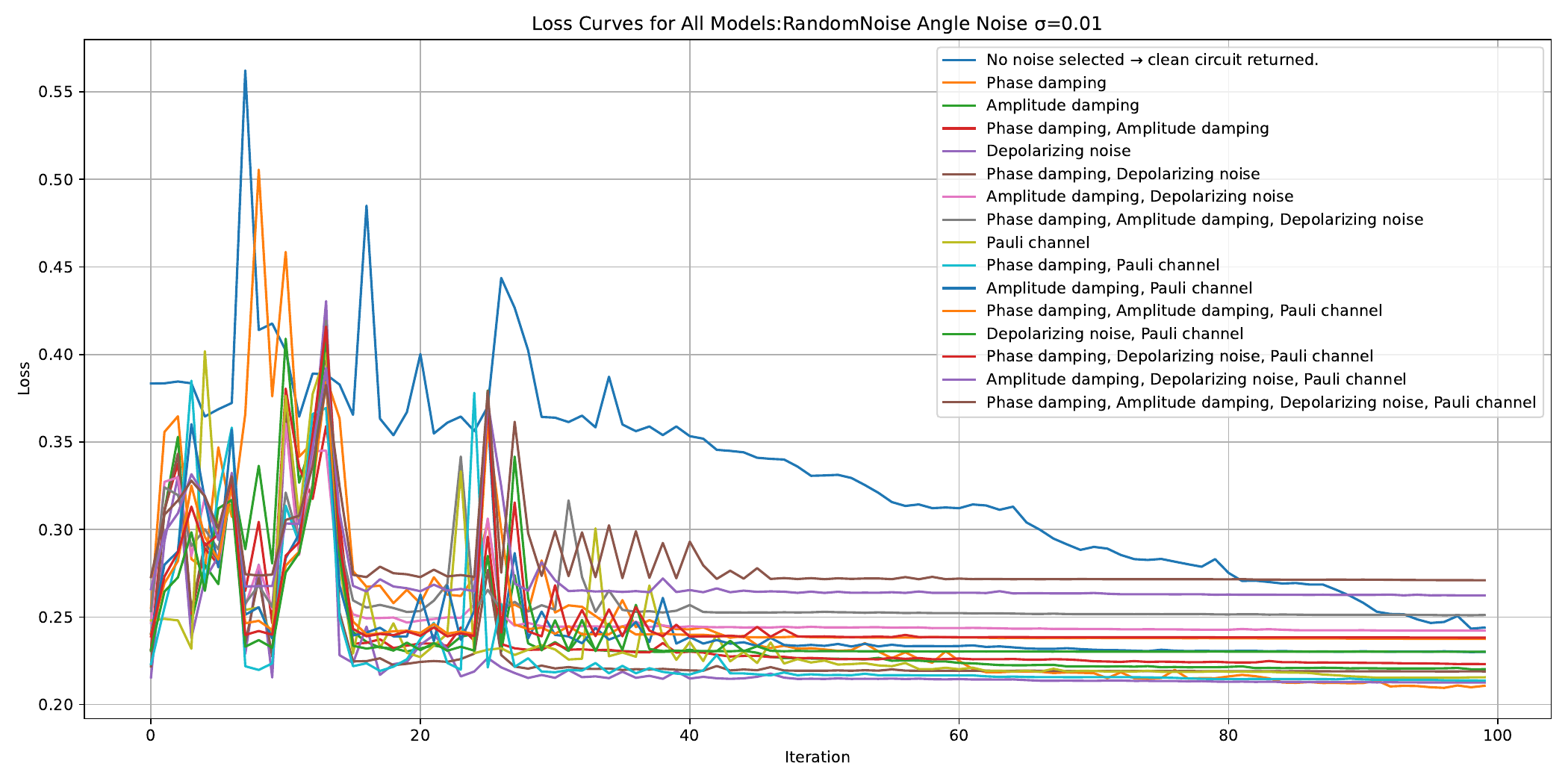}
{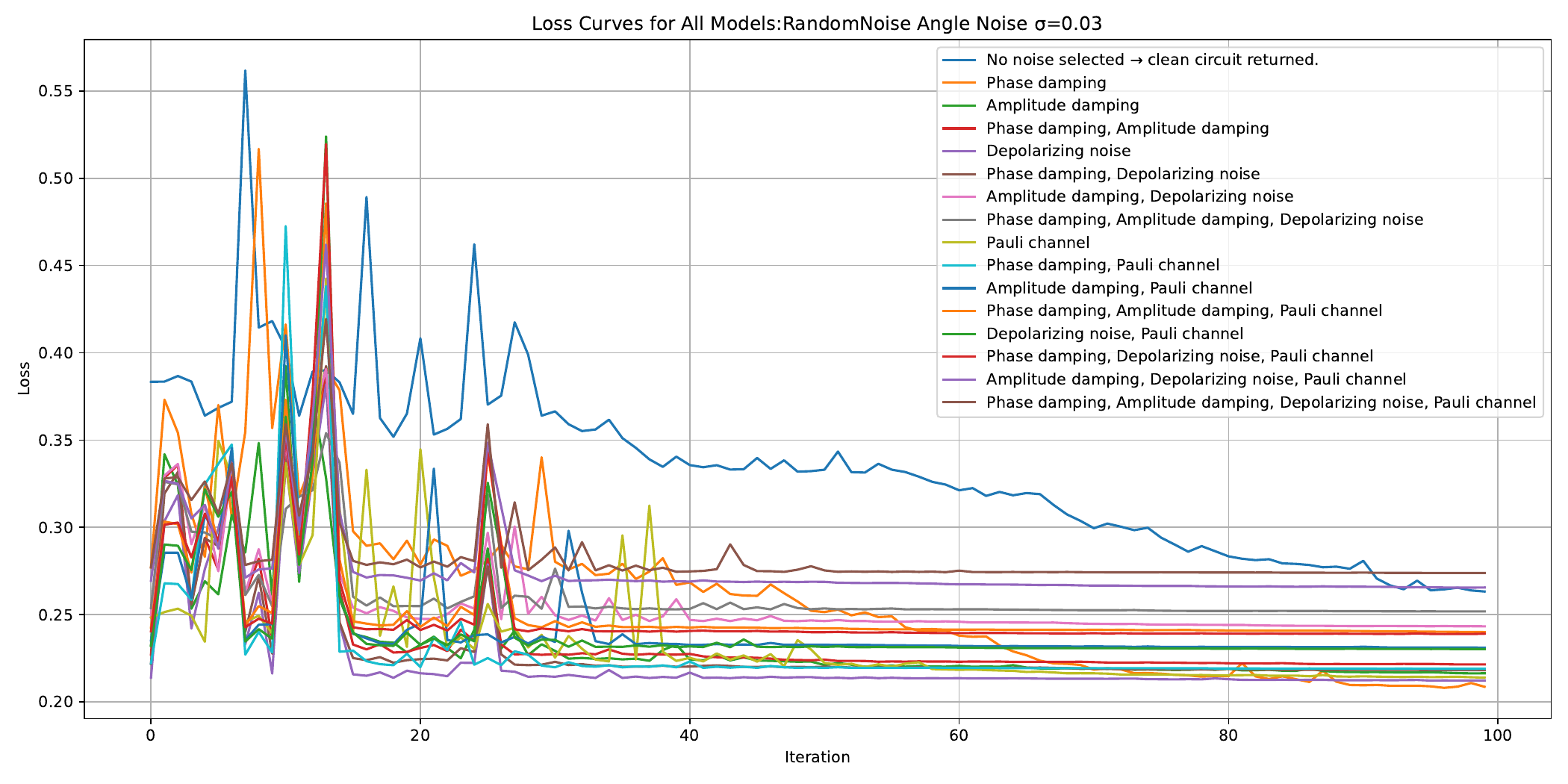}
{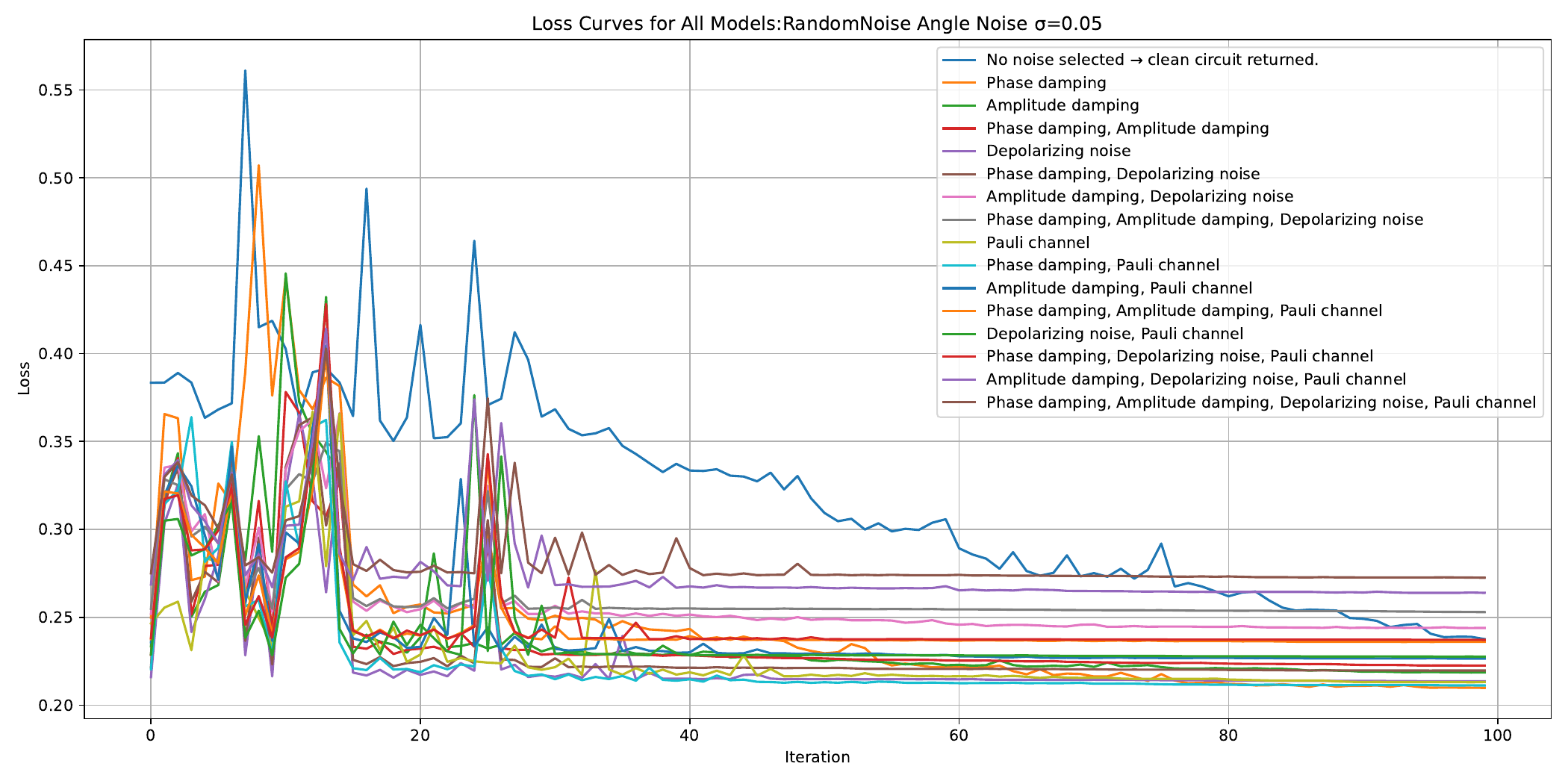}
{Loss convergence curves under \textbf{random noise}.}
{fig:loss_Random_grouped}

\FourPanelFigure
{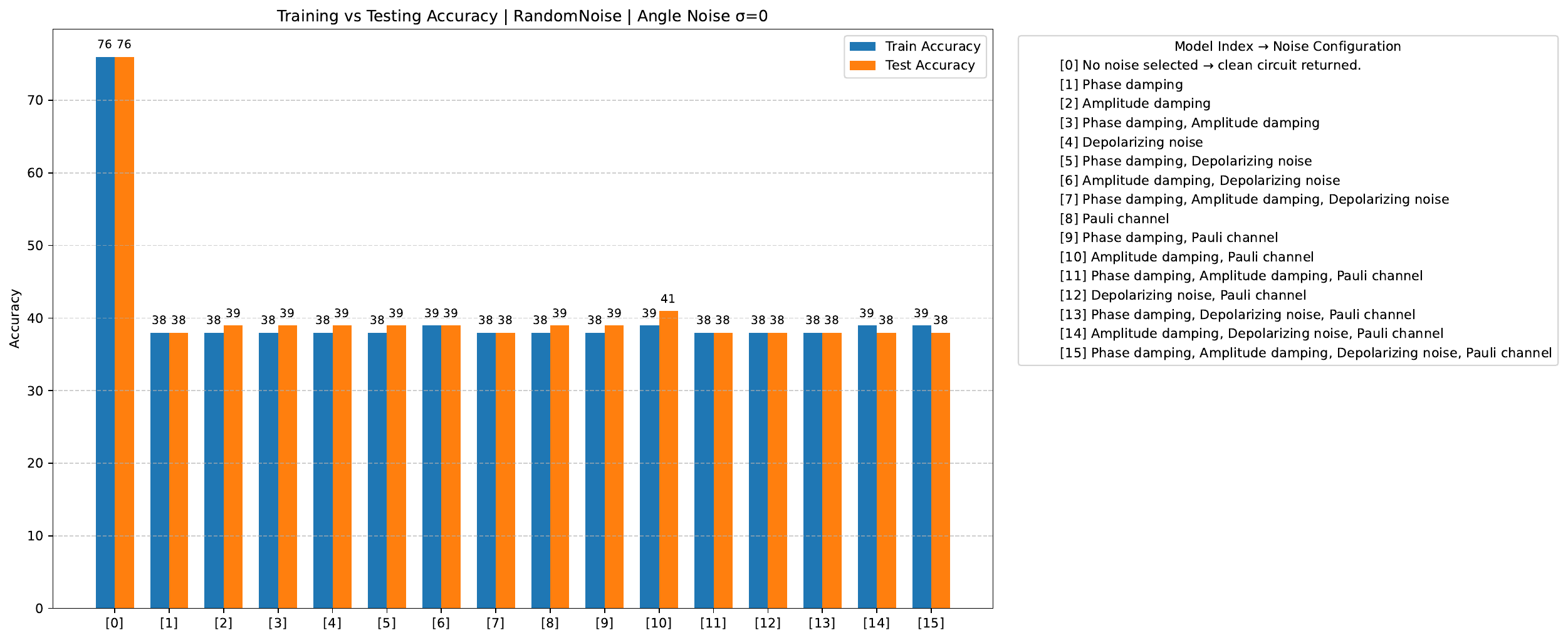}
{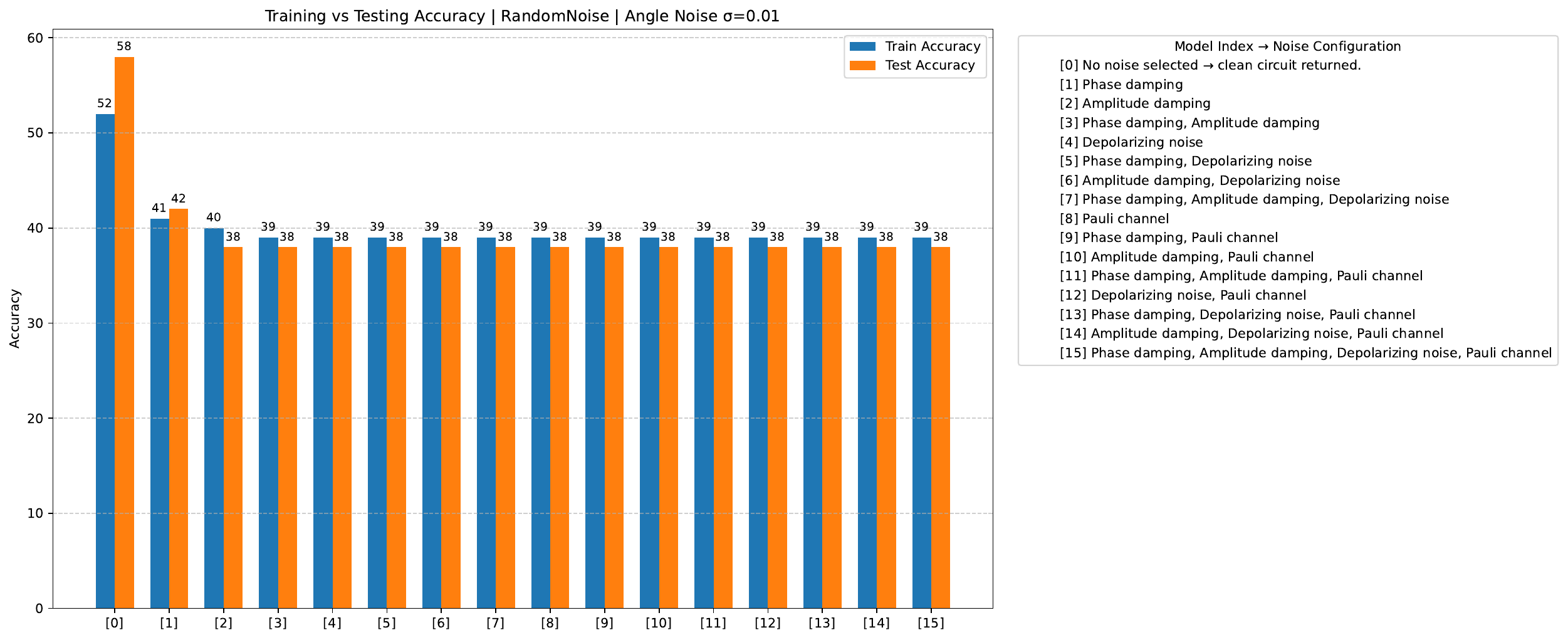}
{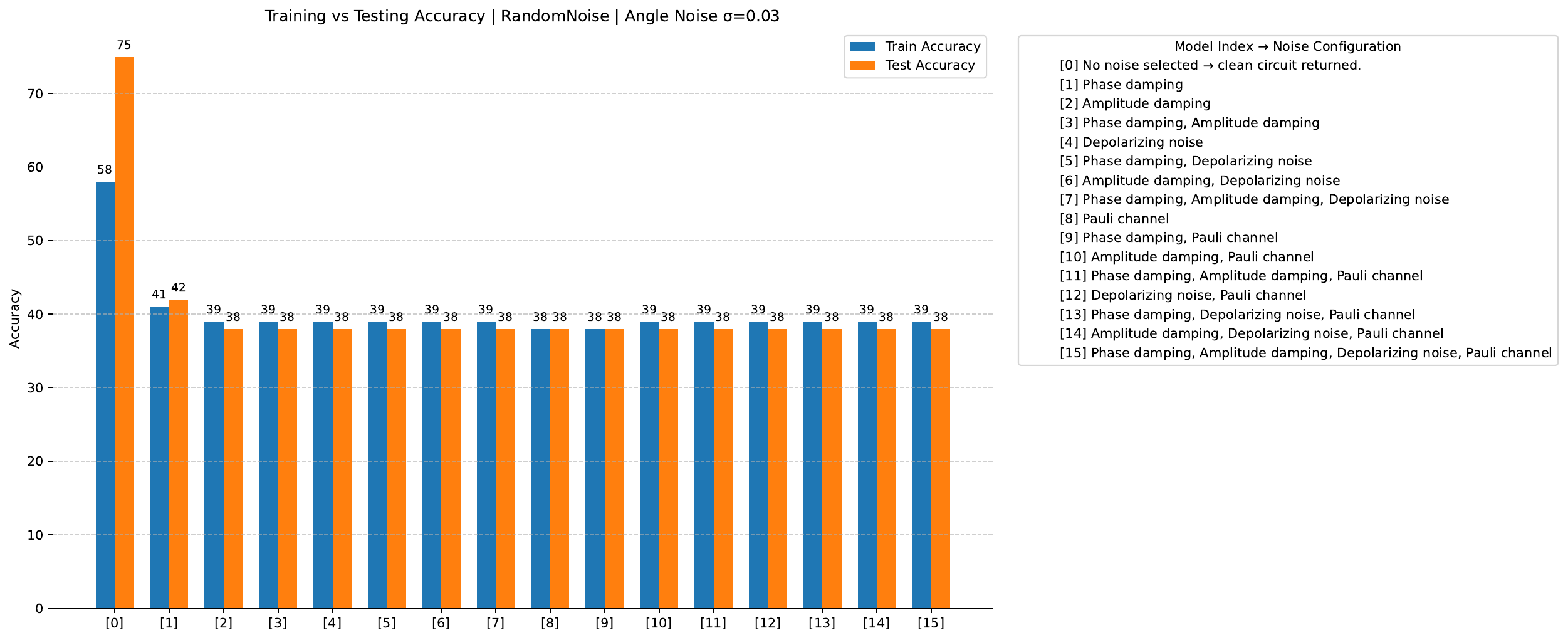}
{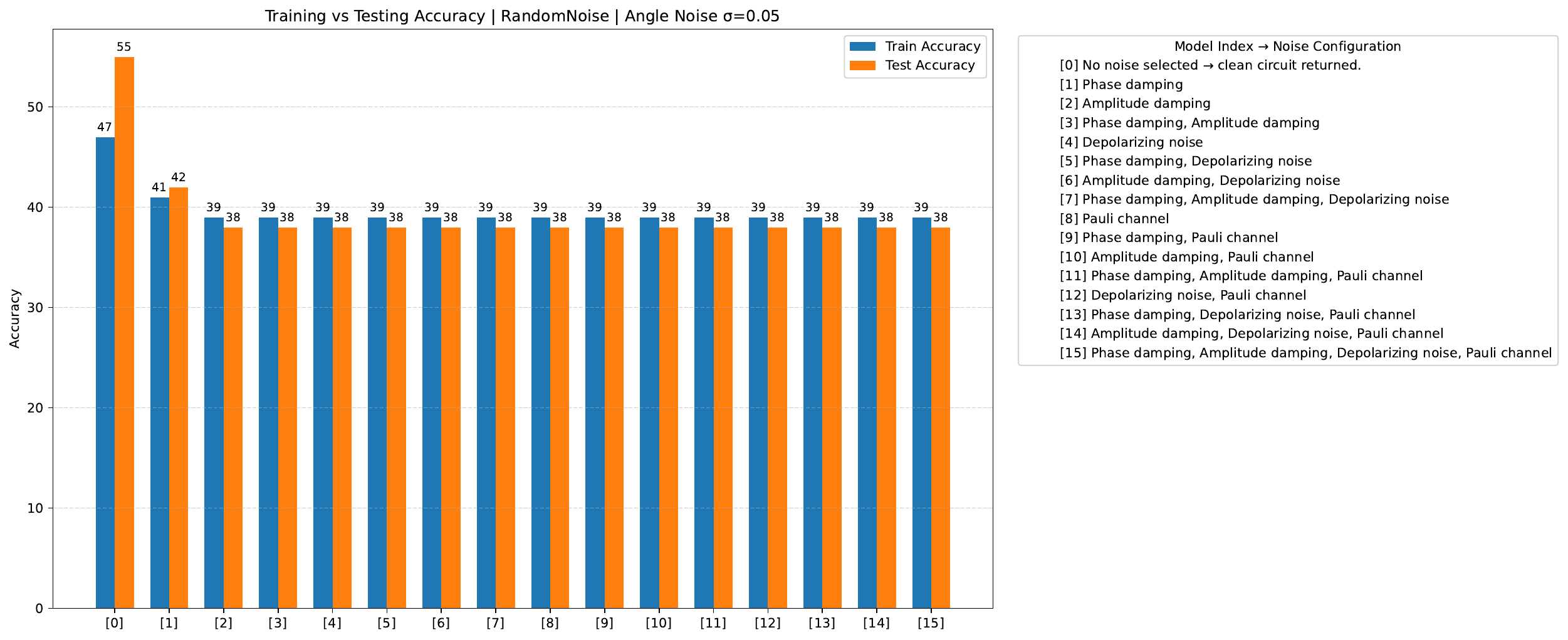}
{Accuracy comparison under \textbf{random noise}.}
{fig:acc_Random_grouped}

\end{appendix}
\section*{Acknowledgment}
This work was conducted as part of the Qiskit Advocate Mentorship Program 2025 run by IBM Quantum.
\bibliographystyle{IEEEtran}
\bibliography{PaperSection/Documents/references}
\begin{IEEEbiography}[{\includegraphics[width=1in,height=1.25in,clip,keepaspectratio]{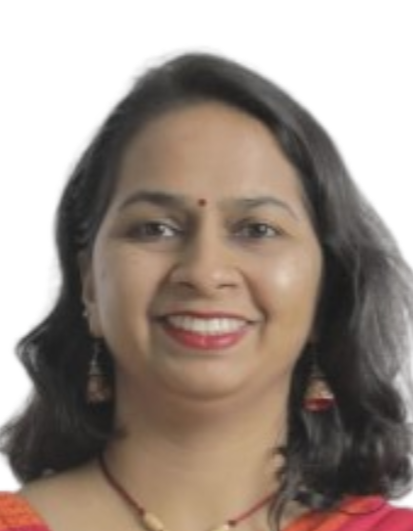}}]{Bhavna Bose} received the M.Tech. degree in Computer Engineering (Gold Medalist) from Veermata Jijabai Technological Institute (VJTI), Mumbai. She is currently pursuing the Ph.D. degree in Quantum Computing.
She is an Assistant Professor with SVKM’s NMIMS Mukesh Patel School of Technology Management and Engineering, Mumbai. She had prior academic experience at VJTI and Malla Reddy Engineering College and industry experience at Wipro Technologies.
Her research interests include quantum computing, quantum machine learning, generative adversarial networks, intrusion detection systems, and data imbalance handling techniques. She is actively involved in curriculum development, faculty development programmess, and student-centric quantum initiatives.
\end{IEEEbiography}
\begin{IEEEbiography}[{\includegraphics[width=1in,height=1.25in,clip,keepaspectratio]{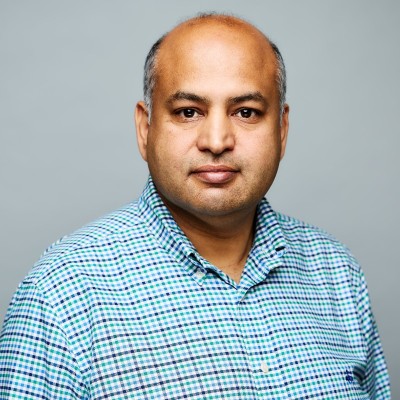}}]{Dr. Muhammad Faryad} is an associate professor of physics at LUMS. He joined LUMS in July 2014. Before that, he was a postdoctoral research scholar at the Pennsylvania State University from 2012 to 2014. He obtained his MSc and MPhil degrees in electronics from the Quaid-i-Azam University in 2006 and 2008, respectively, with certificates of merit in both degrees. He obtained his PhD degree in engineering science and mechanics from the Pennsylvania State University in 2012 with the best dissertation award by the university. He was awarded the Galleino Denardo award by the Abdus Salam International Center of Theoretical Physics (ICTP) in 2019 and the Early Career Achievement award by the department of engineering science and mechanics at the Pennsylvania State University in 2021.
\end{IEEEbiography}
\EOD
\end{document}